
The TeX file follows:

\magnification = \magstep1
\baselineskip = 12 pt

\hskip 8 cm
U.T.F. 332 - July 1994

\vskip 2 cm

\centerline{\bf VACUUM CORRELATIONS AT GEODESIC DISTANCE}
\centerline{\bf IN QUANTUM GRAVITY}

\vskip 4 cm

\centerline{Giovanni Modanese}
\centerline{Gruppo Collegato I.N.F.N. - Trento}
\centerline{Dipartimento di fisica dell'Universit\`a}
\centerline{I-38050 POVO (TN)}

\vskip 4 cm

\centerline{\bf Abstract}

\bigskip

The vacuum correlations of the gravitational field are highly non-trivial
to be defined and computed, as soon as their arguments and indices do not
belong to a background but become dynamical quantities. Their knowledge
is essential however in order to understand some physical properties
of quantum gravity, like virtual excitations and the possibility of
a continuum limit for lattice theory. In this review the most recent
perturbative and non-perturbative advances in this field are presented.
(To appear on Riv. Nuovo Cim.)

\vfill\eject

\centerline{\bf CONTENTS}

\bigskip
\noindent 1. \ INTRODUCTION

\medskip
\noindent 1.1 \ Vacuum correlations in the absence of gravity ... 3

\noindent 1.2 \ Gravitational correlations in a physical gauge ... 5

\noindent 1.3 \ Invariant gravitational correlations ... 6

\bigskip
\noindent 2. \ THE MANDELSTAM COVARIANTS

\medskip
\noindent 2.1 \ The original definition ... 8

\noindent 2.2 \ The computation of Tsamis and Woodard ... 9

\bigskip
\noindent 3. \ CLASSICAL RADIAL GAUGE AND APPLICATIONS

\medskip
\noindent 3.1 \ Main features ... 11

\noindent 3.2 \ Inversion formulas ... 14

\noindent 3.3 \ Translation of the origin ... 16

\noindent 3.4 \ Solutions of (2+1)-gravity ... 18

\noindent 3.5 \ Motion of test particles in a fluctuating field ... 23

\bigskip
\noindent 4. \ RADIAL PROPAGATORS

\medskip
\noindent 4.1 \ The gauge projectors method ... 27

\noindent 4.2 \ Electrodynamics. Regularity properties ... 32

\noindent 4.3 \ Gravitational propagators ... 36

\noindent 4.4 \ Relation to the Mandelstam covariant. Significant components
... 39

\bigskip
\noindent 5. \ GAUGE-INVARIANT CORRELATIONS

\medskip
\noindent 5.1 \ Geometrical definitions ... 42

\noindent 5.2 \ Dynamics and perturbation scheme ... 44

\noindent 5.3 \ Two-point functions ... 45

\noindent 5.4 \ Wilson loops ... 47

\bigskip
\noindent 6. \ FURTHER PROPERTIES OF THE WILSON LOOPS

\medskip
\noindent 6.1 \ Geometrical and physical definitions ... 50

\noindent 6.2 \ The dumbbell correlation function ... 52

\noindent 6.3 \ Geodesic corrections ... 54

\noindent 6.4 \ The static potential energy ... 57

\bigskip
\noindent 7. \ LATTICE GRAVITY

\medskip
\noindent 7.1 \ $(R+R^2)$-gravity ... 59

\noindent 7.2 \ Regge calculus ... 60

\noindent 7.3 \ The simulations of Hamber ... 61

\noindent 7.4 \ The gauge approach and others ... 63

\bigskip
\noindent REFERENCES ... 66

\vfill\eject

\centerline{\bf 1. INTRODUCTION.} \bigskip

The aim of this paper is that of giving an introduction to the issue of
vacuum correlations in quantum gravity and a review of the main advances
recently made in this field. The whole subject is quite new and in rapid
development, so the present account can only be considered as provisional.

{}From the physical point of view, this is a very interesting
field of research, since it concerns the physical modes of
propagation of the gravitational field and their quantization,
the structure of the vacuum state and
the general problem of the observables in quantum gravity.

It is known that a completely
consistent theory of quantum gravity has not been established yet.
Some of the autors we shall mention in this paper really work in this
direction: the simplicial quantum gravity of Hamber (Chapter 7), for instance,
aims to put Einstein's gravity on a solid quantum basis by simulating
numerically its non-perturbative behaviour; also the original purpose
of the work of Tsamis and Woodard (Chapter 2) was that of finding the ``true''
observables of quantum gravity, free of ultraviolet divergences.

Our philosophy throughout this paper, however, will be mainly
that of regarding quantum gravity as an effective
quantum field theory, which has General Relativity as its classical limit, but
which could go over to some more fundamental theory at very short distances.
Our approach will be in some sense ``phenomenological'', as we believe that
knowledge of the properties of the gravitational
vacuum correlations and other observables can help in guiding
the fundamental research.

{}From the technical point of view, the issue is confronted either in a
suitable physical gauge (the Mandelstam covariant and the radial gauge, Ch.s
2, 3, 4) and in gauge-invariant form (Ch.s 5, 6, 7).
In the following of this Introduction, after giving some general
definitions, we shall survey both approaches giving an overall outline
of the paper. A more detailed outline for each chapter can be found
at the beginning of the chapter itself.

\bigskip \noindent
{\bf 1.1  Vacuum correlations in the absence of gravity.} \bigskip

The vacuum correlation functions are very important quantities in quantum field
theory. It was shown by Wightman many years ago
[Wightman, 1956] that the knowledge of all
the correlation functions of a scalar field $\phi(x)$ gives a complete
information about the quantum dynamics of the field. More recently, in the
lattice formulations of euclidean quantum field theory, the field correlations
characterize the ``phases'' of the equivalent statistical system. Typically,
they behave like $\langle \phi(x) \phi(y) \rangle \sim e^{-|x-y|/\xi}$ outside
the region of phase transition, and like $\langle \phi(x) \phi(y) \rangle \sim
|x-y|^{-m}$ in correspondence of the transition (where the continuum limit is
recovered; see Chapter 7).

In general, for vacuum correlations of a quantum field theory whose fields are
denoted by $\Phi_A(x)$, $\Phi_B(x)$ ... we mean the expectation value on
the vacuum state of the product of the fields:
$$\eqalignno{
  G_{AB...}(x,y...) & = \langle \Phi_A(x) \Phi_B(y) ... \rangle_0 =
  {1 \over z} \int d[\Phi] \, e^{(i/\hbar) S[\Phi]} \Phi_A(x) \Phi_B(y) ...
  &(1.1)
\cr}$$
In this expression the indices $A,\, B,\, ...$ are Lorentz indices (possible
``internal'' indices are understood) and the coordinates $x,\, y,\, ...$ are
the coordinates of Minkowski space.

In the case of free fields -- for instance, scalar fields -- the
function $G(x,y)$ is computed exactly and is called the Pauli-Jordan
distribution
$$\eqalignno{
  \langle \phi(x) \phi(y) \rangle_0 &= {1 \over (2\pi)^3} \int
  d^4k \, e^{-ikx} \theta(k^0) \delta(k^2-m^2) .
\cr}$$
 From this expression one can verify that also in a ``trivial'' system
like a free field the vacuum correlations exibit interesting
properties: for instance, they do not vanish when $(x-y)$ is
spacelike, although all the observables in $x$ and $y$ commute in
this case.

Intuitively, from the statistical point of view, the product of two real random
variables vanishes on the average if the two variables are independent, and
have
thus the same probability to assume sign + or --, with various amplitudes. But
since the field $\Phi$ propagates, its value in $y$ depends on that in $x$, and
averaging on all the configurations like in (1.1) we find that the
correlation increases like $|x-y|^{-m}$ when the two points get closer and
closer. Of course, in the functional integral (1.1) all the field
configurations are contained, not just the classical
configurations which minimize the action; but if $\Phi$
oscillates too rapidly at small distances in some configuration, the kinetic
term of the action is very big and that configuration is strongly suppressed.

As we saw before, the function $G$ possesses in general, in the absence of
gravity, some Lorentz indices. For instance, for the correlation of two
vector fields we have
$$\eqalignno{
  G_{\mu \nu}(x,y) &= \langle V_\mu(x) V_\nu(y) \rangle_0 .
\cr}$$
This means that the field $V_\mu(x)$ and the field $V_\nu(y)$ can be
``rotated'' by a Lorentz matrix -- the same in $x$ and in $y$ -- and the
correlation function also transforms correspondingly; for instance,
the correlation $\langle V_1(x) V_2(y) \rangle_0$ becomes, in a rotated
reference frame, $\langle {V'}_2(x') {V'}_3(y') \rangle_0$.
What allows us to compare in a consistent way two components of the
field in two distinct points is the ``rigid'' structure of Minkowski
space. Also it is obvious -- still in the absence of gravity -- that
the distance $|x-y|$ is a fixed quantity, that does not depend on the
field $V$ itself.

If we allow a local invariance in our theory, the situation changes
remarkably. This can be seen already in the usual gauge theories, based
on ``internal'' symmetries. In this case, the two fields $A_\mu(x)$
and $A_\nu(y)$ which have to be correlated gauge-transform independently in $x$
and
$y$. This reflects itself in the fact that the correlation function is
gauge-dependent. However, the coordinates $x$ and $y$ and the Lorentz
indices are still referred to a rigid background.

In the case of gravity, the analogue of the local gauge transformations
are the coordinates transformations. There is no fixed background and
the only way to give a physical meaning to a correlation function
(except for the pure topological content) is that of comparing the
fields in two different points by parallel transport. As the distance
of the points, the geodesic distance must be considered. But either in the
evaluation of the parallel transport and of the geodesic distance
the same dynamical field enters, whose correlations we are looking for. It is
clear that this will introduce in general new technical difficulties.

\bigskip \noindent
{\bf 1.2  Gravitational correlations in a physical gauge.} \bigskip

If we are interested in all the ``components'' of the correlation functions, we
have to compute them in a suitable preferred physical gauge. A first attempt to
the computation of this kind of ``physical Green functions''
for gravity was done by Tsamis and Woodard (Section 2.2).
They computed the vacuum correlations of the Mandelstam
covariant (Section 2.1) to one-loop perturbative order, in the hope that these
quantities, being in principle observable, would have been free of divergences.
Unfortunately, as a result of their hard work (one of the biggest perturbative
computations in quantum gravity) they found that the usual divergences are
still there, and some new ones appear, which are peculiar of the quantity under
consideration.

On the other hand, the Mandelstam formalism cannot be considered in general as
a
gauge-fixing and thus many standard techniques of QFT cannot be applied to this
case. Independently of Tsamis and Woodard, Toller [1988] suggested to try to
define the fields themselves with reference to geodesics and parallel
transport, in order
to give a direct physical significance to their correlation. This led to the
classical radial gauge (Chapter 3), which, on account
of its simplicity and of the field-potential inversion formula (Section 3.2),
has proved to be useful for other purposes too (Sections 3.4, 3.5).

In this gauge the fields are defined in such a way that the vierbein
$e^a_\mu(\xi)$ in a point $\xi$ coincides with that which is parallel
transported from the origin to $\xi$ along a straight line. Moreover, this line
is the (locally unique) geodesic which connects the origin to the point $\xi$
and the geodesic distance from the origin is given by $|\xi|$.
Thus in this gauge the correlation function $\langle e^a_\mu(0) e^b_\nu(\xi)
\rangle$ automatically satisfies the criteria we enunciated at the end
of Section 1.1.

Inserting the radial gauge fixing in some version of the functional integral of
gravity, one may compute the corresponding propagator. This propagator gives
automatically the lowest-order physical correlation function and allows in
principle to compute this correlation to any order.

As expected, the propagator in the radial gauge is not trivial
(including in fact the geodesics and parallel transport) and contains
some singularities (Sections 4.2, 4.3). By means of a suitable
projection procedure, however, it is possibile to gain an insight into the
singularities and to eliminate them.
It is believed now that the ``additional'' divergences which appear in the
mentioned work of Tsamis and Woodard could be eliminated through a similar
procedure. There is no special connection between these
divergences and the usual ultraviolet divergences of quantum gravity.

Summarizing, we have in the radial gauge a complete solution
to the problem of computing perturbatively the vacuum
correlations at geodesic distance. The limitations are those
inherent to perturbation theory in quantum gravity (i.e., non
renormalizability for the Einstein theory, or non unitarity
for the $(R+R^2)$-theory), and those due to the mathematical
complexity of the radial propagator. On the other hand, some
problems typical of algebraic gauges, like the need for
special prescriptions in order to define uniquely the propagator
[see for instance Bassetto, Nardelli and Soldati, 1991; Gaigg
and Kummer, 1990], do not affect the radial gauge. Also, in
four dimensions the four-particles interaction vertex is likely
to vanish in radial gauge since the four fields should be
at the same time orthogonal to $\xi$ and between themselves.

\bigskip \noindent
{\bf 1.3  Invariant gravitational correlations.} \bigskip

The formal advances in the determination of the radial propagators
do not say us much on some physical features which characterize
the gravitational vacuum correlations. In particular, we may ask:
which are the main differences with respect to the other gauge
fields? Which are the modes of the field which are really
correlated? Is the gravitational interaction truly due to the
exchange of virtual particles, and thus connected to the vacuum
fluctuations in the same way like the other fundamental interactions
do?

In Chapters 5 and 6, the first attempts to answer these questions
are presented. The crucial quantities under investigation are the
Wilson loops of the connection, which are scalar quantities and
give a sensible information about the geometry of spacetime.
The computations we present here start from a flat background and
take into account corrections up to order $\hbar$. This is justified
on length scales much larger than Planck's length; at smaller
distances, non-perturbative methods must be employed (see Chapter 7).

In Chapter 5 the Wilson loops of the gravitational connection are
defined (Section 5.1) and computed to order $\hbar$ in Einstein's
theory (Section 5.4). The invariant two-point functions of the
Riemann curvature are computed too (Section 5.3).

The Wilson loops vanish and this peculiar result, when analyzed
in terms of the relevant gauge group for the euclidean theory,
$SO(4)$ (Section 6.1), shows that the functional integral of
quantum gravity does not contain to order $\hbar$ any configuration
with localized curvature. Such a behaviour contrasts strongly with
that of usual gauge fields, and may be interpreted as the absence,
to this order, of true virtual gravitons (those which appear in the
diagrams for the cross sections or similar represent excitations
of the metric which do not carry curvature!). It can be shown
that this result holds more generally than just for Einstein gravity,
but relies on the flat background.

Chapter 6 contains also some more material on the Wilson loops:
the correlation function of two loops to leading order and its
``geodesic corrections'' (Sections 6.2, 6.3). In Section 6.4 the formula for
the
static potential energy in quantum gravity is presented.

Finally, Chapter 7 is dedicated to lattice gravity, with special
regard to the ``quantum Regge calculus'' of Hamber. The outcome of
the recent Montecarlo simulations is very promising: the transition
between the two phases of the lattice system appears to be of
second order, and this has been confirmed by the behavior of the
(simple) correlation functions which have been computed up to now.

\bigskip \noindent
{\bf Notations.} \bigskip

We refer mainly to spacetime dimension 4; otherwise the
dimension is denoted by $N$ and assumed to be bigger than 2.
The units are such that $c=\hbar=1$; in some formulas where
the classical contributions need to be distinguished from the
quantum ones, $\hbar$ is written explicitly. The symbol $\kappa$
(or sometimes $\ell$) denotes Planck's length $\sqrt{16 \pi G}$.
In some of the reviewed papers the metric has euclidean
signature, in other it is minkowskian; we have used both
notations, although for gravity the correspondence between the
euclidean theory and the theory on Minkowski space is not
completely established yet [see for instance Mazur and Mottola,
1990].

The derivatives of the
form $\partial_\mu$ are taken with respect to the whole
argument of the function. A suffix like $[ab]$ denotes
antisymmetrization (without the factor ${1 \over 2}$).
The notation ``$o(\ )$'' means ``of higher order than'';
``$O(\ )$'' means ``of the same order as''.

For recent general reviews on quantum gravity and references see for
instance Alvarez [1989] or the lectures of the Les Houches Summer
School, 1992.

\vfill\eject

\centerline{\bf 2. THE MANDELSTAM COVARIANTS.} \bigskip

In this first, ``historical'' chapter we shall first introduce the
general classical Mandelstam covariants according to the original
definition [Mandelstam, 1962], with a special emphasis on their
geometrical meaning (Section 2.1); then we shall present in short
the perturbative computation, due to Tsamis and Woodard,
of a special kind
of two-point functions inspired to Mandelstam's covariants (Section 2.2).

\bigskip \noindent
{\bf 2.1  The original definition.}\bigskip

In the early Sixties it was already clear to a few quantum theorists that in a
true theory of quantum gravity the gravitational field had to play the hard
role of furnishing coordinates to spacetime while being at the same time a
quantum object.

In the mentioned paper, Mandelstam dicussed a technique for defining spacetime
and its point without using coordinates, but replacing them with ``paths'',
which may be themselves influenced by the (eventually quantized) gravitational
field.

His pioneering work, although technically heavy in some points, had a great
influence on the subsequent developments of gauge theories. Also, it is very
remarkable that through his paths-based formalism Mandelstam succeeded in
[1968] in finding the correct Feynman rules for gravity and Yang-Mills
theories;
these rules were derived at the same time by De Witt [1967 b, c] and by Faddeev
and
Popov, whose functional-integral technique has become later the standard one.

In the Mandelstam formalism the coordinates of spacetime do not appear. A point
is defined as the end of a path, which in turn is specified by an infinite list
of ``small steps'' to be taken in sequence by an observer, with reference to
his local frame. After each infinitesimal step, the frame is parallel
transported to the new location, and so on.

Thus the fields are not functions of four coordinates $(x_0, \, x_1, \, x_2, \,
x_3)$, but functionals of ``paths'' $P$. On the other hand, the paths $P$ are
purely geometric, intrinsic objects, while the coordinates $(x_0, \, x_1, \,
x_2, \, x_3)$ are subject in General Relativity to a vast class of
tranformations without
change of the physical results.

The basic gravitational field is the anholonomic Riemann curvature
$R_{abcd}(P)$. The components of $R$ are referred to the parallel transported
local reference frame, and are then coordinate-independent. The fields can be
canonically quantized and their correlation functions defined, through a
pretty complicated procedure.

In a recent work of Teitelboim [1993] a more accurate mathematical definition
of
the classical Mandelstam's theory is given, also showing how the field
equations in path space may be derived from an action principle and how is it
possible to decide consistently ``when two paths end at the same point'' -- a
concept which was undefined in the original Mandelstam's theory.

An interesting possibility within the Mandelstam formalism is that of fixing
the paths through some prescription. A natural prescription, corresponding to
straight lines in flat spacetime, is that of taking geodesic paths. In fact,
the direction of a geodesic line, as observed in the free-falling local system,
is constant, so the ``infinite'' list defining the path $P$ is very much
simplified. This geometrical construction is explicitly implemented -- starting
from a given central point -- in the radial gauge (Chapters 3, 4), which
thus establishes a connection between a special case of Mandelstam formalism
and the usual formalism of General Relativity or Einstein-Cartan theory.

\bigskip \noindent
{\bf 2.2 \ The computation of Tsamis and Woodard.}

The starting point of this work [1992] was that of considering the correlation
of two Mandelstam-like fields, placed at the ends of a geodesic line of fixed
length. This correlation was written by inserting the suitable matrices
of parallel transport and evaluated perturbatively to one-loop order.

The motivation for this computation originated in the troubles of
quantum gravity with the unrenormalizable ultraviolet divergences.
Some parallels
with QED suggested that these divergences could cancel when the true
``physical Green's functions'' are computed, instead of generic quantities
which are subject to coordinates transformations and are thus
unphysical.

The paper of Tsamis and Woodard is very remarkable from the ``phylosophical''
point of
view and for its quantum-field content.

``Instead of changing the fundamental dynamical principle'', they
proposed ``to alter the way in which physical questions are asked.
This idea is hardly new, nor is it unique to quantum gravity [...].
The particular detail we propose to treat more carefully is the
physical coordinate system. We shall do this by introducing a new
interpolating field, the Mandelstam covariant [...]. The
Mandelstam covariant is not a local, invertible redefinition of the
usual field [...]. It follows that the Green's functions deriving
from the Mandelstam covariant need not interpolate the usual S-matrix,
nor contain the usual ultraviolet divergences. Indeed we will
prove that the usual result {\it does} change. Unfortunately it
does so in the direction of greater divergence; however, the
reason for this seems to be correctable ....''

And going on with the citation: ``The issue for us is not so much
that local Green's functions depend upon the gauge but rather that
this gauge dependence makes them unwieldy experiments. In fact
none of them is a reliable probe of quantum gravity, because they call for
measurements that can never actually be made [...] and must we
be very concerned over the fact that the theory predicts a
divergent result for them?''

Tsamis and Woodard write the Mandelstam covariant as
$$\eqalignno{
  {\cal R}_{abcd}[e](V,x) &=  {\bar{M}}^k_a[e](V,x) {\bar{M}}^l_b[e](V,x)
  {\bar{M}}^m_c[e](V,x) {\bar{M}}^n_d[e](V,x)
  {\bar{R}}_{klmn}(\chi[e](1,V,x))
\cr}$$
where the bar over the Riemann tensor denotes that its indices are
local Lorentz:
$$\eqalignno{
  {\bar{R}}_{abcd}(z) &= e_{\alpha a}(z) e^\beta_b(z)
  e^\gamma_c(z) e^\delta_d(z) R^\alpha_{\beta \gamma \delta}(z) .
\cr}$$
By $\chi[e](\tau,V,x)$ the geodesic path is meant which carries the
Mandelstam observer from his origin $x$ to the point at which he measures
the curvature. $\tau$ is the geodesic parameter and $V$ the initial
direction. It is always assumed that the fundamental dynamical
variable is the vierbein $e$. $M$ is the matrix of the parallel
transport:
$$\eqalignno{
  {\bar{M}}^a_b[e](V,x) &= e^a_\alpha(\chi(1)) M^\alpha_\beta[e](V,x)
  e^\beta_b(x) \cr
  M^\alpha_\beta[e](V,x) &= {\rm P \ exp} \ \left(
  - \int_0^1 d\tau {\dot{\chi}}^\mu(\tau) \Gamma^\alpha_{\mu \beta}
  (\chi(\tau)) \right).
\cr}$$

It can be proved that ${\cal R}$ can not be related to $R$ by any local
field transformation. The reason for this is essentially the
appearence in ${\cal R}$ of the non-local matrix $M$.

The one-loop computation of the correlations of ${\cal R}$ is extremely long.
The expectation values are organized into 30 ``$k$-point functions''
which were evaluated using the dimensional regularization. The
``$V$-ordering'' corrections (corresponding to the path ordering)
were then added, and the final result expressed as a linear combination
of 8-index ``master tensor structures''.

At each order the result breaks up into ``uncorrected terms''
(corresponding to the naive flat space perturbations), ``length
corrections'', ``index corrections'' and ``length and index corrections''.
This is precisely what is expected from the physical features of the
gravitational correlation functions (see the Introduction, and Section 6.3
for a concrete example).

The ``additional singularities'' which affect the result have the same
form like those of the ``$\langle P^0 \, P^0 \rangle$ propagator''
of radial gauge. For this reason, it is very likely that they could
be eliminated, in principle, by a suitable projection procedure
(compare Section 4.2, 4.3). In practice, however, the algebraic
structure of the Mandelstam covariant employed by Tsamis and Woodard is
too complicated
to allow this. For a comparison with the radial-gauge formalism,
see the beginning of Section 4.4.

\vfill\eject

\centerline{\bf 3. CLASSICAL RADIAL GAUGE AND APPLICATIONS.} \bigskip

In this chapter an introduction to the classical radial gauge
is given. In Section 3.1 the gauge condition is defined (in the
vierbein formalism) and its attainability is shown. In Section
3.2 we write the formulas which allow to express the radial vierbein
and the radial connection in terms of the Riemann tensor and
of the torsion (the so-called inversion formulas). In Section
3.3 the transformation properties of the radial fields under
rotation or translation of the central vierbein are given.
In Section 3.4 we illustrate in short a remarkable application
of the radial gauge to (2+1)-gravity and finally, in Section
3.5, an application to the description of the geodesic motion
of test particles in a fluctuating gravitational field.

\bigskip \noindent
{\bf 3.1 \ Main features.} \bigskip

The radial gauge $x_\mu A_\mu(x) = 0$
has been proposed many years ago by Fock [1937] and Schwinger [1952, 1989; for
a complete bibliography see M.\ and Toller, 1990] for electrodynamics or
Yang-Mills theory. Its most interesting
property is the so called fields-potentials inversion formula. Let us suppose
that the field strength $F_{\mu \nu}(x)$ is known. In order to compute from
$F_{\mu \nu}(x)$ the vector potential $A_\mu(x)$ in radial gauge, one must
solve
the system
$$\eqalignno{
        & x_\mu A_\mu(x)=0 \cr
        & \partial_{\mu} A_\nu(x)-\partial_{\nu} A_\mu(x)
        + ig[A_\mu(x),A_\nu(x)] = F_{\mu \nu}(x).
\cr}$$
Its general solution is
$$\eqalignno{
  A_\mu(x)&=\int_0^1 d\lambda \, \lambda x_\nu F_{\nu \mu}(\lambda x)
  + A^{Hom}_\mu(x),
\cr}$$
where the homogeneous part $A^{Hom}_\mu(x)$ is solution of the system
$$\eqalignno{
        & x_\mu A^{Hom}_\mu(x)=0 \cr
        & \partial_{\mu} A^{Hom}_\nu(x)-\partial_{\nu} A^{Hom}_\mu(x)
        + ig[A^{Hom}_\mu(x),A^{Hom}_\nu(x)] = 0 ;
\cr}$$
that is, we have
$$\eqalignno{
  A^{Hom}_\mu(x)&= {\partial f(x) \over \partial x^\mu} , &(3.1)
\cr}$$
being $f$ an arbitrary homogeneous function of degree $0$. If one wants
$A_\mu(x)$ to be regular at the origin, $f$ must be set equal to zero. In this
case, the gauge condition can be regarded as ``complete'', in the sense that no
residual gauge is present. Until the introduction of the general radial
projectors (Section 4.2), we shall stay here in the hypotesis that all fields
are
regular at the origin; thus our inversion formula will be simply given by the
``0-1 projector''
$$\eqalignno{
  A_\mu(x)&=\int_0^1 d\lambda \, \lambda x_\nu F_{\nu \mu}(\lambda x). &(3.2)
\cr}$$
Let us now consider the case of gravitation. We recall that in order to
describe the
gravitational field it is sometimes convenient to use, instead of the metric
tensor $g_{\mu \nu}(x)$, the vierbein fields $e^a_\mu(x)$ and $e^\mu_a(x)$,
which are defined by
$$\eqalignno{
  & e^a_\mu(x) e^b_\nu(x) \, \delta_{ab} = g_{\mu \nu}(x); &(3.3) \cr
  & e^\mu_a(x) e^a_\nu(x) = \delta^\mu_\nu; \qquad
  e^\mu_a(x) e^b_\mu(x) = \delta^b_a. &(3.4)
\cr}$$
The holonomic indices $\mu$, $\nu$,... and the ``internal'' indices $a$, $b$,
... range, in general, between $1$ and $N$, where $N$ is the dimension of
spacetime; the $N$ vectors $\{ e_1(x),...,e_N(x) \}$ can be thought to
represent a local reference frame at any point $x$. Condition (3.3)
determines $e^a_\mu(x)$ up to a local rotation; the gauge
potentials $\Gamma^{ab}_\mu$ and field strengths $R^{ab}_{\mu \nu}$
corresponding to this local invariance are related
to the Christoffel symbol and to the Riemann tensor by the
formulas\footnote\dag{The connection $\Gamma^a_{b \mu}$
is viewed as a variable independent
of the vierbein; so this formalism is often referred to as ``first order'',
while the metric formalism is defined to be of the ``second order''.
As we shall illustrate below, it is possible to define a radial gauge condition
also
in the second order formalism. We do not dwell here upon the relationship
between the two formalisms and the role of the
torsion [see, for instance, Hehl et al., 1976]. We notice, however,
that the radial gauge in the first order formalism is meaningful also in spaces
without metric connection and for gauge groups different from the Poincar\'e
group.}

$$\eqalignno{
  & \Gamma^\mu_{\rho \sigma} = \Gamma^a_{b \, \sigma} \, e^b_\rho \, e^\mu_a
  + e^\mu_a \, \partial_\sigma e^a_\rho; &(3.5) \cr
  & R^\mu_{\nu \rho \sigma} = e^\mu_a \, e^b_\nu \,
  R^a_{b \rho \sigma}. &(3.6)
\cr}$$

In this formalism, two gauge fixing conditions are needed, in order to
eliminate
the freedom to make diffeomorphisms (i.e., coordinate transformations), and
the freedom for local rotations of the vierbein. Recently a radial gauge
condition has been
proposed also for gravity. It has the form [M.\ and Toller, 1990]
$$\eqalignno{
  & \xi^\mu \, \Gamma^a_{b \, \mu}(\xi) = 0, &(3.7a) \cr
  & \xi^\mu \, e^a_\mu(\xi) = \xi^\mu \, \delta^a_\mu. &(3.7b)
\cr}$$
The gauge-fixing conditions (3.7a) and (3.7b) have a simple
geometrical interpretation. Condition (3.7a) means that the tetrads are
parallel transported from the origin along the straight lines of the form
$$\eqalignno{
  & \{ s \xi, \ 0 \leq s \leq 1\} ; &(3.8)
\cr}$$
condition (3.7b) means that these lines are auto-parallel, i.e. they are
geodesic lines (in the absence of torsion). The coordinates $\xi$ coincide then
with the well-known ``normal coordinates'' [see for ex. Kobayashi and Nomizu,
1969].

It is possible to give explicit formulae for the calculation of the normal
coordinates and of the parallel-transported tetrads of the radial gauge.
To this end, we consider an arbitrary
coordinate system $x^\mu$ and we indicate by $x_0^\mu$ the coordinates of the
origin $P_0$. We assume that the holonomic coefficients of the connection $\hat
\Gamma^\mu_{\nu \rho}(x)$ are known. We indicate by $\hat e^\mu_a$ the
components of the tetrads with respect to the natural holonomic basis
determined by the coordinates $x^\mu$. The components of the dual tetrads in
the natural basis determined by the normal coordinates $\xi^\mu$ can be
computed  by means of the formula
$$\eqalignno{
  e_\mu^a(\xi) &=  {\partial x^\rho(\xi) \over \partial \xi^\mu}
  \, \hat e^a_\rho(\xi).
\cr}$$
Since the tetrads are parallel-transported along the line (3.8), we have
$$\eqalignno{
  {d\hat{e}_b^\mu(\lambda) \over{d\lambda}} &=
  - \hat\Gamma^\mu_{\sigma \tau}[x(\lambda)] \,
  \hat{e}_b^\sigma(\lambda) \, \hat{e}_a^\tau(\lambda) \, v^a &(3.9)
\cr}$$
and the fact the the vector with constant anholonomic components $v^a$  is
tangent to the line (3.8) is expressed by the equation
$$\eqalignno{
  {dx^\mu(\lambda) \over d\lambda} &=
  \hat{e}_a^\mu(\lambda) \, v^a . &(3.10)
\cr}$$

The equations (3.9) and (3.10), with the initial conditions
$x^\mu(0) = x^\mu_0$ and $\hat e_a^\mu(0) = e_a^\mu(P_0)$, determine the
quantities $x^\mu$ and $\hat{e}_b^\mu$ as functions of $\lambda$ and of $v^a$.
However, it is easy to see that they depend on a particular combination of
these variables, namely on the normal coordinates $\xi^a$=$\lambda v^a$. If we
are considering a metric space and a metric connection, we can choose an
orthonormal initial tetrad $e_a(0)$  and all the parallel-transported tetrads
are automatically orthonormal.

It is convenient to transform these differential equations with their
initial conditions into the following pair of coupled integral equations
$$\eqalignno{
  x^\mu(\xi) &= x_0^\mu +
  \xi^a \int_0^1  \hat{e}_a^\mu(\lambda \xi)\, d\lambda, &(3.11) \cr
  \hat{e}_b^\mu(\xi) &= \hat{e}_b^\mu(0)-\xi^a \int_0^1
  \hat\Gamma^\mu_{\sigma \tau}[x(\lambda \xi)] \,
  \hat{e}_b^\sigma(\lambda \xi) \, \hat{e}_a^\tau(\lambda \xi)\, d\lambda.
  &(3.12)
\cr}$$
It is possible to solve these equations perturbatively to any desired order in
$\hat\Gamma$, by substituting at each step the lower order solution in the left
hand side integrals. This shows that the gauge condition is ``attainable'' in
the sense that given the fields in an arbitrary gauge, it is always possible to
compute the corresponding fields in the radial gauge. A simple application of
these equations is the computation to lowest order in $\hat{\Gamma}$ of the
coordinate transformation $x^\mu(\xi)$. One obtains
$$\eqalignno{
  x^\mu(\xi) &= \xi^\mu - \xi^\rho \xi^\sigma \int_0^1 dt \, (1-t)
  \, {\hat{\Gamma}}^\mu_{\rho \sigma}(t\xi) + o(\hat{\Gamma}^2). &(3.13)
\cr}$$
This formula will be used in Section 3.5 and 6.3.

Note that if we change the initial conditions by performing a Lorentz
transformation $L$ of the tetrad $e_a(0)$, (or more generally a transformation
of the gauge group) we get a new solution obtained from the old one by means of
the same Lorentz transformation $L$ applied to all the tetrads and to the
normal coordinates. This property is not trivial, because the parallel
transport is generally non-commutative. A translation of the origin $P_0$
affects the normal coordinates and the parallel-transported tetrads in a more
complicated way, which will be illustrated in Section 3.3.

The radial gauge conditions (3.7a) and (3.7b)  can be regarded, in a
sense, as an operational prescription which permits one to locate the measuring
instruments in a neighbourhood of the observer, who lies at the origin $P_0$.
In fact a simple way to explore this neigbourhood is to send from the origin
many ``space-probes'' carrying clocks, gyroscopes and measuring instruments. A
space-probe will be launched with four-velocity $v^a$ with respect to the given
tetrad $e_a(0)$ and, if $\tau$ is the proper time measured by the clock, $\xi^a
= \tau v^a$ are the normal coordinates (in the absence of torsion).  Of course,
in Minkowski spacetime only the interior of the future cone can be explored in
this way.

Any space-probe will be able, in principle, to measure by local experiments the
anholonomic components of the fields at any point $\xi$. Moreover, let us go on
with our {\it Gedankenexperiment} and suppose that any space-probe emits all
the time some peculiar light signals along the axis of its local frame; the
signals are received and recorded by the neighboring space-probes. By
collecting all the data, the mentioned ``central observer'' will thus be able
to compute also the vierbein $e^a_\mu(\xi)$.

\medskip
Finally, we remind that the radial gauge can be introduced also in the
so-called second order formalism, that is, as a condition on the metric tensor
[Menotti, M.\ and Seminara, 1993]. In this case it is defined by
$$\eqalignno{
  \xi^\mu g_{\mu \nu}(\xi) &= \xi^\mu \eta_{\mu \nu} . &(3.14)
\cr}$$

A number of properties of the radial gauge in the tetrad formalism still
hold in the metric formalism. We enumerate them without proof:

\noindent -- if $g_{\mu \nu}(\xi)$ is regular at the origin, then we have
$g_{\mu \nu}(0)=\eta_{\mu \nu}$;

\noindent -- coordinates $\xi$ satisfying (3.14) are normal coordinates;

\noindent -- the gauge is attainable, under some regularity hypotesis.

An equation analogous to the inversion formulas (3.17), (3.19)
holds just in the linearized theory. If we denote by $R^L$ the
linearized Riemann tensor, we have
$$\eqalignno{
  h_{\nu\beta}( x) &= -2 \,  x^\alpha  x^\mu \int^1_0
  d\tau \, \tau \int^1_0 d\lambda \, \lambda^2 \,
  R^{L}_{\mu \nu\alpha \beta}(\lambda \tau  x) = \cr
  &= -2 \,  x^\alpha  x^\mu \int^1_0 d\lambda \, \lambda (1-\lambda)
  \, R^{L}_{\mu \nu \alpha \beta}(\lambda  x),
\cr}$$
provided $|R^L(x)|<|x|^{-2+\varepsilon}$ as $x \to 0$.

\bigskip \bigskip \noindent
{\bf 3.2  Inversion formulas.} \bigskip

We shall now derive from the radial gauge conditions (3.7a) and
(3.7b) two formulas analogous to (3.2), which give the field
potentials $\Gamma^a_{b \mu} (\xi)$ and $e^a_\mu (\xi)$ in terms of the Riemann
tensor $R^a_{b \mu \nu}(\xi)$ and the torsion tensor $S^a_{\mu \nu}(\xi)$.

We would like first to make the following observation. Remember that in this
chapter we work in the hypotesis that the fields are regular (and
differentiable) at the origin. Thus, taking the derivative of eq.s (3.7a)
and (3.7b) we obtain
$$\eqalignno{
  \Gamma^a_{b \mu}(0) &= 0, \qquad e^a_\mu(0)=\delta^a_\mu .
\cr}$$
These equations show that in the radial gauge the gauge potentials at the
origin take the values they have in a flat space. In other words, it is
possible to eliminate the gravitational field at a given point. This is a
formulation of the equivalence principle which is valid also in the presence of
torsion, when it is not possible to eliminate the holonomic connection
coefficients at a given point [see the references in M.\ and Toller, 1990].

Actually, one of the motivations which led to the formulation of the radial
gauge was the need of a generalization of the equivalence principle to
10-dimensional theories defined on group manifolds. To this end, it is crucial
for the gauge to be ``local'', in the sense that the inversion formulas should
involve only the fields in a neighbourhood of the origin. The formulas
of this subsection match this condition. In the following chapters we
shall see that, giving up the regularity of the fields at the origin and
exploiting the residual gauge, it is possible to write down some
formulas which are similar to those of this chapter, but contain integrals on
the domain $(1-\infty)$. In that case, the ``locality'' of the gauge condition
is lost.

\medskip
We remind that the Riemann tensor and the torsion tensor are given by the
expressions
$$\eqalignno{
  & R^a_{b \mu \nu} = \partial_\mu \Gamma^a_{b \nu}
  - \partial_\nu \Gamma^a_{b \mu} + \Gamma^a_{c \mu}
  \Gamma^c_{b \nu} - \Gamma^a_{c \nu} \Gamma^c_{b \mu}, &(3.15) \cr
  & S^a_{\mu \nu} = \partial_\mu e^a_\nu - \partial_\nu e^a_\mu +
  e^b_\nu \Gamma^a_{b \mu} - e^b_\mu \Gamma^a_{b \nu}. &(3.16)
\cr}$$
Multiplying (3.15) by $\xi^\mu$ we have from (3.7a)
$$\eqalignno{
  \xi^\mu R^a_{b \mu \nu} (\xi) &= \xi^\mu \partial_\mu \Gamma^a_{b \nu} (\xi)
  + \Gamma^a_{b \nu} (\xi).
\cr}$$
If we now put $\xi \rightarrow \lambda \xi$, we obtain
$$\eqalignno{
  {d \over {d \lambda}}
  \big( \lambda \Gamma^a_{b \nu} (\lambda \xi) \big) &=
  \lambda \xi^\mu R^a_{b \mu \nu} (\lambda \xi)
\cr}$$
and integrating we finally have
$$\eqalignno{
  \Gamma^a_{b \nu} (\xi) &=
  \xi^\mu \int_0^1 R^a_{b \mu \nu} (\lambda \xi) \lambda \,
  d\lambda. &(3.17)
\cr}$$
In a similar way, by multiplying (3.15) by $\xi^\mu$ and taking into
account condition (3.7b) we obtain
$$\eqalignno{
  \xi^\mu S^a_{\mu \nu} (\xi) &= \xi^\mu \partial_\mu e^a_\nu (\xi) +
  e^a_\nu (\xi) - \delta^a_\nu - \xi^b \Gamma^a_{b \nu} (\xi) ;
\cr}$$
by the same procedure we obtain
$$\eqalignno{
  {d \over {d \lambda}}
  \big( \lambda (e^a_\nu(\lambda \xi) - \delta^a_\nu )\big) &=
  \lambda \xi^{b} \Gamma^{a}_{b \nu}(\lambda \xi) +
  \lambda \xi^\mu S^{a}_{\mu \nu}(\lambda \xi)
\cr}$$
and
$$\eqalignno{
  e^a_\nu (\xi) &= \delta^a_\nu +
  \int_0^1  [\xi^b \Gamma^a_{b \nu}
  (\lambda \xi) + \xi^\mu S^a_{\mu \nu} (\lambda \xi)] \lambda \,
  d\lambda. &(3.18)
\cr}$$
By substituting (3.17) into (3.18) we have
$$\eqalignno{
  e^a_\nu (\xi) &= \delta^a_\nu +
  \xi^\mu \xi^{b} \int_0^1
  R^a_{b \mu \nu} (\lambda \xi) (1 - \lambda) \lambda \, d\lambda
  + \xi^\mu \int_0^1 S^a_{\mu \nu}
  (\lambda \xi) \lambda \, d\lambda. &(3.19)
\cr}$$
The equations (3.17) and (3.19) are the analogue of (3.2). The
gauge potentials they give satisfy the gauge conditions (3.7a) and
(3.7b) as a consequence of the antisymmetry of $R^a_{b \mu \nu}$ and
$S^a_{\mu \nu}$ with respect to the indices $\mu$, $\nu$. However, they  solve
(3.15) and (3.16) only if the functions $R^{a}_{b \mu \nu}(\xi)$ and
$S^{a}_{\mu \nu}(\xi)$ satisfy some conditions. These conditions have been
derived and used in the Yang-Mills case [Durand e Mendel, 1982] and a similar
treatment can be given in the case under consideration. If we substitute
(3.17) and (3.19) into (3.15), (3.16), after a long
calculation we find that they are equivalent to the following projected Bianchi
identities
$$\eqalignno{
  & \xi^\mu \sum_{\{\mu \nu \rho\}} (\partial_\mu R^{a}_{b \nu \rho} +
  \Gamma^{a}_{c \mu}
  R^{c}_{b \nu \rho} - \Gamma^{c}_{b \mu} R^{a}_{c \nu \rho} ) = 0, \cr
  & \xi^\mu \sum_{\{\mu \nu \rho\}} (\partial_\mu S^{a}_{\nu \rho} +
  \Gamma^{a}_{b \mu}
  S^{b}_{\nu \rho} - e_\mu^{b} R^{a}_{b \nu \rho}) = 0,
\cr}$$
in which the potentials $e^b_\mu$ and $\Gamma^a_{b \mu}$ are replaced by the
integral expressions (3.17), (3.19). We have indicated by
$\sum_{\{\mu \nu \rho\}}$ the sum over the cyclic permutations of the indices
$\mu$, $\nu$, $\rho$.

These conditions have a non local character, since they are expressed by
integro-differential equations. Also the Einstein field equations $e^\mu_a \,
R^a_{b \mu \nu}=0$ contain the potential $e^a_\mu$ and therefore take a non
local character if we want to express them in terms of the curvature alone.  In
conclusion, if we try to use the inversion formulas to eliminate the gauge
potentials from the theory (like in Mandelstam's formalism), we get very
complicated non local field equations.

\bigskip \noindent
{\bf 3.3  Translation of the origin.} \bigskip

If we adopt the radial gauge, the geometry of the space-time manifold in a
neighbourhood of the origin $P_0$ is completely described  by the functions
$e^a_\nu(\xi)$, $\Gamma^a_{b \nu}(\xi)$, which have to satisfy the gauge
conditions (3.7a), (3.7b) and the field equations of the theory we
are considering. In a similar way, the matter fields are completely described
by their components with respect to the tetrads, expressed as functions of the
normal coordinates $\xi$.   The only arbitrary elements in this description are
the choice of the origin $P_0$ and of the tetrad $e_a(P_0)$. Thus the fields in
the radial gauge can be considered as ``observable'' as the fields in Minkowski
space-time described by their components with respect to a given inertial
coordinate frame. In fact, also in this case the values of the field components
depend on the choice of the origin and of the directions of the coordinate
axes.

The transformation properties of the fields with respect to the Poincar\'e
group permit to compute the components in the new reference frame in terms of
the old components. For instance, if $V^a$ is a vector field, we have
$$\eqalignno{
  V^{\prime a}(x') &= [L^{-1}]^{a}_{b} V^{b}(x), \qquad
  x = L x' + x_0. &(3.20)
\cr}$$
In the following we generalize this formula to a curved space-time with radial
gauge, namely we derive the explicit form of the transformations of the fields
caused by a translation of the origin. We consider a radial gauge with origin
at the point $P_0$ , coordinates $\xi^\mu$ and tetrads $e_a(\xi)$ and we start
from the point $P_1$ with coordinates $\xi_1^\mu$ and from the tetrad
$$\eqalignno{
  e'_{a}(0) &= L_{a}^{b} e_{b}(\xi_1)
\cr}$$
to build a new radial gauge with coordinates $\xi^{\mu '}$ and tetrads
$e'_{a}(\xi')$. If, for simplicity, we take $L = 1$, the coordinates and the
tetrads are connected by
$$\eqalignno{
  \xi^\mu &= \Xi^\mu(\xi_1, \xi'), \qquad
  e'_{a}(\xi') =
  \Omega_{a}^{b}(\xi_1, \xi') e_{b}(\xi).
\cr}$$
It is easy to see that for general values of $L$ these relations are modified
as follows
$$\eqalignno{
  \xi^\mu &= \Xi^\mu(\xi_1,  L \xi'), \qquad
  e'_{a}(\xi') =  L_{a}^{b}
  \Omega_{b}^{c}(\xi_1, L \xi') e_{c}(\xi).
\cr}$$

The transformation property of a vector field is
$$\eqalignno{
  V^{a '}(\xi') &=
  [\Omega^{-1}(\xi_1, L \xi') L^{-1}]^{a}_{b}V^{b}(\xi) &(3.21)
\cr}$$
and tensors of arbitrary order transform in a similar way. In Poincar\'e  or
Euclidean gauge theories it is easy to write also the transformation properties
 of spinor fields. In a flat space-time we have
$$\eqalignno{
  \Xi^\mu(\xi_1, \xi') &= \xi_1^\mu + \xi^{\mu '}, \qquad
  \Omega^{b}_{a}(\xi_1, \xi') = \delta^{b}_{a} &(3.22)
\cr}$$
and the transformation property (3.21) takes the form (3.20). It
follows from the definitions that (3.22) holds for general spaces when
$\xi_1$ and $\xi'$ are proportional.

The gauge potentials, which describe the geometry, transform in the following
way (for $L = 1$):
$$\eqalignno{
  & e^{\prime a}_\mu(\xi')  =
  {\partial \Xi^\nu(\xi_1, \xi') \over \partial \xi^{\prime \mu}}
  [\Omega^{-1}(\xi_1, \xi')]_{b}^{a}
  e^{b}_\nu(\xi), &(3.23) \cr
  & \Gamma^{\prime a}_{b \mu}(\xi') =
  [\Omega^{-1}(\xi_1, \xi')]_{c}^{a}
  \left[ \Omega_{b}^{d}(\xi_1, \xi')
  {\partial \Xi^\nu(\xi_1, \xi') \over \partial \xi^{\prime \mu}}
  \Gamma^{c}_{d \nu}(\xi)
  +{\partial \Omega^{c}_{b}(\xi_1, \xi') \over \partial \xi^{\prime \mu}}
  \right]. &(3.24)
\cr}$$
Note that these transformations do not form a group because the quantities
$\Xi^\mu(\xi_1, \xi')$ and $\Omega^b_a(\xi_1, \xi')$ can depend also on the
initial point $P_0$. They can be computed by means of a method similar to that
used in Section 3.1. The result are the following integral equations
$$\eqalignno{
  & \Xi^\nu(\xi_1, \xi') = \xi^\nu_1 + \xi^{\prime a}
  \int_0^1 \Omega^{b}_{a}(\xi_1, \lambda \xi')
  e^\nu_{b}(\Xi(\xi_1, \lambda \xi')) \, d \lambda, \cr
  & \cr
  & \Omega^{b}_{a}(\xi_1, \xi') = \delta^{b}_{a} + \cr
  & \qquad - \xi^{\prime d}
  \int_0^1 \Omega^{e}_{d}(\xi_1, \lambda \xi')
  e^\nu_{e}(\Xi(\xi_1, \lambda \xi'))
  \Gamma^{b}_{c \nu}(\Xi(\xi_1, \lambda \xi'))
  \Omega^{c}_{a}(\xi_1, \lambda \xi') \, d \lambda.
\cr}$$
It is useful to consider an infinitesimal displacement $\xi_1$ of the origin
and put
$$\eqalignno{
  & \Xi^\nu(\xi_1, \xi') =  \xi_1^\nu + \xi^{\prime \nu}
  + \xi^\mu_1 A^\nu_\nu(\xi') + O(\xi_1^2), \cr
  & \Omega_{a}^{b}(\xi_1, \xi') =  \delta_{a}^{b}
  + \xi^\mu_1 B^{b}_{a \mu}(\xi') + O(\xi_1^2).
\cr}$$
 From eq.\ (3.22), which holds when $\xi_1$ and $\xi'$ are proportional, we
get the conditions
$$\eqalignno{
  \xi^\mu A^\nu_\mu(\xi) &= 0, \qquad \xi^\mu B^{b}_{a \mu}(\xi) = 0.
\cr}$$
Formulas (3.21), (3.23) and (3.24) take the form
$$\eqalignno{
  V^{\prime a}(\xi) - V^{a}(\xi) &=
  \xi_1^\rho \delta_\rho V^{a} + O(\xi_1^2), \qquad
  e^{\prime a}_\mu(\xi) - e^{a}_\mu(\xi) =
  \xi_1^\rho \delta_\rho e^{a}_\mu + O(\xi_1^2),
\cr}$$
$$\eqalignno{
  \Gamma^{\prime a}_{b \mu}(\xi) - \Gamma^{a}_{b \mu}(\xi) &=
  \xi_1^\rho \delta_\rho \Gamma^{a}_{b \mu} + O(\xi_1^2),
\cr}$$
where we have introduced suitable definitions for $\delta_\rho V^a$,
$\delta_\rho e^a_\mu$, $\delta_\rho \Gamma^a_{b \mu}$ [M.\ and Toller, 1990];
for instance,
$$\eqalignno{
  \delta_\rho V^{a} &=  -B^{a}_{b \rho} V^{b}
  + (\delta_\rho^\nu +A_\rho^\nu) {\partial V^{a} \over
  \partial \xi^\nu}.
\cr}$$

We may easily obtain integral equations for the quantities $A$ and $B$, which
can be solved perturbatively for small values of \ $e^\mu_{a} - \delta^\mu_{a}$
\ and $\Gamma^{b}_{a \mu}$. The formulas obtained in this subsection turn also
out to be useful in applying the radial gauge to the problem of non-singular
homogeneous and isotropic random fields.

\bigskip \noindent
{\bf 3.4 \ Solutions of (2+1)-gravity.} \bigskip

In this section we give a brief account of the applications
of the radial gauge to (2+1)-gravity due to Menotti and Seminara.

In the last years the amount of field-theoretical work about 2- and
3-dimensional models has greatly increased. The motivations
for this interest are various. In general, lower dimensional
theories are more easily solved in a complete and rigorous
mathematical fashion than higher dimensional ones.
Some important and general effects, like
spontaneous breaking of gauge symmetry, anomalies, solitons,
have first been discovered in two dimensions. For another class
of phenomena, the generalization to higher dimensions is
very difficult or impossible; there exists, however, some real
physical system to which the theory can be applied. This is
the case, for instance, of the high-temperature superconductivity
and the quantum Hall effect.

In the case of (2+1)-dimensional gravity\footnote\dag{For
an introduction to the classical and
first-quantization aspects see Jackiw [1989]; for the
second-quantization aspects see Witten [1988], and references.}
the main interest of the theory resides in its connection
to the topological theories and to the problem
of the cosmic strings in 4 dimensions [Vilenkin, 1985]. Starting
from the work of Deser, Jackiw and 't Hooft [1984] a number of papers
has been devoted to the classical solutions in (2+1) dimensions,
and some new solution techniques have been developed [Clement, 1985, 1990;
Deser and Jackiw, 1989; Grignani and Lee, 1989].

All the known solutions, and a lot of new ones, have been
re-obtained by Menotti and Seminara [1991 a, 1992] using
the radial gauge. In fact, we recall that in radial gauge
the tetrads and the connection can be written -- in the absence
of torsion -- as integrals of the Riemann's tensor
$R^{ab}_{\mu \nu}$ (eq.s (3.17), (3.19)).
Moreover, in (2+1) dimensions the Einstein's equations
state a linear dependence between the tensor $R^{ab}_{\mu \nu}$
and the energy-momentum tensor $T^\mu_a$\footnote\ddag{Because
of this, there is no field propagation
in empty space. This shows that (2+1)-gravity is physically very different
from the ``true'' (3+1)-dimensional gravity.}.
If the energy-momentum tensor is given, we can therefore write the
metric as an integral of $T^\mu_a$. Of course, in order the
whole procedure to be consistent, $T^\mu_a$ cannot be
arbitrary; it must satisfy a constraint relation which
is related to the Bianchi identities. We shall derive it here
following a slightly different approach from
that of Menotti and Seminara. Let us consider the Einstein
action in the first-order formalism, in any dimension $N \geq 3$:
$$\eqalignno{
  S &= {\rm const.} \int d^N x \, R^{ab}_{\mu \nu} \, e^{c_1}_{\rho_1} \,
  e^{c_2}_{\rho_2} \, ... \, e^{c_{N-3}}_{\rho_{N-3}} \
  e^d_\sigma \ \varepsilon_{a b c_1 c_2 ... c_{N-3} d} \
  \varepsilon^{\mu \nu \rho_1 \rho_2 ... \rho_{N-3} \sigma} ,
\cr}$$
where $R^{ab}_{\mu \nu}$ is defined in terms of the connection
$\Gamma^{ab}_{\mu}$ like in (3.15). Coupling the
vierbein to an energy-momentum source $T^\sigma_d$ and
minimizing $S$ we obtain Einstein's equations in the form
$$\eqalignno{
  R^{ab}_{\mu \nu} \, e^{c_1}_{\rho_1} \,
  ... \, \varepsilon_{a b c_1 c_2 ... c_{N-3} d} \
  \varepsilon^{\mu \nu \rho_1 \rho_2 ... \rho_{N-3} \sigma} &=
  {\rm const.} \ T^\sigma_d ,
\cr}$$
or
$$\eqalignno{
  R^{ab}_{\mu \nu} \, e^{c_1}_{\rho_1} \,
  ... \, \varepsilon_{a b c_1 c_2 ... c_{N-3} d} \
  \varepsilon^{\mu \nu \rho_1 \rho_2 ... \rho_{N-3} \sigma} \,
  e_{\sigma e} &= {\rm const.} \ T_{de} . &(3.25a)
\cr}$$
Now, it is generally believed \footnote\dag{See,
however, the paper of Boulware e Deser [1976].}
that in gravity it is not possible to ``give the conserved
source $T^\sigma_d$ and solve for the fields $e^{a}_\mu$ and
$\Gamma^{ab}_{\mu}$ ''. In fact, the conservation of $T^\sigma_d$
cannot be checked without any knowledge of the geometry. So
in general one must consider the simultaneous evolution of
the fields and the sources must (ADM formalism).
Let us then look at the ``inverse problem'': given the field,
we want to find the conserved sources that produce it. By virtue
of the Bianchi identities, the tensor $T^\sigma_d$ defined by
(3.25a) is covariantly conserved. We just need to impose
the symmetry, obtaining the constraint
$$\eqalignno{
  \left\{
  R^{ab}_{\mu \nu} \, e^{c_1}_{\rho_1} \,
  ... \, \varepsilon_{a b c_1 c_2 ... c_{N-3} d} \
  \varepsilon^{\mu \nu \rho_1 \rho_2 ... \rho_{N-3} \sigma} \,
  e_{\sigma e} \right\}_{[de]} &= 0 . &(3.25b)
\cr}$$

The task of finding solutions of (3.25b) in any dimension
is very difficult. Let us take, however, $N=3$;
eq.\ (3.25b) becomes
$$\eqalignno{
  \left\{ R^{ab}_{\mu \nu} \, \varepsilon_{abc} \,
  \varepsilon^{\mu \nu \rho} \, e_{\rho d} \right\}_{[cd]} &= 0 . &(3.26)
\cr}$$
We now fix the radial gauge. The most general structure of a
radial connection is
$$\eqalignno{
  \Gamma^{ab}_\mu(\xi) &= \varepsilon^{abc} \,
  \varepsilon_{\mu \rho \sigma} \, \xi^\rho \, A^\sigma_c(\xi) , &(3.27)
\cr}$$
where $A^\sigma_c$ is an arbitrary tensor field. From (3.27)
we have
$$\eqalignno{
  R^{ab}_{\mu \nu} \, \varepsilon_{abc} \, \varepsilon^{\mu \nu \rho} &=
  2 \, A^\rho_c + \xi^\mu \partial_\mu \, A^\rho_c -
  \xi^\rho \left( \partial_\mu A^\mu_c - {1 \over 2} \,
  \varepsilon_{abc} \, \varepsilon_{\alpha \mu \nu} \, \xi^\alpha \,
  A^{\mu a} \, A^{\nu b} \right) .
\cr}$$
We recall (eq.\ (3.18)) that in radial gauge
the vierbein can be expressed as
$$\eqalignno{
  e_{\rho d}(\xi) &= \delta_{\rho d} +
  \int_0^1 d\lambda \, \lambda \, \xi_c \, \Gamma^c_{\rho d}(\lambda \xi) .
\cr}$$
Then the symmetry condition (3.26) becomes
$$\eqalignno{
  & \varepsilon_{ab\rho}
  \left[ 2 \, A^{b\rho} + \xi^\mu \partial_\mu \, A^{b\rho} - \xi^\rho
  \left( \partial_\mu A^{b\mu} - {1 \over 2} \, \varepsilon^{mnb}
  \, \varepsilon_{\sigma \mu \nu} \, \xi^\sigma \, A^\mu_m \, A^\nu_n \right)
  \right] + \cr
  & \qquad + \varepsilon_{\rho \sigma \nu} \, \xi^\sigma
  \left( 2 \, A^{b\rho} + \xi^\mu \partial_\mu \, A^{b\rho} \right)
  \int_0^1 d\lambda \, \lambda^2
  \left( \xi_a A^\nu_b(\lambda \xi) - \xi_b A^\nu_a(\lambda \xi) \right)
  = 0 . &(3.28)
\cr}$$
This is the constraint equation given by Menotti and Seminara.

\bigskip \noindent
{\it Some solutions of the constraint in (2+1) dimensions.} \bigskip

Let us first define the scalar operators
$$\eqalignno{
  D_n &= \xi_\mu \, {\partial \over \partial \xi^\mu} + n ;
  \qquad D_\chi = \chi^\mu \, {\partial \over \partial \xi^\mu} .
\cr}$$
The operator $D_n$ is invertible, for $n>0$, provided it acts
on functions which are regular at the origin. In fact, if
$D_n \, f =0$, then $f$ is an homogeneous function of degree
$-n$. The inverse of $D_n$ has the form
$$\eqalignno{
  {D_n}^{-1} \, f(\xi) &= \int_0^1 d\lambda \,
  \lambda^{n-1} f(\lambda \xi). &(3.29)
\cr}$$
We shall consider two simple ``Ansaetze'' for $A^a_\mu$
[Menotti and Seminara, 1991 a], which make vanish the term with
the integral of the constraint (3.28) \footnote\dag{In the
absence of this term, the holonomic index of
$T$ is lowered, in practice, just with $\delta^a_\mu$
instead of the whole vierbein; this could be viewed as
a linear approximation. In fact, the other quadratic term
in $A$ vanishes too.} :

\noindent (1) $A^a_\mu(\xi) = \xi^a \, f_\mu(\xi)$,
  with $f_\mu(\xi)$ arbitrary;

\noindent (2) $A^a_\mu(\xi) = \chi_\mu \, \rho^a(\xi)$,
  with $\rho^a(\xi)$ arbitrary.

Substituting the first ansatz into (3.28) we obtain
$$\eqalignno{
  D_4 \, f^\rho(\xi) &= \xi^\rho \, F(\xi) , &(3.30)
\cr}$$
where $F$ is any function. From (3.29), (3.30)
it follows that $f^\rho(\xi)$
is proportional to $\xi^\rho$; we then conclude from (3.27)
that the connection is identically zero. So the first Ansatz
has to be rejected.

Substituting the second Ansatz, we get the equation
$$\eqalignno{
  \chi^\mu \, D_2 \, \rho^a - \xi^\mu \, D_\chi \, \rho^a &=
  \chi^a \, D_2 \, \rho^\mu - \xi^a \, D_\chi \, \rho^\mu . &(3.31)
\cr}$$
Multiplying both sides of (3.31) by $\chi_\mu$ we obtain
(if $\chi^2 \neq 0$) the vector equation
$$\eqalignno{
  \left[ \chi^2 \, D_2 - (\vec{\chi} \, \vec{\xi}) \, D_\chi \right]
  \vec{\rho} &=
  \vec{\chi} \, D_2 \,(\vec{\chi} \, \vec{\rho}) -
  \vec{\xi} \, D_\chi (\vec{\chi} \, \vec{\rho}) .
\cr}$$
In a suitable coordinate system, the operator in square brackets
becomes an operator of the form $D_2$, and is thus invertible;
it is sufficient to choose as new coordinates
$z_1=(\vec{\chi} \, \vec{\xi})$,  $z_2=(\vec{a} \, \vec{\xi})$,
$z_3=(\vec{b} \, \vec{\xi})$, being $a$ and $b$
two vectors which are orthogonal to $\chi$ and between themselves.
We then obtain for $\rho$ the structure
$$\eqalignno{
  \vec{\rho} &= f \, \vec{\chi} + g \, \vec{\xi} . &(3.32)
\cr}$$
It can be shown [Menotti and Seminara, 1991 a]
that $\rho$ has the same structure also in the case
$\chi^2=0$. Substituting (3.32) in
(3.31) we obtain the following differential equation for
the functions $f$ and $g$:
$$\eqalignno{
  D_4 \, g &= - D_\chi \, f &(3.33)
\cr}$$

A simple and interesting solution of (3.33) can be found
by setting
$$\eqalignno{
  g &= 0 ; \qquad f = f \left[ (\vec{a} \, \vec{\xi}) , (\vec{b} \, \vec{\xi})
  \right] ,
\cr}$$
where $a$ and $b$ are two vectors orthogonal to $\chi$.
It can be easily shown that the corresponding vierbein
is given by
$$\eqalignno{
  e^a_\mu(\xi) &= \delta^a_\mu + {1 \over 2} \ell^2
  \varepsilon^{abc} \, \varepsilon_{\alpha \mu \nu} \,
  \xi_b \, \xi^\nu \, \chi_c \, \chi^d \, \phi(\xi) ,
\cr}$$
where
$$\eqalignno{
  \phi(\xi) &= \int_0^1 d\lambda \, \lambda \, (1-\lambda) \,
  f(\lambda \xi) .
\cr}$$
 From this, through (3.3), we obtain the metric
$$\eqalignno{
  g_{\mu \nu}(\xi) &= \eta_{\mu \nu} +
  \ell^2 \, \varepsilon_{\rho \sigma \mu} \, \varepsilon_{\alpha \beta \nu}
  \, \chi^\rho \, \xi^\sigma \, \chi^\alpha \, \xi^\beta \, \phi(\xi)
  \times \cr
  & \left\{ 1 + {1 \over 4} \, \ell^2 \left[
  (\xi^\beta \xi_\beta) \, (\chi^\gamma \chi_\gamma)
  - (\xi^\gamma \chi_\gamma)^2 \, \right] \, \phi(\xi) \right\} . &(3.34)
\cr}$$

It is easy to verify that $\chi$ is a Killing vector for the
metric (3.34). We thus go over to the reference system where
$\chi$ assumes its simplest form. Let us consider the cases
of time-like $\chi$ and space-like $\chi$. (For the case of
light-like $\chi$ see [Menotti and Seminara, 1991 a].)

\noindent
(1) $\chi$ {\it time-like}. Setting $\chi=(1,0,0)$
the metric becomes
$$\eqalignno{
  g_{00} &= 1; \qquad g_{0i} = 0;
\cr}$$
$$\eqalignno{
  g_{ij}(\xi) &= - \delta_{ij} \pm \ell^2 \, \xi_i \, \xi_j  \, \phi(\xi) \,
  \left( 1 - {1 \over 4} \, \ell^2 (\xi_1^2 + \xi_2^2) \, \phi(\xi)
  \right) ,
\cr}$$
where the + sign holds for the diagonal and the
-- sign for the off-diagonal components. Going over to polar
coordinates we obtain
$$\eqalignno{
  g_{\mu \nu} &= {\rm diag} \, \left\{
  1, \ -1, \ -r^2 \, \left[ 1 - {1 \over 2} \, \ell^2
  r^2 \, \tilde{\phi}(r,\theta) \right]^2 \right\} ,
\cr}$$
where
$$\eqalignno{
  \tilde{\phi}(r,\theta) &= \int_0^1 d\lambda \, \lambda \,
  (1-\lambda) \, f(\lambda r \cos \theta, \lambda r \sin \theta) . &(3.35)
\cr}$$

Eq.\ (3.35) generalizes the usual conical metric
to an extended distribution of matter which does not possess
axial symmetry. Defining the defect angle by the ratio
between the radius and the circumference, we obtain
$$\eqalignno{
  \Delta \theta &= {1 \over 2} \ell^2 \, M ,
\cr}$$
where
$$\eqalignno{
  M &= \int_0^{2\pi} d\theta \int_0^\infty dr \, r \,
  f(r \cos \theta , r \sin \theta) = \int d^2x \, \sqrt{g} \, T_{00} .
\cr}$$

Suitably choosing $f$ we obtain the static point-like solutions
of Deser, Jackiw and 't Hooft [1984] and the string solutions
of Grignani and Lee [1989], Clement [1985, 1990],
Deser and Jackiw [1989].

\medskip
\noindent
(2) $\chi$ {\it space-like}. Setting $\chi=(0,0,1)$ and
performing the hyperbolic transformation
$$\eqalignno{
  \xi^1 &= S \, {\rm cosh} \, T ; \qquad
  \xi^0 = S \, {\rm sinh} \, T ,
\cr}$$
the metric takes the form
$$\eqalignno{
  g_{\mu \nu} &= {\rm diag} \, \left\{ \mp  S^2 \,
  \left[ 1 \mp {1 \over 2} \, \ell^2 \, S^2 \, \tilde{\phi}(S,T)
  \right]^2 , \ -1,\ -1 \right\} ,
\cr}$$
where the -- sign holds inside the light cone, and the + sign
outside. This is a generalization of Rindler's metric.

\medskip
The formulas we have written in this section are just an
example of the solutions that can be found using the radial
gauge. Moreover, it is possible to define a ``reduced
radial gauge'' [Menotti and Seminara, 1991 a], which is
useful in the analysis of stationary problems and allows a complete
solution of the problem of timelike closed curves in
(2+1)-gravity [Menotti and Seminara, 1993]. It is even
possible to solve the constraint equation (3.28)
in general form [Menotti and Seminara, 1992].

\bigskip \noindent
{\bf 3.5 \ Motion of test particles in a fluctuating field.} \bigskip

In this section we shall present another application of the radial gauge,
namely to the case of test particles moving in a weak quantized
gravitational field (in four dimensions).

We start recalling that,
from an operational point of view, the definition of Minkowski space-time
is based on the possibility of building up an orthogonal network of rods
and clocks. If a gravitational field is present this possibility is lost,
but the equivalence between inertial and gravitational mass still allows a
geometrical formulation of the theory. The idea that fields
and geometry are intimately related has proved to be one of the most
useful and fruitful concepts of physics. We shall thus assume that in
the presence of a {\it classical} gravitational field space-time is
described by a differentiable manifold M, endowed with a metric $g$.

We recall that it is possible to define on M a ``geodesic structure''
[see for ex. Kobayashi and Nomizu, 1969]. This structure is in some
sense the physical manifestation
of the geometrical framework of the theory: namely it is invariant with
respect to coordinate transformations, and it can be operationally
tested by observing the trajectories of many free-falling test particles.
Owing to the equivalence principle, the mass of such particles has no
importance, as long as they do not perturbate the field. The motion
of test particles in an external field has been studied in its various
aspects by many authors [see for instance Papapetrou, 1951; Souriau, 1974;
Toller, 1983;
Toller and Vaia, 1984].

In this section we want to show -- as an application of the
radial gauge -- that a weak quantized gravitational
field introduces a difficulty in the definition of
the geodesic structure of space-time at very small distances: namely
the vacuum correlations of the field will influence the motion of a test
particle, and this influence will depend on the {\it size} of the particle
itself.
Such a phenomenon is not difficult to be intuitively understood.
In fact, the size $L$ of a particle
is most properly defined by diffraction
experiments. This means that when the particle travels in a fluctuating
gravitational field its motion will be affected just by the fluctuations
of wavelenght $\lambda$ such that $\lambda > L$. The
point we would like to clarify is the average effect of these
fluctuations on the geodesic structure measured by the particle. The
conclusion is that such an effect is
non vanishing and proportional to $\ell_{Planck}/L$ [M., 1992c];
this confirms that a geometric theory based on the equivalence
principle is not operationally meaningful at distances comparable
with $\ell_{Planck}$, since the geometry depends on the size of the
test particles.

The calculation
relies on eq.\ (3.19), which gives the tetrad
$e^a_\mu(\xi)$ in radial gauge as an integral of
$R^a_{b  \mu \nu}(\xi)$ (we assume that no torsion is present).
Writing $R^a_{b  \mu \nu}(\xi)$
in terms of the Riemann's tensor through (3.6),
we obtain the equation
$$\eqalignno{
  e^a_\nu(\xi) &= \delta^a_\nu + \xi^\mu \xi^b \int_0^1 ds \, s
  (1-s) R^\rho_{\sigma \mu \nu}(s \xi) \, e^a_\rho(s \xi) \,
  e^\sigma_b(s \xi).
\cr}$$
This is an integral equation for $e^a_\nu(\xi)$, which can be iteratively
solved to the desired order in $R$. After two iterations we find
$$\eqalignno{
  & e^a_\mu(\xi) = \delta^a_\mu + \delta^a_\nu \, \xi^\alpha \xi^\beta
  \int_0^1 ds \, s (1-s) R^\nu_{\alpha \beta \mu}
  (s \xi) + \cr
  & \ \ + \delta^a_\nu \, \xi^\alpha \xi^\beta \xi^\rho \xi^\sigma
  \int_0^1 ds \, s^3 (1-s) \int_0^1 dt \, (1 - t)
  R^\nu_{\alpha \beta \gamma}(s \xi) \, R^\gamma_{\rho \sigma \mu}
  (s t \xi) + O(R^3) \cr
  & &(3.36)
\cr}$$

The crucial observation now is that the Riemann tensor of a weak euclidean
field
$g_{\mu \nu}=\delta_{\mu \nu}+h_{\mu \nu}$ is invariant with respect to
gauge transformations. Thus we can substitute in
(3.36) the Riemann tensor $R(\xi)$ in the radial gauge with that
in the Feynman-De Witt gauge, say $R^F(x)$. To this end we also must
express $\xi$ in terms of $x$, but this will just affect the term
with one single $R$. The function
$x^\mu(\xi)$ is given by (3.13).
Finally, eq.\ (3.36) can be rewritten as
$$\eqalignno{
  e^a_\mu(\xi) &= \delta^a_\mu + \delta^a_\nu \, \xi^\alpha \xi^\beta
  \int_0^1 ds \, s (1-s) {R^F}^\nu_{\alpha \beta \mu}
  (s \xi) + \cr
  & - \int_0^1 ds \, s^3 (1-s) \int_0^1 dt \, (1 - t)
  \, f^a_\mu(s,t,\xi) + o(R^3), &(3.37)
\cr}$$
where
$$\eqalignno{
  f^a_\mu(s,t,\xi) &=
  \delta^a_\nu \, \xi^\alpha \xi^\beta \xi^\rho \xi^\sigma
  \left[
  {R^F}^\nu_{\alpha \beta \mu,\gamma}(x) \, {\Gamma^F}^\gamma_{\rho \sigma}(y)
  - {R^F}^\nu_{\alpha \beta \gamma}(x) \, {R^F}^\gamma_{\rho \sigma \mu}(y)
  \right]_{x=s \xi, \ y=s t\xi}  &(3.38)
\cr}$$

Eq. (3.37) describes the geodesic motion of the parallel-transported
tetrad. Note that $e^a_\mu(\xi)$ still satisfies the gauge condition
(3.7b). Let us now suppose that the gravitational field consists of
weak fluctuations
quantized around a flat background. We can find the ``average motion''
of the tetrad by averaging (3.37) on the vacuum state. To this end
we just need to replace the quantities in square brackets with their
Feynman-De Witt propagators. The term linear in $R$ will vanish in this
approximation. One thus starts from the propagator of the metric
in dimension 4 [De Witt, 1967 c; Veltman, 1976]
$$\eqalignno{
  \left< h^F_{\mu \nu}(x) \, h^F_{\rho \sigma}(y) \right>_0 &=
  - {\kappa^2 \over (2\pi)^2} \, {P_{\mu \nu \rho \sigma} \over (x-y)^2},
  &(3.39)
\cr}$$
where $\kappa=\sqrt{16\pi G}$ is the Planck length and
$$\eqalignno{
  P_{\mu \nu \rho \sigma} &= \delta_{\mu \rho} \delta_{\nu \sigma}
  + \delta_{\mu \sigma} \delta_{\nu \rho} -
  \delta_{\mu \nu} \delta_{\rho \sigma}. &(3.40)
\cr}$$
Then one exploits the linearized expressions for the connection and
the curvature
$$\eqalignno{
  & \Gamma^\alpha_{\mu \nu}={1 \over 2} \delta^{\alpha \beta} (\partial_\nu
h_{\mu \beta}
  + \partial_\mu h_{\nu \beta} - \partial_\beta h_{\mu \nu}), &(3.41) \cr
  & R^\alpha_{\beta \mu \nu}={1 \over 2} \delta^{\alpha \gamma}
  (\partial_\gamma \partial_\mu h_{\beta \nu} - \partial_\beta \partial_\mu
h_{\gamma \nu}
  - \partial_\gamma \partial_\nu h_{\beta \mu} + \partial_\beta \partial_\nu
h_{\gamma \mu})
  &(3.42)
\cr}$$
and the formula
$$\eqalignno{
  \xi^\alpha \xi^\beta \partial_\alpha \partial_\beta \partial_\mu \partial_\nu
{1 \over \xi^2}
  &= {40 \over \xi^6} (4\xi_\mu \xi_\nu - \xi^2 \delta_{\mu \nu}). &(3.43)
\cr}$$
In this way one finds, after a long but straightforward calculation
$$\eqalignno{
  \left< f^a_\mu(s,t,\xi) \right>_0 &=
  {160 \, \ell^2 \over (2\pi)^2} \, {1 \over s^7 \, (1-t)^2}
  \, {1 \over \xi^2} \, \left( \delta^a_\mu - {\xi^a \xi_\mu \over \xi^2}
\right).
  &(3.44)
\cr}$$

As we explained above,
we must now take care of the finite dimension of the tetrad.
It is apparent that (3.44) will
give a divergent contribution after integration in $s$ and $t$.
This happens because the arguments of the fields in (3.38) coincide
in some points. However, if we cut off the modes of the field with
wavelenght smaller than $L$, the propagator (3.39) can be
replaced by
$$\eqalignno{
  \left< h^F(x) \, h^F(y) \right>_0 & \sim {1 \over (x-y)^2+L^2}; &(3.45)
\cr}$$
we implement this condition restricting the integration in $s$ and $t$
to a domain such that
$$\eqalignno{
  |s\xi - s t \xi| & > L, &(3.46)
\cr}$$
that is
$$\eqalignno{
  s (1-t) & > |\zeta|^{-1}, \qquad {\rm where} \ \ \zeta={\xi \over L}>1.
\cr}$$
Executing the integration we finally find
$$\eqalignno{
  \left< e^a_\mu(\zeta) \right>_0 &= \delta^a_\mu
  + {\ell^2 \over L^2} \, \phi(|\zeta|)
  \left( \delta^a_\mu - {\zeta^a \zeta_\mu \over \zeta^2} \right)
  + o(\ell^2), &(3.47)
\cr}$$
where $\phi(|\zeta|)$ is an analytic function which vanishes at $\zeta=1$
and whose form depends on the details of the regulator.
Note that $\zeta$ expresses the distance in the new unit length $L$.
In order our procedure to be meaningful, $\zeta$ must be not too large.

Equation (3.47) is the main result. It shows
that the ``physical'' tetrads, parallel transported in the vacuum,
differ from those of flat space by a correction proportional to the square of
$\ell_{Planck}/L$. Our position has been, of course, to consider
the Feynman-De Witt gauge just as a way to fix the dynamical components
of the field, whereas the true coordinates have to be
constructed in an operational way. If one accepts this point of view,
one concludes from eq.\ (3.47) that at very small distances,
when $L$ becomes comparable with $\ell$, the size of the test
particle influences its average motion in a fluctuating gravitational
field.

\vfill\eject

\centerline{\bf 4.\ RADIAL PROPAGATORS.}
\bigskip

In this chapter we shall introduce a technique [Menotti and Seminara, 1991
b; Menotti, M.\ and Seminara, 1993] which allows to obtain the propagator of a
gauge field in a arbitrary ``sharp'' gauge (that is, a gauge fixing
obtained by insertion of a delta function in the functional
integral), starting from the Feynman gauge. This technique, called the
``projectors method'' is very general, since it works in arbitrary dimension
and can be easily specialized to electrodynamics, Yang-Mills theory and
Einstein gravity in the first- or second-order formalism.
It also allows to select
the various solutions of the propagator equation, which are usually connected
by a residual gauge transformation and have different regularity properties. In
the case of the radial gauge, as we shall see, there are three propagators,
which differ by the behaviour at the origin and at infinity.
In Section 4.1 we shall explain the method in generic form,
with emphasis on the algebraic properties; in the
Section 4.2 we shall apply it to electrodynamics,
specifying also in details for illustrative purposes
the convergence and regularity properties,
and in Section 4.3 we shall apply it to gravity.
Finally, in Section 4.4, after comparing the radial gauge vacuum correlations
with those of the Mandelstam covariant, we shall compute explicitly
their components in a simplified case, in order to distinguish
the components which are physically meaningful.

\bigskip \noindent
{\bf 4.1 \ The gauge projectors method.} \bigskip

Let us consider a generic gauge field $A(x)$. $A$ can be an electromagnetic
field, or a Yang-Mills field, or finally the gravitational field, in the first
or second order formalism. Let the linearized equation of motion for $A(x)$ in
the presence of an external source be written in the form
$$\eqalignno{
  K_x \, A(x) &= - J(x),
\cr}$$
where $K$ is a linear, non-invertible, hermitean ``kinetic'' operator and
$J(x)$ is an external source coupled to $A$. The gauge transformations
of $A$ have the form
$$\eqalignno{
  A(x) & \rightarrow A(x) + C_x \, f(x),
  \qquad f(x) \ {\rm any \ function}.
\cr}$$
Here $C_x$ is another linear operator, which has the properties
$$\eqalignno{
  K_x \, C_x &= 0 \qquad {\rm (gauge \ invariance)}
\cr}$$
and
$$\eqalignno{
  C^\dagger_x \, K_x &= 0 \Rightarrow C^\dagger_x \, J(x) = 0
  \qquad {\rm (source \ conservation)}. &(4.1)
\cr}$$

We assume that we can  always add to the kinetic operator $K$ an operator
$K^{\cal F}$ which makes $K$ invertible, as it happens, for example, in the
Feynman gauge. We assume $K^{\cal F}$ to be of the form ${\cal F}^\dagger {\cal
F}$, as deriving from a quadratic gauge fixing $\int d x \, {\cal F}^2(A(x))$.
${\cal F}$ is meant to be a linear operator on $A$. The propagator $G^{\cal F}$
corresponding to this gauge fixing satisfies
$$\eqalignno{
  (K_x + K^{\cal F}_x) \, G^{\cal F}(x,y) &= - \delta(x-y) &(4.2)
\cr}$$
and has the following property
$$\eqalignno{
  \int dy \, K^{\cal F}_x \, G^{\cal F}(x,y) \, J(y) &= 0
  \qquad {\rm if} \ \ C^\dagger_y \, J(y) = 0. &(4.3)
\cr}$$
In other words, $K^{\cal F}_x$ vanishes when applied to the  fields
generated by physical sources. In fact, applying $C^\dagger$
to (4.2) we get
$$\eqalignno{
  C^\dagger_x \, K_x \, G^{\cal F}(x,y) +
  C^\dagger_x \, {\cal F}^\dagger_x \, {\cal F}_x \, G^{\cal F}(x,y)
  &= C^\dagger_x \, \delta(x-y).
\cr}$$
But using (4.1) we also have
$$\eqalignno{
  C^\dagger_x \, {\cal F}^\dagger_x \, {\cal F}_x \, G^{\cal F}(x,y) &=
  C^\dagger_x \, \delta(x-y)
\cr}$$
and integrating on a conserved source $J$ we obtain
$$\eqalignno{
  \int dy \, C^\dagger_x \, {\cal F}^\dagger_x \, {\cal F}_x \,
  G^{\cal F}(x,y) \, J(y) &= 0
  \qquad {\rm if} \ \ C^\dagger_y \, J(y) = 0. &(4.4)
\cr}$$
We notice that ${\cal F} \, C$ is the kinetic ghost operator and as
such invertible. Then from (4.4) we get
$$\eqalignno{
  \int dy \, {\cal F}_x \, G^{\cal F}(x,y) \, J(y) &= 0
\cr}$$
and multiplying by ${\cal F}^\dagger_x$ we finally prove (4.3).

\medskip
Let us now impose on $A$ a generic ``sharp'' gauge condition ${\cal G}$
$$\eqalignno{
  A^{\cal G}(x) &= \{ A(x) : \ {\cal G}(A(x))=0 \}. &(4.5)
\cr}$$
${\cal G}$ is meant to be a linear function of $A$ and of its derivatives.
The field $A^{\cal G}(x)$ can be obtained from a generic field $A(x)$
through a (generally non-local) projector $P^{\cal G}$
$$\eqalignno{
  A^{\cal G}(x) &= P^{\cal G}[A(x)] = A(x) + C_x \, F^{\cal G} [A(x)]. &(4.6)
\cr}$$
We require this projector to be insensitive to any ``previous gauge''
of the field, namely to satisfy
$$\eqalignno{
   P^{\cal G}[C_x \, f(x)] &= 0, \qquad {\rm for \ any} \ f(x). &(4.7)
\cr}$$
We shall now prove the following properties of the
adjoint projector ${P^{\cal G}}^\dagger$.

\medskip
\noindent {\it (1) ${P^{\cal G}}^\dagger$ produces conserved
sources.}

\noindent For, integrating (4.7) on a current $J(x)$ we have
$$\eqalignno{
  \int dx \, P^{\cal G}[C_x \, f(x)] \, J(x) &= 0;
\cr}$$
by definition of the adjoint projector, this means that
$$\eqalignno{
  \int dx \, (C_x \, f(x)) \, {P^{\cal G}}^\dagger[J(x)] &= 0
\cr}$$
and integrating by parts we have
$$\eqalignno{
  \int dx \, f(x) \, C^\dagger_x \, {P^{\cal G}}^\dagger[J(x)] &= 0
\cr}$$
or, due to the abitrariness of $f(x)$
$$\eqalignno{
  C^\dagger_x {P^{\cal G}}^\dagger[J(x)] &= 0 \qquad
  {\rm for \ any} \ J(x).
\cr}$$

\noindent {\it (2) ${P^{\cal G}}^\dagger$ leaves
a conserved source unchanged.}

\noindent Let us suppose that $J$ is conserved. Using (4.6)
we have for a generic $A$
$$\eqalignno{
  \int dx \, A(x) \, {P^{\cal G}}^\dagger [J(x)] &=
  \int dx \, A(x) \, J(x) + \int dx \, \{ C_x \, F^{\cal G}[A(x)] \} \, J(x) .
\cr}$$
Integrating by parts the second term on the r.h.s. and using
(4.1) and the arbitrariness of $A$ we have
$$\eqalignno{
  {P^{\cal G}}^\dagger[J(x)] &= J(x) \qquad {\rm if} \ C^\dagger_x \, J(x) = 0.
  \qquad q.e.d.
\cr}$$

\bigskip
The equation of motion obtained varying the action under the constraint
(4.5) is
$$\eqalignno{
  {P^{\cal G}}^\dagger[K_x \, A^{\cal G}(x)] &= - {P^{\cal G}}^\dagger[J(x)].
\cr}$$
 From (4.1) and Property 2 we have
$$\eqalignno{
  K_x \, A^{\cal G}(x) &= - {P^{\cal G}}^\dagger[J(x)],
\cr}$$
or for the propagator
$$\eqalignno{
  K_x \, G^{\cal G}(x,y) &= - {P^{\cal G}}^\dagger[\delta(x-y)], &(4.8)
\cr}$$
where the meaning of the r.h.s. is
$$\eqalignno{
  \int dy \, {P^{\cal G}}^\dagger[\delta(x-y)] \, J(y) &=
  {P^{\cal G}}^\dagger[J(x)].
\cr}$$
Next we show that a solution of (4.8) is
$$\eqalignno{
  G^{\cal G}(x,y) &= \left< P^{\cal G}[A^{\cal F}(x)] \,
  P^{\cal G}[A^{\cal F}(y)] \right>_0, &(4.9)
\cr}$$
where $A^{\cal F}$ denotes the field in the original gauge.
Integrating on a source $J(y)$ we have
$$\eqalignno{
  & \int dy \, K_x \left< P^{\cal G}[A^{\cal F}(x)] \,
  P^{\cal G}[A^{\cal F}(y)] \right>_0 \, J(y) = \cr
  & = \int dy \, K_x \left< P^{\cal G}[A^{\cal F}(x)] \,
  A^{\cal F}(y) \right>_0 \, {P^{\cal G}}^\dagger[J(y)] = \cr
  & = \int dy \, K_x \left< \{ A^{\cal F}(x) +
  C_x \, F^{\cal G}[A^{\cal F}(x)] \} \,
  A^{\cal F}(y) \right>_0 \, {P^{\cal G}}^\dagger[J(y)] = \cr
  & = \int dy \, K_x \left< A^{\cal F}(x) \,
  A^{\cal F}(y) \right>_0 \, {P^{\cal G}}^\dagger[J(y)] = \cr
  & = \int dy \, (K_x + K^{\cal F}_x) \, G^{\cal F}(x,y) \,
  {P^{\cal G}}^\dagger[J(y)] -
  \int dy \, K^{\cal F}_x \, G^{\cal F}(x,y) \, {P^{\cal G}}^\dagger[J(y)]
  = \cr
  & = - \int dy \, \delta(x-y) \, {P^{\cal G}}^\dagger[J(y)]. &(4.10)
\cr}$$
In the last step we have used (4.3) and Property 1. Note however that $G^{\cal
G}$ defined in (4.9) remains unchanged if we replace the original field with
any other gauge equivalent field.

Given two different projectors $P_1$ and $P_2$ which project on the same
gauge (which as a rule differ for different boundary conditions), one has
$$\eqalignno{
  P_1 \, P_2 &= P_1; \qquad P_2 \, P_1 = P_2 , &(4.11)
\cr}$$
due to (4.7). Thus also $P_{12}=\alpha P_1 + (1-\alpha) P_2$ is a projector on
the considered gauge and one can write down the $P_{12}$-projected Green
function equation (we omit the suffix ${\cal G}$)
$$\eqalignno{
  K_x \, G(x,y) &= P^\dagger_{12}[\delta(x-y)]. &(4.12)
\cr}$$
It is immediate to verify that a solution of (4.12) is also  given by
$$\eqalignno{
  \alpha \, \left< P_2[A(x)] \, P_1[A(y)] \right>_0 &+
  (1-\alpha) \left< P_1[A(x)] \, P_2[A(y)] \right>_0. &(4.13)
\cr}$$
Namely, repeating the same procedure of eq.\ (4.10) we have
$$\eqalignno{
  K_x \, \alpha \, \left< P_2[A(x)] \, P_1[A(y)] \right>_0 &=
  \alpha \, K_x \, \left< A(x) \, P_1[A(y)] \right>_0 = \alpha \, P^\dagger_1
  [\delta(x-y)].
\cr}$$
Acting similarly with
the $(1-\alpha)$ term in (4.13), we get (4.12). We notice that
for $\alpha =  {1 \over 2}$, \  (4.13) is symmetric in the exchange
of the field arguments.

We close this section writing in explicit form the operators
which appear in electrodynamics and linearized Einstein theory.

\bigskip \noindent
{\it Electrodynamics and linearized Yang-Mills theory.}

\medskip
This is the most simple case. We have the following identifications
$$\eqalignno{
  & A \rightarrow A_\mu; \cr
  & J \rightarrow J_\mu; \cr
  & C f \rightarrow \partial_\mu \, f; \cr
  & K \rightarrow \delta_{\mu \nu} \,
  \partial^2 - \partial_\mu \partial_\nu; \cr
  & K^{\cal F} \rightarrow \partial_\mu \partial_\nu,
\cr}$$
where $K^{\cal F}$ is the operator which is produced by the usual
Feynman gauge fixing $ {1 \over 2}(\partial^\mu A_\mu)^2$.

\bigskip \noindent
{\it Linearized Einstein gravity in the second-order formalism.}

\medskip
In this case we have
$$\eqalignno{
  & A \rightarrow h_{\mu \nu}; \cr
  & J \rightarrow T_{\mu \nu}; \cr
  & C f \rightarrow \left( \delta_{\alpha \sigma}
  \partial_\rho + \delta_{\alpha \rho} \partial_\sigma \right)
  f_\alpha ; \cr
  & K \rightarrow
  K_{\mu \nu \rho \sigma} =  {1 \over 4} [2 \, \delta_{\rho \sigma} \,
  \partial_\mu \partial_\nu +
  2 \, \delta_{\mu \nu} \, \partial_\rho \partial_\sigma + \cr
  & \qquad \qquad \qquad -
  ( \delta_{\mu \rho} \, \partial_\nu \partial_\sigma +
  \delta_{\nu \rho} \, \partial_\mu \partial_\sigma
  + \delta_{\mu \sigma} \, \partial_\nu \partial_\rho +
  \delta_{\nu \sigma} \, \partial_\mu \partial_\rho ) + \cr
  & \qquad \qquad \qquad + (\delta_{\mu \rho} \, \delta_{\nu \sigma}
  + \delta_{\mu \sigma} \, \delta_{\nu \rho}
  - 2 \, \delta_{\mu \nu} \, \delta_{\rho \sigma}) \, \partial^2 ] ;  \cr
  & K^{\cal F} \rightarrow
  K^{\cal F}_{\mu \nu \rho \sigma} =
   {1 \over 4} [ - ( 2 \, \delta_{\rho \sigma} \,
  \partial_\mu \partial_\nu +
  2 \, \delta_{\mu \nu} \, \partial_\rho \partial_\sigma ) +
  \cr
  & \qquad \qquad \qquad
  + ( \delta_{\mu \rho} \, \partial_\nu \partial_\sigma +
  \delta_{\nu \rho} \, \partial_\mu \partial_\sigma
  + \delta_{\mu \sigma} \, \partial_\nu \partial_\rho +
  \delta_{\nu \sigma} \, \partial_\mu \partial_\rho) ] .
\cr}$$
Here $K^{\cal F}$ is the operator produced by the harmonic gauge fixing
$$\eqalignno{
   {1 \over 2} & \left( \partial^\mu h_{\mu \nu} -
   {1 \over 2} \partial^\nu h^\mu_\mu \right)^2. &(4.14)
\cr}$$

\bigskip \noindent
{\it Linearized Einstein gravity in the first-order formalism.}
\medskip

The quadratic part of the lagrangian has the form
$$\eqalignno{
  L^{(2)} &= - (
  \delta^{\mu \nu \gamma}_{abc} \, \partial_\mu \, \Gamma^{ab}_\nu
  \, \tau^c_\gamma + \delta^{\mu \nu}_{ab}
  \, \Gamma^a_{c \mu} \, \Gamma^{cb}_\nu
  + T^\mu_a \, \tau^a_\mu + \Sigma^\mu_{ab} \, \Gamma^{ab}_\mu ) ,
\cr}$$
where
$$\eqalignno{
  & \tau^a_\mu = e^a_\mu - \delta^a_\mu \cr
  & \delta^{\mu \nu}_{ab} = \delta^\mu_a \delta^\nu_b -
  \delta^\mu_b \delta^\nu_a \cr
  & \delta^{\mu \nu \gamma}_{abc} = \delta^\mu_a \delta^{\nu \gamma}_{bc}
  - \delta^\mu_b \delta^{\mu \gamma}_{ac} +
  \delta^\mu_c \delta^{\nu \gamma}_{ab}
\cr}$$
and $T^\mu_a$ and $\Sigma^\mu_{ab}$ are the energy-momentum source and
the spin-torsion source, respectively. The gauge transformations
have the form
$$\eqalignno{
  & Cf \rightarrow
  \pmatrix{
    0 & \ \ \delta^{ab}_{cd} \, \partial_\mu \cr
    & \cr
    \partial_\mu & \ \ - \delta_{\mu b} \, \delta^{ab}_{cd}
  \cr}
  \pmatrix{
    \Lambda^a \cr
    \cr
    \theta^{cd}
  \cr} .
\cr}$$
The field equations are given by
$$\eqalignno{
  & K \, A = J \rightarrow
  \pmatrix{
     {1 \over 2} (\delta_{c \sigma} \delta^{\nu \sigma \mu}_{dab}
    - \delta_{d \sigma} \delta^{\nu \sigma \mu}_{cab}) & \ \
    - \delta^{\lambda \gamma \mu}_{rab} \, \partial_\lambda  \cr
    & \cr
    - \delta^{\mu \lambda \nu}_{abc} \, \partial_\mu & 0
  \cr}
  \pmatrix{
    \Gamma^{cd}_{\nu} \cr
    \cr
    \tau^r_\gamma
  \cr} =
  \pmatrix{
    \Sigma^\mu_{ab} \cr
    \cr
    T^\mu_d
  \cr}
\cr}$$
and the gauge-fixing term has the form
$$\eqalignno{
  & K^{\cal F} \, A \rightarrow
  \pmatrix{
  0 & \ \ 0 \cr
  & \cr
  0 & \ \ 4K^{\cal F}_{a \mu b \nu} \, \delta^{\nu \gamma}
  + 2\beta \, (\delta_{ab} \delta_\mu^\gamma - \delta_{\mu b} \delta^\gamma_a)
  \cr}
  \pmatrix{
  \Gamma^{ab}_{\nu} \cr
  \cr
  \tau^b_\gamma
  \cr},
\cr}$$
and is produced by the harmonic gauge fixing (4.14) with
$h_{\mu \nu}=\tau^a_\mu \delta_{a \nu} + \tau^a_\nu \delta_{a \mu}$,
to which the symmetric gauge fixing $  {\beta \over 2}
(\tau^a_\mu \delta_{a \nu} - \tau^a_\nu \delta_{a \mu})^2$ has been
added.

\bigskip \noindent
{\bf 4.2 \ Electrodynamics. Regularity properties.} \bigskip

In this Section we shall apply the projectors method to the case of
electrodynamics or linearized Yang-Mills theory in radial gauge [Menotti and
Seminara, 1991 b]. We start from the inversion formula (3.2), writing it in the
form
$$\eqalignno{
  A^0_\mu(x) &= P^0[A]_\mu(x) =
  x_\rho \int_0^1 d\lambda \, \lambda \, F_{\rho \mu}(\lambda x) . &(4.15)
\cr}$$
Let be $r=|x|$. The integral in (4.15) converges if $|F(x)|<r^{-2+\varepsilon}$
as $r \to 0$. We notice that if the field $A_\mu(x)$ is such that
$|A(x)|<r^{-1+\varepsilon}$ for $r \to 0$, then (4.15) can be rewritten as
$$\eqalignno{
  A^0_\mu(x) &= P^0[A]_\mu(x) = A_\mu(x) -
  {\partial \over \partial x_\mu} \int_0^1
  d\lambda \, x_\rho \, A_\rho(\lambda x). &(4.16)
\cr}$$
This shows that the general form (4.6) of the propagator is maintained,
provided the fields are not too much singular. There is, however, a limiting
case in which (4.16) is not true any more: namely, when we start from a field
which is already radial and differs from $A^0_\mu(x)$ by a residual gauge
transformation of the form (3.1). For instance, let us take as starting field
the radial field $A_\mu(x)=-x_\rho \int_1^\infty d\lambda \, \lambda \, F_{\rho
\mu}(\lambda x)$. Suppose that $F_{\mu \nu}(x)$ is finite at the origin. It is
easy to see that $|A(x)| \sim r^{-1}$ as $r \to 0$, while $F$ keeps finite.
This happens because the singular part of $A$ is a pure gauge, which does not
contributes to $F$. The exact significance of this detail will become more
clear as we proceed with our discussion. In practice, we assume the following
definition of $P^0$.

\medskip
\noindent {\it Definition of $P^0$:}
$$\eqalignno{
  A^0_\mu(x) & \equiv P^0[A]_\mu(x) \equiv
  x_\rho \int_0^1 d\lambda \, \lambda \, F_{\rho \mu}(\lambda x) . &(4.17)
\cr}$$
In a similar way, we can define a projector in which the integration limits 0
and $\infty$ are exchanged.

\medskip
\noindent {\it Definition of $P^\infty$:}
$$\eqalignno{
  A^\infty_\mu(x) & \equiv P^\infty[A]_\mu(x) \equiv
  - x_\rho \int_1^\infty d\lambda \, \lambda \, F_{\rho \mu}(\lambda x).
&(4.18)
\cr}$$

It is easy to verify that the difference between $A^0$ and $A^\infty$ is a
residual gauge term of the form (3.1).

It is apparent from (4.17), (4.18) that $P^0$ and $P^\infty$ are projection
operators, namely we have $(P^0)^2=P^0$, $(P^\infty)^2=P^\infty$. In addition,
the following properties hold (compare (4.11))
$$\eqalignno{
  P^0 \, P^\infty & = P^0 ; \qquad P^\infty \, P^0 = P^\infty .
\cr}$$

The definition domains of $P^0$ and $P^\infty$ are respectively given by

\noindent
$\{ F: |F(x)| < r^{-2+\varepsilon}\ {\rm for} \ r \to 0 \}$ and
$\{ F: |F(x)| < r^{-2-\varepsilon}\ {\rm for} \ r \to \infty \}$.
Restricting such domains respectively to

\noindent
$\{ F: |F(x)| < r^{-1+\varepsilon}\ {\rm for} \ r \to 0 \}$ and
$\{ F: |F(x)| < r^{-3-\varepsilon}\ {\rm for} \ r \to \infty \}$,

\noindent
we find that the fields $A^0$ and $A^\infty$ can be characterized in a simple
way through their asymptotic behaviour. Namely, we find that

\noindent - $A^0_\mu(x)$ vanishes at the origin and decreases like
$r^{-1}$ at infinity;

\noindent - $A^\infty_\mu(x)$ grows like $r^{-1}$ at the origin
and vanishes like $r^{-1-\varepsilon}$ at infinity.

In fact, let $\hat{x}$ be the unit versor of $x$. We have
$$\eqalignno{
  A^0_\mu(x) &= x_\rho \int_0^1 d\lambda \, \lambda \,
  F_{\rho \mu}(\lambda r \hat{x}) =
  {x_\rho \over r^2} \int_0^r dt \, t \, F_{\rho \mu}(t \hat{x}) . &(4.19)
\cr}$$
In the limit $r \to 0$ we obtain (in the domain above)
$$\eqalignno{
  A^0(x) & <  {1 \over r} \left[ t^{1+\varepsilon} \right]^r_0
  \to 0 \qquad {\rm for} \ r \to 0 .
\cr}$$
On the other hand, in the limit $r \to \infty$, the integral converges and
$A^0$ has the form
$$\eqalignno{
  A^0_\mu(x) &=  {1 \over r} \, H^0_{\mu}(x) \qquad {\rm for} \ r \to \infty ,
\cr}$$
where $H^0_\mu(x)$ is a homogeneous function of degree 0. Analogous formulas
hold for $A^\infty$. In the limit $r \to 0$, $A^\infty$ is regular, in the
sense that it vanishes faster than $r^{-1}$. At the origin we have instead
$$\eqalignno{
  A^\infty_\mu(x) &= -
  {1 \over r} H^0_\mu(x) \qquad {\rm for} \ r \to 0 . &(4.20)
\cr}$$

 From eq.s (4.19) - (4.20) we see that if we compute a radial Wilson loop (fig.
1) going to infinity, we obtain the same result both using $A^0$ and
$A^\infty$. In the first case the effective contribution is localized at
infinity, in the second case at the origin. Finally, if we define the mixed
projector
$$\eqalignno{
  P^S &= {1 \over 2} (P^0 + P^\infty) ,
\cr}$$
we find that the contribution of the field $A^S$ to the radial loop is splitted
into two parts: one half at the origin, one half at infinity.

For the computation of the adjoint projectors ${P^0}^\dagger$,
${P^\infty}^\dagger$ and ${P^S}^\dagger$ we refer to the mentioned work of
Menotti and Seminara.

By means of the projectors $P^0, P^\infty, P^S$ we can now write down
three symmetric
radial propagators, denoted by $G^0$, $G^\infty$ and $G^S$:
$$\eqalignno{
  G^0_{\mu\nu}(x,y) & =  \left<
  P^0[A]_\mu(x) \, P^0[A]_\nu(y) \right>_0 = \cr
  & =  x^\rho y^\sigma \int^1_0 d\lambda \, \lambda
  \int^1_0 d\tau \, \tau
  \left< F_{\rho \mu}(\lambda x) \, F_{\sigma \nu}(\tau y) \right>_0 ; \cr
  & \cr
  G^\infty_{\mu\nu}(x,y) & = \left<
  P^\infty[A]_\mu(x) \, P^\infty[A]_\nu(y) \right>_0 = \cr
  & = x^\rho y^\sigma \int_1^\infty d\lambda \, \lambda
  \int_1^\infty d\tau \, \tau
  \left< F_{\rho \mu}(\lambda x) \, F_{\sigma \nu}(\tau y) \right>_0 ; \cr
  & \cr
  G^S_{\mu\nu}(x,y) & = {1 \over 2} \left<
  P^0[A]_\mu(x) \, P^\infty[A]_\nu(y) \right>_0
  +  {1 \over 2} \left< P^\infty[A]_\mu(x) \, P^0[A]_\nu(y) \right>_0 = \cr
  & = -  {1 \over 2} x^\rho y^\sigma \int_1^\infty d\lambda \, \lambda
  \int^1_0 d\tau \, \tau
  \left< F_{\rho \mu}(\lambda x) \, F_{\sigma \nu}(\tau y) \right>_0 \cr
  & - {1 \over 2} x^\rho y^\sigma \int_0^1 d\lambda \, \lambda
  \int_1^\infty d\tau \, \tau
  \left< F_{\rho \mu}(\lambda x) \, F_{\sigma \nu}(\tau y) \right>_0 .
\cr}$$
To compute them explicitly we make use of the formula
$$\eqalignno{
  x^\rho \, y^\sigma \left<
  F_{\rho \mu}(x) \, F_{\sigma \nu}(y) \right>_0 &=
  x^\rho \, y^\sigma D_{\mu\rho\nu\sigma} \, {\cal D}(x-y),
\cr}$$
where
$$\eqalignno{
  D_{\mu\rho\nu\sigma} &=
  \partial_{\mu\rho} {\partial \over \partial x^\nu}
  {\partial \over \partial y^\sigma}
  +\partial_{\nu\sigma} {\partial \over \partial x^\mu}
  {\partial \over \partial y^\rho}-
  \partial_{\mu\sigma} {\partial \over \partial x^\nu}
  {\partial \over \partial y^\rho}-
  \partial_{\nu\rho} {\partial \over \partial x^\mu}
  {\partial \over \partial y^\sigma}
\cr}$$
and ${\cal D}(x-y)$ is the Feynman propagator. It can be shown that the
integrals of $G^0$ converge in dimension $N<4$, while those of $G^\infty$
converge for $N>4$; finally, the integrals of $G^S$ converge for $N>3$. Let us
compute explicitly, for instance, $G^S$. Using the formula
$$\eqalignno{
  \int_0^1 \, d\rho  {1 \over [(\rho x - y)^2]^{ {N \over 2}-1}}
  &= \int_0^\infty d\rho \,  {1 \over [(\rho x - y)^2]^{ {N \over 2}-1}} -
  \int_0^1 d\rho \,  {\rho^{N-4} \over [(x - \rho y)^2]^{ {N \over 2}-1}}
\cr}$$
we can write the integral of interest in the form
$$\eqalignno{
  D(x,y) & = -  {\Gamma( {N \over 2}-1) \over 4 \pi^{N/2}} \,
   {1 \over N-4} \left\{ \int_0^1 d\rho \, \rho^{N-4}
  \left[  {1 \over [(\rho x - y)^2]^{ {N \over 2}-1}} +
  \right. \right. \cr
  & \left. \left. +  {1 \over [(x - \rho y)^2]^{ {N \over 2}-1}}
  \right]
  - \int_0^\infty d\rho \,  {1 \over [(\rho x - y)^2]^{ {N \over 2}-1}}
  \right\} . &(4.21)
\cr}$$
The first integral which appears in (4.21) can be rewritten as follows
$$\eqalignno{
  I_1 &= \int^1_0 d\rho \, {\rho^{N-4} \over
  [\rho^2 X^2-2 \rho X Y \cos\theta +Y^2]^{N/2-1}} ,
\cr}$$
where $X=|x|$, $Y=|y|$, and performing the change of variables $\rho= { {Y
\over d}\zeta \over  {Y \over d}\zeta+1}$, with $d=\sqrt{X^2+Y^2-2 X Y
\cos\theta}\ $, we obtain [Erd\'ely et al., 1953]
$$\eqalignno{
  & Y^{-1}d^{3-N}\int^\infty_0 d\zeta {\zeta^{N-4} \over \left [\zeta^2+2 \zeta
  \left( { Y-X\cos\theta \over d}\right )+1\right ]^{N/2-1}}= \cr
  & \cr
  &= {1 \over N-3}Y^{-1}d^{3-N} \left (4\sin^2{\phi_x}\right)^{ {3-N \over 4}}
  \Gamma\left ({N-1 \over 2}\right)
  {\rm P}^{ {3-N \over 2}}_{ {3-N \over 2}}\left (\cos{\phi_x}\right) ,
\cr}$$
being $\cos\phi_x={Y-X\cos\theta \over d}$. Here $\theta$ is the angle between
$x$ and $y$, and $\phi_x$ is the angle opposite to $x$ in the euclidean
triangle of sides $x,\, y,\, d$. Expressing ${\rm P}^{ {3-N \over 2}}_{ {3-N
\over 2}}$ in terms of hypergeometric functions, we have
$$\eqalignno{
  I_1 &= {1 \over N-3} \, {d^{3-N} \over Y^{-1}}
  \left(\cos^2 {\phi_x \over 2}\right )^{ {3-N \over 2}}
  {}_2 F_1 \left(  {N-3 \over 2} ,  {5-N \over 2} ;  {N-1 \over 2} ;
  \sin^2 {\phi_x \over 2} \right).
\cr}$$

In a similar way we can compute
$$\eqalignno{
  I_2 &= \int^\infty_0 d\rho \, {1 \over [\rho^2 X^2-2 \rho X Y \cos\theta
  +Y^2]^{N/2-1}} ,
\cr}$$
obtaining
$$\eqalignno{
  I_2 &= {1 \over N-3} \, {Y^{3-N} \over X^{-1}}
  \left(\sin^2 {\theta \over 2}\right )^{ {3-N \over 2}}
  {\ }_2 F_1 \left(  {N-3 \over 2} ,  {5-N \over 2} ;  {N-1 \over 2} ;
  \cos^2 {\theta \over 2} \right).
\cr}$$
The behaviour of $[D(x,y)+D(y,x)]$ as $X\to 0$, and $y$ and $\theta$ are
constant, is obtained from $I_1$ and $I_2$ by noticing that, in this limit,
$\phi_x\to0$ and $\phi_y\to \pi-\theta$. We then have
$$\eqalignno{
  & D(x,y)+D(y,x) \simeq  \cr
  & \simeq {1 \over N-3} \,
  \left(\sin^2 {\theta \over 2}\right )^{ {3-N \over 2}}
  {\ }_2 F_1 \left(  {N-3 \over 2} ,  {5-N \over 2} ;  {N-1 \over 2} ;
  \cos^2 {\theta \over 2} \right) \times \cr
  & \left({ Y^{3-N} \over X^{-1}}-{X^{3-N} \over Y^{-1}} \right) .
\cr}$$
This is also the behaviour for $x$ and $\theta$ fixed, while $Y\to\infty$.
Owing to the symmetry of $[D(x,y)+D(y,x)]$, we have that for $Y\to 0$, $x$ and
$\theta$ fixed
$$\eqalignno{
  & D(x,y)+D(y,x) \simeq \cr
  & \simeq - {1 \over N-3} \,
  \left(\sin^2 {\theta \over 2}\right )^{ {3-N \over 2}}
  {\ }_2 F_1 \left(  {N-3 \over 2} ,  {5-N \over 2} ;  {N-1 \over 2} ;
  \cos^2 {\theta \over 2} \right) \times \cr
  & \left( {Y^{3-N} \over X^{-1}}-{X^{3-N} \over Y^{-1}} \right)
\cr}$$
and the same holds for $X\to \infty$, $y$ and $\theta$ fixed. Hence we see that
the behaviour for $X\to \infty$ and $X\to 0$ is the same, except for the sign.
Being $D_{\mu\nu\rho\sigma}$ a zero degree operator, this also holds for the
whole propagator $G^S_{\mu\nu}(x,y)$.

\bigskip \noindent
{\bf 4.3 \ Gravitational propagators.} \bigskip

In Section 4.1 we have developed in general form the projectors method, which
provides an algebraic way for constructing the propagator in any ``sharp''
gauge, provided it is known in a Feynman-like gauge. Then
in Section 4.2 the projectors and propagators of electrodynamics
were explicitly written, taking care of their
existence and regularity properties.

The reason for which it is desirable to impose the radial gauge in gravity as a
sharp gauge (that is, to insert a delta into the functional integral) is that
the geometrical meaning of the radial gauge can be formally preserved, at the
quantum level, only if the fields admitted in the functional integration
satisfy the gauge condition.

The functional integral of gravity in the first order formalism is given
by\footnote\dag{It is well known that the rigorous definition of these
integrals is
affected by the unboundedness of the euclidean action and by the arbitrariness
of the functional integration measure [see Menotti, 1990]. We shall not dwell
on such problems here.}
$$\eqalignno{
  z &= \int d[\Gamma^{ab}_\mu] \, d[e^a_\mu] \, \mu[e^a_\mu] \,
  e^{- \ell^{-2} \int d^Nx \, R^{ab} \wedge e^c ... \varepsilon_{abc...}}.
\cr}$$

The gauge fixing term is $\delta(x^\mu \Gamma^{ab}_\mu)
\delta(x^\mu(e^a_\mu-\delta^a_\mu))$. Note that it preserves the symmetry
between vierbein and connection.

As it happens in Yang-Mills theory, the ghosts associated to the local Lorentz
symmetry formally decouple, while those associated to the diffeomorphisms
survive (in the first- as well as in the second order formalism).

We remind that in the functional integral approach the correlation functions
are computed by averaging the products of fields like $O(x)O(y)$ on all
geometries weighted by the exponential of the gravitational action. Fixing the
gauge to the radial gauge gives $x$, $y$,... a well defined meaning, as the
points that acquire geodesic coordinates $x$, $y$,... in each of the geometries
we are summing over. Hence, when the points lie on a straight line through the
origin, their correlation is automatically geodesic.

\medskip
Next we write the expressions for the radial projectors which are the
analogue of the $P^0$, $P^\infty$ and $P^S$ of the preceding Section.

$P^0$ is given by (3.17) and (3.19) substituting to
$R^{ab}_{\mu\nu}$ and $S^a_{\mu\nu}$ their linearized expressions
$$\eqalignno{
  {R^L}^{ab}_{\mu\nu}(x) &= \partial_\mu \Gamma^{ab}_\nu (x)
  -\partial_\nu \Gamma^{a b}_\mu (x) , \cr
  {S^L}^a_{\mu\nu}(x) &= \partial_\mu \tau^a_\nu (x) -\partial_\nu
  \tau^{a}_\mu (x) +\Gamma^{ab}_\mu (x) \delta_{b\nu} -
  \Gamma^{ab}_\nu (x) \delta_{b\mu} ,
\cr}$$
which are invariant under linearized gauge transformations.
We thus have
$$\eqalignno{
P^{0} \pmatrix{
\Gamma^{a}_{b,\mu}(x) \cr
\cr \cr
\tau^{a}_{\mu}(x)
\cr}
&=
\pmatrix{
x^\nu\displaystyle{\int}^1_0 d\lambda \, \lambda
R^{La}_{b\nu\mu}(\lambda x) \cr
\cr
x^\nu x^b\displaystyle{\int}^1_0 d\lambda \,\lambda \, (1-\lambda) \,
R^{La}_{b\nu\mu}(\lambda x)
+x^\nu \displaystyle{\int}^1_0 d\lambda \,\lambda \, S^{La}_{\nu\mu}(\lambda x)
\cr} &(4.22)
\cr}$$

$$\eqalignno{
P^{\infty}
\pmatrix{
\Gamma^{a}_{b,\mu}(x) \cr
\cr \cr
\tau^{a}_{\mu}(x)
\cr}
&=
\pmatrix{
-x^\nu\displaystyle{\int}^\infty_1 d\lambda \, \lambda \,
R^{La}_{b\nu\mu}(\lambda x) \cr
\cr
-x^\nu x^b\displaystyle{\int}^\infty_1 d\lambda \, \lambda (1-\lambda)\,
R^{La}_{b\nu\mu}(\lambda x)
-x^\nu \displaystyle{\int}^\infty_1 d\lambda \, \lambda \,
S^{La}_{\nu\mu}(\lambda x)
\cr}. &(4.23)
\cr}$$
The projector $P^0$ produces radial fields which are
regular at the origin (i.e. behaving
like the original fields) and gives a connection $\Gamma^{ab}_\mu (x)$ behaving
like ${1/r}$ at infinity. On the other hand $P^\infty$ projects on a radial
field which is regular at infinity and gives a connection behaving like ${1/r}$
at the origin.

$P^S$, which allows us to construct a finite propagator in $N>2$, treats the
origin and infinity in symmetrical way by giving the same ${1/r}$ behavior in
the two limits with opposite coefficients. This can be shown in a similar way
as we did for electrodynamics [Menotti, M.\ and Seminara, 1993].

The adjoint projectors can also be easily computed and their expressions can be
found in [Menotti, M.\ and Seminara, 1993], like many other details that we
shall omit in the following.

When we construct by these projectors the propagators
$$\eqalignno{
  & \langle P^0(\Gamma,\tau) P^0(\Gamma,\tau)\rangle_0
\cr}$$
and
$$\eqalignno{
  & \langle P^\infty(\Gamma,\tau) P^\infty(\Gamma,\tau)\rangle_0
\cr}$$
we find that the first one is always divergent,
while the second one diverges for $N\le 4$. Thus we shall construct the
solution for the $P^S$ projected Green's function equation, namely
$$\eqalignno{
  & \pmatrix{
  {1 \over 2}(\delta_{am}^{\mu\nu}\delta_{bn}-
  \delta_{an}^{\mu\nu}\delta_{bm}) &
  & \delta^{\lambda\mu\sigma}_{abd}\partial_\lambda \cr
  \cr
  \delta^{\mu\lambda\nu}_{pmn}\partial_\lambda & & 0
  \cr}
  \pmatrix{
  G^{mn,rs}_{\nu,\gamma} & & G^{mn,g}_{\nu,\beta} \cr
  \cr
  G^{d,rs}_{\sigma,\gamma} & & G^{d,g}_{\sigma,\beta}
  \cr} = \cr
  & \cr
  & \qquad \qquad = - {P^S}^\dagger
  \pmatrix{
  \delta^\mu_\gamma \delta^{rs}_{ab} & &  0 \cr
  \cr
  0 & & \delta^\mu_\beta \delta^g_p
  \cr}
  \delta^N(x-y) . &(4.24)
\cr}$$
by use of the now familiar procedure. One shows that the propagator
$$\eqalignno{
  {1 \over 2} \left< P^{0}
  \pmatrix{
  \Gamma^{a}_{b,\mu}(x) \cr
  \cr
  \tau^{a}_{\mu}(x)
  \cr}
  P^{\infty}
  \pmatrix{
  \Gamma^{a'}_{b',\mu '}(y) , \,
  \tau^{a'}_{\mu '}(y)
  \cr} \right>
  &+ (P^0 \leftrightarrow P^\infty) &(4.25)
\cr}$$
is a convergent symmetric radial solution of (4.24) for all $N>2$. The explicit
form of solution (4.25) of the Green' s function equation (4.24) is easily
computed  by using (4.22) and (4.23) where the correlators between Riemann and
torsion two-forms, which are invariant under linearized gauge transformations,
can be obtained using e.g. the usual symmetric harmonic gauge.

Let be $M^{ab,cd}_{\mu,\nu}(x,y)$ the ultra-local part of the propagator of the
anholonomic connection (that is, the part which is proportional to
$\delta^N(x-y)$). $M$ is given by the expression [Menotti and Pelissetto, 1987]
$$\eqalignno{
  M^{ab,cd}_{\mu,\nu}(x,y) &=
  - {i \over 4} \left( \delta_{\mu \nu} \, \delta^{ab,cd} +
  \delta^{ab}_{\mu \gamma} \, \delta^{cd}_{\nu \gamma} -
  \displaystyle{{2 \over N-2}} \, \delta^{ab}_{\nu \gamma} \,
  \delta^{cd}_{\mu \gamma} \right) \, \delta^N(x-y) .
\cr}$$
($\delta^{ab,cd}$ is an antisymmetric symbol in the pairs
$(a,b)$, $(c,d)$.)

We have
$$\eqalignno{
  \langle {R^L}^{ab}_{\mu\nu}(x) \, {R^L}^{cd}_{\rho\sigma}(y)\rangle &=
  \langle {R^L}^{ab}_{\mu\nu}(x) \, {R^L}^{cd}_{\lambda\rho}(y)\rangle^{II}+
  \left[ \partial_{\mu}\partial_{\rho} M^{ab,cd}_{\nu,\sigma}(x,y)
  \right]_{[\mu \nu],[\rho \sigma]} ; &(4.26)
\cr}$$
$$\eqalignno{
  \langle {S^L}^a_{\mu\nu}(x) \, {S^L}^b_{\rho\sigma}(y)\rangle &=
  \left[ M^{a c,b d}_{\mu,\rho}(x,y) \, \delta_{c\nu} \, \delta_{d\sigma}
  \right]_{[\mu \nu],[\rho \sigma]} ; &(4.27)
\cr}$$
$$\eqalignno{
  \langle {R^L}^{ab}_{\mu\nu}(x) \, {S^L}^c_{\rho\sigma}(y)\rangle &=
  \left[\partial_{\mu} M^{ab,cd}_{\nu,\rho}(x,y) \,
  \delta_{d\sigma} \right]_{[\mu \nu],[\rho \sigma]} . &(4.28)
\cr}$$

In (4.26), $\langle\ \rangle^{II}$ denotes the correlator in the second order
formalism:
$$\eqalignno{
  \left< R_{\mu \nu \alpha \beta}(x) \, R_{\rho \sigma  \lambda
  \gamma}(y) \right>_0 &=
  {1 \over 4} \delta^{\mu' \nu'}_{\mu \nu} \times \cr
  & \left(
  \delta^{\rho' \nu'}_{\rho \sigma} \delta^{\alpha' \beta'}_{\alpha \beta}
  \delta^{\lambda' \beta'}_{\lambda \gamma}
  + \, \delta^{\lambda' \nu'}_{\lambda \gamma}
  \delta^{\alpha' \beta'}_{\alpha \beta} \delta^{\rho' \beta'}_{\rho \sigma}
  - {2 \over N-2} \,
  \delta^{\alpha' \nu'}_{\alpha \beta} \delta^{\rho' \sigma'}_{\rho \sigma}
  \delta^{\lambda' \sigma'}_{\lambda \gamma} \right) \times \cr
  & \partial_{\mu'} \partial_{\alpha'} \partial_{\rho'} \partial_{\lambda'}
  {\cal D}(x-y) .
\cr}$$

The ultra-local nature of the correlators (4.27) and (4.28) reflects the
well known fact that the torsion does not propagate in the Einstein-Cartan
theory.

The integrals over  $\lambda$ and $\tau$ of the $\langle R R \rangle^{II}$
correlators are similar to those given for electrodynamics (Section 4.2). The
contact terms generated by  $M$ give rise to integrals in $\lambda$ and $\tau$
which are convergent. In fact the generic integral is of the form
$$\eqalignno{
& \int^1_0 d\lambda \, \lambda^A \int^\infty_1 d\tau \, \tau^B
\delta^N(\lambda x-\tau y)= \cr
& \qquad = \int^1_0 d\lambda \, \lambda^A \int^\infty_1 d\tau \,
\tau^B \delta (\lambda |x|-
\tau |y|)(\lambda |x|)^{-N+1} \delta^{N-1}(\Omega_x-\Omega_y)= \cr
& \qquad = \delta^{N-1}(\Omega_x-\Omega_y) \Theta (|x|-|y|){1 \over A+B-N+2}
\times \cr
& \qquad \left ( |y|^{-B-1}|x|^{B-N+1}-|y|^{A-N+1}|x|^{-A-1} \right).
\cr}$$

We notice that in three dimensions the correlator (4.26) is identically zero,
as can be explicitly verified by observing that the matrix component (2,1) of
eq.\ (4.24) takes the form $\langle R^L(x)\Gamma(y)\rangle \equiv 0$, which
implies $\langle R^L(x) R^L(y)\rangle\equiv 0$. This means in turn, through
(4.22) and (4.23), that $\langle\Gamma(x)\Gamma(y)\rangle\equiv 0$. Hence
$\langle \tau(x) \tau(y)\rangle$ vanishes, except for singular contributions
which arise when the origin, $x$ and $y$ are collinear; such contributions are
given by (4.26), (4.27) and (4.28), that is, by the torsion-Riemann and
torsion-torsion correlators.

In the second order formalism, the Riemann-Riemann correlator, which in
Min\-kow\-ski space corresponds to the $T^*$ product, is not identically zero
even in dimension $N=3$. Namely, if this were the case, {\it all} propagators
would vanish identically; but this cannot be true, since the collinear
singularity must be present in the propagator $\left<h(x)h(y)\right>$ in order
to produce the conical defect which is typical of 3D gravity.

A merit of the radial gauge is to show explicitly the absence of propagation in
the three-dimensional theory. The harmonic gauge -- on the contrary --
propagates a pure gauge field.

\bigskip \noindent
{\bf 4.4 \ Relation to the Mandelstam covariant.
Significant components.} \bigskip

We recall that the ``Mandelstam covariant'' ${\cal R}_{abcd}(x,P)$
(Section 2.1) is the curvature tensor observed in the local
reference frame parallel transported to a point from a fixed origin
$x$ along the path $P$. In the version of Tsamis and Woodard (Section 2.2)
the path $P$ is assumed to be geodesic.

It is then clear that in the radial gauge the (geodesic) Mandelstam
covariant referred to the origin of the coordinates can be written as
$$\eqalignno{
  {\cal R}_{abcd}(\xi) &= e^\mu_c(\xi) e^\nu_d(\xi) R_{ab \mu \nu}(\xi) ,
\cr}$$
where $R_{ab \mu \nu}$ is given by eq.\ (3.15). One verifies from
this expression that ${\cal R}$ is a scalar under coordinates
transformations.

Thus the (geodesic) Mandelstam covariant is a composite field which can be
constructed by radial fields. Its correlations can be written
using the propagators of $e^\mu_a(\xi)$ and $\Gamma^a_{\mu b}(\xi)$
given in Section 4.3, and the resulting analytical structure will
be the same, including hypergeometric functions etc. The index structure
is quite complicated however, and it is more interesting instead to show
explicitly the components of the radial vierbein-vierbein
correlation function, in order to gain a feeling of its behavior.

Denoting as usual the vierbein by $e_{a \mu}=\delta_{a \mu} + \tau_{a \mu}$
and all the integration procedure of the propagator (see Section 4.3)
by $\int d[\lambda] \int d[t]$, we may write
$$\eqalignno{
  \langle \tau_{a \mu}(x) \, \tau_{c \rho}(y) \rangle &=
  \int d[\lambda] \int d[t] \left[ x^\nu x^b y^\sigma y^d
  \langle R_{ab \mu \nu}(x) R_{cd \rho \sigma}(y) \rangle
  \right]_{x=\lambda x; \ y=ty} . &(4.29)
\cr}$$
In this formula, $R_{ab \mu \nu}(x)$ denotes the usual ``holonomic''
components of the (linearized) Riemann tensor, that is,
$R_{ab \mu \nu}(x)=\delta^\alpha_a \delta^\beta_b
R_{\alpha \beta \mu \nu}(x)$, and similarly for $R_{cd \rho \sigma}$.
Also, ``ultralocal'' correlations are disregarded (compare eq.\ (4.26)).

According to our usual assumptions, the two vierbein fields have to be
connected by a geodesic of length $\xi$; this can be realized for
instance by choosing the radial coordinates $x$ and $y$ in the
following way:
$$\eqalignno{
  x_0 &= \left( - {\xi \over 2}, \, 0, \, 0, \, 0 \right) ; \cr
  y_0 &= \left( {\xi \over 2}, \, 0, \, 0, \, 0 \right) . &(4.30)
\cr}$$
This singles out the direction 1 and causes $\tau$ to vanish by
radiality when one of its indices is equal to one; the other three
directions, however, are completely equivalent. The expression in
square brackets of (4.29) reduces then to
$$\eqalignno{
  & {1 \over 8} \xi^4 \langle R_{a 1 \mu 1}(x_0) \, R_{c 1 \rho 1}(y_0)
  \rangle &(4.31)
\cr}$$
and there are three possibilities of choosing the indices $\{ a \mu \}$
and $\{ c \rho \}$ of the two vierbein fields:

\noindent (a) like in $\langle \tau_{22} \tau_{22} \rangle$ or
$\langle \tau_{23} \tau_{23} \rangle$ ...; this represents the correlation
between the same component of the same vector of the vierbein;

\noindent (b) like in $\langle \tau_{22} \tau_{23} \rangle$ or
$\langle \tau_{22} \tau_{32} \rangle$ ...; this is a correlation between
different components of the same vector or between the same
component of different vectors of the vierbein;

\noindent (c) like in $\langle \tau_{22} \tau_{33} \rangle$ or
$\langle \tau_{22} \tau_{44} \rangle$ ...; that is, a correlation between
different components of different vectors, but chosen in such a
way that $a=\mu$ and $c=\rho$.

It is straightforward to insert these choices of indices in (4.31)
and to evaluate the propagator for each choice. Also, the integrals
in $\lambda$ and $t$ are trivial in this case and the dependence
on $\xi$ factorizes as $\xi^{-2}$, as expected on dimensional
grounds. The three cases (a), (b) and (c) give different results,
and it is interesting to look at them keeping in mind that each
vector of the vierbein is parallel-transported between $x_0$ and
$y_0$ in any configuration of the functional integral used to produce
the averages. It is found the following:

\noindent The correlation (a), which can also be expressed as
$\langle \tau_{ij} \tau_{ij} \rangle$ (no sum), does not vanish. It is in fact
the basic correlation function of the radial gauge, as we shall
now explain in short. Namely, it will be shown in Section 5.3 that
the invariant correlation between two vierbein fields includes the
matrix ${\cal U}$ of the parallel transport:
$$\eqalignno{
  G_{\rm Vierbein}(D) &= \langle e^a_\mu(x) {\cal U}^{\mu \mu '}_{a a'}(x,x')
  e^{a'}_{\mu '}(x') \rangle
\cr}$$
(here $D$ is the distance between $x$ and $x'$). But in radial gauge
this matrix reduces to the identity along any ray, thus we have for the
choice of coordinates (4.30)
$$\eqalignno{
  G_{\rm Vierbein}(\xi) &= \langle \tau_{22}(x_0) \tau_{22}(y_0) \rangle +
  \langle \tau_{23}(x_0) \tau_{23}(y_0) \rangle + \langle \tau_{24}(x_0)
\tau_{24}(y_0) \rangle
  + \cr
  &\ + \langle \tau_{33}(x_0) \tau_{33}(y_0) \rangle +
  \langle \tau_{34}(x_0) \tau_{34}(y_0) \rangle + \langle \tau_{44}(x_0)
\tau_{44}(y_0) \rangle .
  &(4.31)
\cr}$$
This is true to any order. We notice that the terms of (4.31) which differ
by the first index of the vierbein are certainly equal, since the
vectors of the vierbein are equivalent. Moreover, if no external
source is present, different components of the same vector are
equivalent too, in the average. One concludes that all the radial
correlations of the form $\langle \tau_{ij} \tau_{ij} \rangle$ are equivalent
on the vacuum and equal to ${1 \over 6}$ of the invariant correlation
$G_{\rm Vierbein}$.

\medskip \noindent The correlation functions (b), also expressible as
$\langle \tau_{ij} \tau_{kj} \rangle$ or $\langle \tau_{ij} \tau_{ik} \rangle$
(no sum),
vanish to lowest order and are in general not significant, because
there is no direct connection (apart from the orthonormality
realtions) between the values of two different components of the same
vierbein vector or the values of the same components of two different
vectors.

\medskip \noindent The correlations (c), also expressible like
$\langle \tau_{ii} \tau_{jj} \rangle$ (no sum; $i \neq j$), do not vanish
to lowest order. Nevertheless, this property is due to the weak-field
approximation and does not hold in general. It is easy to
verify by geometrical arguments that on a almost-flat space the i-th
component of the i-th vierbein vector keeps constant in sign, so its
correlation does not vanish. This ceases to be true on a generic
strongly curved space.

\vfill\eject

\centerline{\bf 5. GAUGE INVARIANT CORRELATIONS.} \bigskip

In the first part of this review (Chapters 2-4) we have shown how
some correlation functions of the gravitational field at
geodesic distance can be computed in a special physical gauge.
By this technique two-point functions
are obtained, which really depend on geodesic coordinates $x$
and $y$ (more exactly, they are hypergeometric functions of
$x$ and $y$ -- see Sections 4.2, 4.3) and have ``parallel transported''
tensorial components which are a generalization of the usual
Lorentz tensorial components (compare Section 3.3).

In this second part (Chapters 5, 6), on the contrary, the most
relevant scalar, or ``gauge invariant'' correlations will be
introduced; by this we mean functionals of the field, whose
corresponding classical quantities are independent of the choice
of the coordinates. The most typical and useful example of this
kind of correlations is the Wilson loop of the Christoffel or
anholonomic connection (Sections 5.1, 5.4).

The physical interpretation of these quantities turns out to be more
clear and interesting than in the case of the gauge-dependent
correlations. On the other hand, the ``geodesic corrections'' are
less easy to be taken into account in this case; we shall
illustrate a special procedure for that in Section 6.3.

This Chapter is organized as follows. In Section 5.1 the matrix
${\cal U}$ of the parallel transport is defined, both in terms
of the Christoffel connection $\Gamma^\alpha_{\mu \nu}$ and of the
gauge, or ``anholonomic'' connection $\Gamma^a_{\mu b}$. In Section 5.2
we specify the dynamical scheme as the traditional perturbative
scheme which starts from the Einstein lagrangian, splits the
gravitational field into a flat ``background'' and a weak,
quantized part. In Section 5.3 the invariant two-point functions
involving the Riemann curvature are computed to leading order.
They vanish as a consequence of the equations of motion. On the
contrary, the correlations of the metric and of the vierbein
do not vanish. Finally, in Section 5.4 we compute to leading
order the Wilson loop of the Christoffel connection and find that
it vanishes too. The physical interpretation of this property
is postponed to Chapter 6.

\bigskip \bigskip \noindent
{\bf G.1 \ Geometrical definitions.} \bigskip

We consider a classical spacetime $M$ described by a metric tensor
$g_{\mu \nu}(x)$ of signature $(-1,\, 1,\, 1,\, 1)$ (the conventions are
those of Weinberg [1972]).

The variation of a vector $V^\alpha$ by an infinitesimal parallel transport
is defined by
$$\eqalignno{
  dV^\alpha &= - \Gamma^\alpha_{\mu \beta}(x) \, V^\beta \, dx^\mu , &(5.1)
\cr}$$
where $\Gamma^\alpha_{\mu \beta}$ is the Christoffel connection
$$\eqalignno{
  \Gamma^\alpha_{\mu \beta} &= {1 \over 2} \, g^{\alpha \gamma} \,
  \left(\partial_\mu g_{\beta \gamma} + \partial_\beta g_{\mu \gamma} -
  \partial_\gamma g_{\mu \beta} \right) . &(5.2)
\cr}$$
Integrating (5.1) we find that the parallel transport of $V$ along a
finite differentiable curve connecting the points $x$ and $x'$ is performed by
the matrix
$$\eqalignno{
  {\cal U}^\alpha_{\beta}(x,x') &= {\rm P} \, \exp \int_{x}^{x'}
  dy^\mu \, \Gamma^\alpha_{\mu \beta}(y), &(5.3)
\cr}$$
where the symbol P means that the matrices
$(\Gamma_\mu)^\alpha_{\beta}=\Gamma^\alpha_{\mu \beta}$ are ordered along
the path. The indices of ${\cal U}^\alpha_\beta(x,x')$ are lowered and
raised by $g_{\alpha \gamma}(x)$ and $g^{\beta \gamma}(x')$, respectively.

When the manifold is curved, the matrix ${\cal U}$ depends not only on the
end points $x$ and $x'$, but also on the path. So,
if $C$ is a smooth closed curve on $M$, we define the loop functional
(or ``holonomy'') ${\cal W}(C)$ as
$$\eqalignno{
  {\cal W}(C) &= -4 + {\rm Tr} \, {\cal U}(C) =
  -4 + {\rm Tr} \, {\rm P} \exp \oint_C dx^\mu \Gamma_\mu(x) . &(5.4)
\cr}$$

The term $-4$ sets the holonomy to zero in the case of a
flat space, when the matrix ${\cal U}$ reduces to an identity matrix.

\medskip
Under a coordinates transformation $x \to \zeta$, the matrix \ ${\cal U}$
transforms in the following way
$$\eqalignno{
  {\cal U}^{\alpha}_{\beta}(x,x') &\to {\cal U}^\alpha_\beta(x,x') \,
  \left[ {\partial {\zeta}^{\gamma} \over \partial {x}^\alpha} \right]_x \,
  \left[ {\partial {x}^\beta \over \partial {\zeta}^{\varepsilon}}
  \right]_{x'}.
\cr}$$
For a closed curve, this transformation, being of the form
$$\eqalignno{
  {\cal U} &\to \Omega {\cal U} \Omega^{-1} ,
\cr}$$
does not affect the trace of ${\cal U}$. So the loop ${\cal W}(C)$ is invariant
with respect to coordinate transformations.

In the first order formalism, the ``anholonomic'' components of a vector
are defined by
$$\eqalignno{
  V^a &= V^\mu \, e^a_\mu(x) .
\cr}$$

The equivalent of (5.1) in terms of the anholonomic connection
$\Gamma^a_{\mu b}$ is
$$\eqalignno{
  dV^a &= - \Gamma^a_{\mu b}(x) V^b dx^\mu &(5.5)
\cr}$$
and the matrix ${\cal U}$ of the finite parallel transport has an expression
which is formally the analogue of (5.3), namely
$$\eqalignno{
  {\cal U}^a_b(x,x') &= {\rm P} \, \exp \int_{x}^{x'}
  dy^\mu \, \Gamma^a_{\mu b}(y) .
\cr}$$

We remind (see eq.\ 3.5)
that the relation between the connections $\Gamma^\alpha_{\mu \beta}$ and
$\Gamma^a_{\mu b}$ is the following
$$\eqalignno{
  \Gamma^\alpha_{\mu \beta} &= e^\alpha_a e^b_\mu \Gamma^a_{\beta b}
  + e^\alpha_a \partial_\beta e^a_\mu
\cr}$$
and that the relation between the matrices ${\cal U}^\alpha_\beta$
and ${\cal U}^a_b$ is
$$\eqalignno{
  {\cal U}^a_b(x,x') &= e^a_\alpha(x) {\cal U}^\alpha_\beta(x,x')
  e^\beta_b(x'). &(5.6)
\cr}$$

It is known that gravity in the vierbein formalism has a local Lorentz
invariance, since the definition of the vierbein,
eq.\ 3.3, is insensitive to a Lorentz rotation of
$e^a(x)$, $e^b(x)$. The connection $\Gamma^a_{\mu b}$ is then completely
analogous to an usual gauge connection, and its Wilson loop
$$\eqalignno{
  {\cal W}(C) &= -4 + {\rm Tr} \, ({\cal U}^a_b)(C)
\cr}$$
is a natural invariant quantity of the theory. But from eq.\ (5.6) we see
that this loop is equal to that defined in (5.4). So the Christoffel
connection $\Gamma^\alpha_{\mu \beta}$ and the anholonomic $\Gamma^a_{\mu b}$
connection have the same loop, denoted by ${\cal W}(C)$. In the computations we
shall employ the connection $\Gamma^\alpha_{\mu \beta}$, which is usually
simpler to deal with.

\medskip
When the exponential in (5.4) is expanded, one obtains terms with 1, 2, 3,
... fields $\Gamma$. We introduce the notation, to be used in the following
$$\eqalignno{
  {\cal U} &= {\bf 1} + \oint_C dx^\mu \, \Gamma_\mu(x) +
  {1 \over 2} \, {\rm P} \oint dx^\mu \oint dy^\nu \, \Gamma_\mu(x) \,
  \Gamma_\nu(y) + ... \cr
   &= {\bf 1} + {\cal U}^{(1)} + {1 \over 2} \, {\cal U}^{(2)} + ... &(5.7)
\cr}$$
and
$$\eqalignno{
  {\cal W} &= -4 + {\rm Tr} \, {\cal U} =
  {\rm Tr} \, {\cal U}^{(1)} + {1 \over 2} \, {\rm Tr} \, {\cal U}^{(2)}
  + ... &(5.8)
\cr}$$

\bigskip \noindent
{\bf 5.2 \ Dynamics and perturbation scheme.} \bigskip

Let us now introduce dynamics through the Einstein action, which has
the form (for any $N>2$)
$$\eqalignno{
  S &= {1 \over \kappa^2} \int d^Nx \, \sqrt{g(x)} \, R(x) .
\cr}$$

A completely consistent quantum theory of this model does not exist
yet. In view of applications to lattice gravity, we shall
refer in the following, as a possible approximation to the full
theory, to the (regularized) functional integral approach (see for
instance [Hawking, 1979; Mazur and Mottola, 1990]; compare also
Chapter 7).

In the perturbative evaluations of this chapter, however, we follow the
``traditional'' approach and regard quantum gravity as an ordinary
field theory on a fixed flat background. The fundamental field is
$\kappa h_{\mu \nu}(x)=g_{\mu \nu}(x)-\delta_{\mu \nu}$, which is
subject to the gauge transformations
$$\eqalignno{
  h_{\mu \nu}(x) &\to h_{\mu \nu}(x) + \partial_\mu f_\nu(x) +
  \partial_\nu f_\mu(x) , &(5.9)
\cr}$$
where $f_\mu(x)$ is an arbitrary function. Eq.\ (5.9) represents
the action on $h$ of a linearized diffeomorphism
${x'}^\mu=f^\mu(x)$. It should be noticed, nevertheless, that the
argument $x$ remains unchanged in the transformation (5.9): this is
what is meant by ``fixed background''. The transformation
of $\Gamma$ which corresponds to (5.9) is
$$\eqalignno{
  \Gamma^\alpha_{\mu \nu}(x) &\to \Gamma^\alpha_{\mu \nu}(x)
  + \partial_\mu \partial_\nu  f^\alpha(x) . &(5.10)
\cr}$$

Keeping the quadratic part of $S$ and adding to it the harmonic
gauge-fixing
$$\eqalignno{
   {1 \over 2} & \left( \partial^\mu h_{\mu \nu} -
   {1 \over 2} \partial^\nu h^\mu_\mu \right)^2 ,
\cr}$$
we obtain the Feynman-De Witt propagator (compare Section 3.5,
where however $h$ differs by a factor $\kappa$)
$$\eqalignno{
  \langle h_{\mu \nu}(x) h_{\rho \sigma}(y) \rangle &=
  - {1 \over (2\pi)^2}  \ { {\delta_{\mu \rho} \delta_{\nu \sigma} +
  \delta_{\mu \sigma} \delta_{\nu \rho} -
  \delta_{\mu \nu} \delta_{\rho \sigma}} \over (x-y)^2 }. &(5.11)
\cr}$$

Due to the non-polynomial character of the lagrangian, there are
infinitely many interaction vertices; the first two ones,
respectively proportional to $\kappa$ and $\kappa^2$, connect 3 and 4
fields $h$. Hence the first few orders of perturbation theory are
formally very similar to those of Yang-Mills theory [see
for instance Veltman, 1976].

As it is known, the Einstein action is not the only action which
describes correctly the macroscopic behaviour of gravity. In
particular, the $(R+R^2)$-action (see Chapter 7) has been proposed
a long time ago as a renormalizable generalization of General
Relativity. Also, in order to make the euclidean Einstein action
bounded from below, a ``stabilized'' euclidean action has been
recently proposed. We shall briefly consider the effects of these
modified kinds of dynamics on the correlation functions in
Section 5.4.

\bigskip \noindent
{\bf 5.3 \ Two-point functions.} \bigskip

An important point to be clarified about the vacuum correlation
functions in gravity is the following: ``which are the most
suitable field quantities to be correlated''? We recall that the
criteria for the existence of gravitational waves [Zakharov, 1973]
are usually based on the propagation of the curvature, which is
believed to be the most physical effect of gravitation. One is thus
led to consider as first candidates the following quantities:

\noindent -
the Riemann tensor $R^\alpha_{\beta \mu \nu}(x)$;

\noindent -
the Ricci tensor $R_{\mu \nu}(x)=R^\alpha_{\mu \alpha \nu}(x)$;

\noindent -
the curvature scalar $R(x)=g^{\mu \nu}(x)R_{\mu \nu}(x)$;

\noindent -
the ``rotation matrix'', or plaquette ${\cal R}^\alpha_\beta(x)=
R^\alpha_{\beta \mu \nu}(x)\sigma^{\mu \nu}$, where $\sigma^{\mu \nu}$
is an infinitesimal surface around $x$ (${\cal R}$ must not be
confused with the ``Mandelstam covariant'' of Chapter 2).

The linearized parts of $R$, $R_{\mu \nu}$ and $R^\alpha_{\beta \mu \nu}$
are respectively given by
$$\eqalignno{
  & R^L = \partial^2 h_\alpha^\alpha - \partial^\alpha \partial^\beta
  h_{\alpha \beta}; \cr
  & R^L_{\mu \nu} = {1 \over 2} (\partial^2 h_{\mu \nu}
  + \partial_\mu \partial_\nu h_\alpha^\alpha
  - \partial_\mu \partial^\alpha h_{\alpha \nu} -
  \partial_\nu \partial^\alpha h_{\alpha \mu}); \cr
  & R^L_{\alpha \beta \mu \nu} = {1 \over 2}
  ( \partial_\alpha \partial_\mu h_{\beta \nu}
  - \partial_\beta \partial_\mu h_{\alpha \nu} -
  \partial_\alpha \partial_\nu h_{\beta \mu}
  + \partial_\beta \partial_\nu h_{\alpha \mu}).
\cr}$$

Let us consider the invariant correlations
$$\eqalignno{
  & G_R(D) = \langle R(x) \, R(x') \rangle_0; \cr
  & \cr
  & G_{Ricci}(D) = \langle R_{\mu \nu}(x) \,
  {\cal U}^{\mu \nu \mu' \nu'}(x,x')
  \, R_{\mu' \nu'}(x') \rangle_0; \cr
  & \cr
  & G_{Riemann}(D) = \langle R^\alpha_{\beta \mu \nu}(x) \,
  {\cal U}^{\beta \mu \nu \beta' \mu' \nu'}_{\alpha \alpha'}(x,x') \,
  R^{\alpha'}_{\beta' \mu' \nu'}(x') \rangle_0; \cr
  & \cr
  & G_{Loop}(D,\sigma,\sigma') = \langle {\cal R}^\alpha_{\beta}(x) \,
  {\cal U}^{\beta \beta'}_{\alpha \alpha'}(x,x') \,
  {\cal R}^{\alpha'}_{\beta'}(x')
  \rangle_0 .
\cr}$$
Here ${\cal U}$ is the matrix of the parallel transport introduced
in Section 5.1, computed along the geodesic joining $x$ to $x'$;
$D$ is the geodesic distance between $x$ and $x'$.

On a flat background and to order
$\kappa^2$ these functions are very simple, since the
matrices ${\cal U}(x,x')$ reduce to identity matrices and the
tensors retain only their linearized parts. It is easy to
verify that for the first three correlation functions one has
[M., 1992b]
$$\eqalignno{
  G_R &\sim G_{Ricci} \sim G_{Riemann} \sim \kappa^2 \, \partial^2
  \delta^4(x-x') + o(\kappa^2). &(5.12)
\cr}$$
Thus these correlation functions vanish if $x$ and $x'$ are distinct,
and they are not indicative of the correlation lenght of the system
-- at least in this regime of weak fields.

The perturbative result above can be related to a general
property of the correlation functions [see also Collins, 1984].
Let us consider a functional integral of the form
$$\eqalignno{
  z &= \int d[\phi(x)] \, e^{-S[\phi(x)]} &(5.13)
\cr}$$
and perform an infinitesimal arbitrary translation
$\delta \phi(x)$ of the integration variable $\phi(x)$.
Since the integration measure is invariant with respect to
such a transformation, we have
$$\eqalignno{
  z &= \int d[\phi(x)] \, e^{-S[\phi(x)]} \,
  e^{-\int dx \, \delta \phi(x) \, E(\phi(x))} ,
\cr}$$
where $E(\phi(x))$ represents the classical equations of motion.
Expanding the second exponential and disregarding terms
$O(\phi^2)$ we obtain
$$\eqalignno{
  z &= \int d[\phi(x)] \, e^{-S[\phi(x)]} \,
  \left( 1 - \int dx \, \delta \phi(x) \, E(\phi(x)) \right) ,
\cr}$$
from which, by comparison with (5.13), we have
$$\eqalignno{
  \int d[\phi(x)] \, e^{-S[\phi(x)]} \,
  \int dx \, \delta \phi(x) \, E(\phi(x))  &= 0 . &(5.14)
\cr}$$
If we choose $\delta \phi(x)$ to be of the form
$\varepsilon \delta^4(x)$, we conclude that
$$\eqalignno{
  \langle  E(\phi(x)) \rangle_0 &= 0 ;
\cr}$$
performing in (5.14) one more translation of the form
$\varepsilon \delta^4(x')$ we obtain by similar arguments
$$\eqalignno{
  \langle  E(\phi(x)) \, E(\phi(x')) \rangle_0 &= 0 ,
\cr}$$
and so on. That is, in a quantum field theory the vacuum
correlations of field equations vanish.

In the case of Einstein gravity, the equations of motion are
$$\eqalignno{
  R_{\mu \nu}(x) &= 0 ;
\cr}$$
thus we have to any order
$$\eqalignno{
  & \langle R_{\mu \nu}(x) \,  R_{\rho \sigma}(x') \rangle_0 = 0 . &(5.15)
\cr}$$
Since in our perturbative approximation for the functions $G$
the indices are contracted, in practice,
just by $\delta_{\mu \nu}$ instead of the full metric $g_{\mu \nu}$, from
(5.15) follows (5.12).

\medskip
Going back to the first-order analysis, we notice that the last
correlation function, $G_{Loop}$, is identical, in this approximation,
to a Wilson loop computed along a dumbbell-like contour (fig.\ 2).
The evaluation of the Wilson loop will be the subject of the
next Section, where we shall see, however, that it vanishes in this
approximation. In conclusion, the only gravitational correlation
functions which do not vanish to lowest order are those of the
metric or of the vierbein
$$\eqalignno{
  & G_{Metric}(D) = \langle g_{\mu \nu}(x) \, {\cal U}^{\mu \nu \mu'
\nu'}(x,x') \,
  g_{\mu' \nu'}(x') \rangle_0; \cr
  & \cr
  & G_{Vierbein}(D) = \langle e^a_\mu(x) \, {\cal U}^{\mu \mu'}_{a a'}(x,x') \,
  e^{a'}_{\mu'}(x') \rangle_0,
\cr}$$
as can be easily checked by a computation similar to that of the
correlation functions of the curvature. They both behave like
(compare also Section 4.4)
$$\eqalignno{
  G(D) & \sim {\kappa^2 \over D^2} .
\cr}$$

\bigskip \noindent
{\bf 5.4 \ Wilson loops.} \bigskip

In the usual gauge theories the Wilson loop of the connection is one of
the most important observable quantities ([Kogut, 1979]; see also the
Section on the static potential energy, 6.4).

Unfortunately, in quantum gravity the Wilson loop
of the Christoffel connection vanishes, to lowest
order, along any contour\footnote\dag{Hamber ([1993]; see
Chapter 7) argues that also in the non-perturbative sector
the Wilson loop ``does not have the same interpretation as in gauge
theories, since it is not associated with the newtonian potential
energy of two static bodies. In ordinary gauge theories at
strong coupling the Wilson loop decays like the area of the
loop, due to the strong independent fluctuations of the gauge
fields at different points in spacetime and ensuing cancellations.
In lattice gravity the situation is quite different since the
connections cannot be considered as independent variables, and
the fluctuations in the deficit angles at different points
in spacetime are strongly correlated.'' \
The reasoning we present here holds in perturbation theory and to lowest
order. See, however, Section 6.4 for
a physical justification of why a closed loop cannot give in gravity
the same result it gives in the gauge theories with ``charges'' of
opposite signs.}.
The proof of this surprising property is straightforward: consider
eq.s (5.7), (5.8) and denote by roman letters the corresponding vacuum
averages. For instance:
$$\eqalignno{
  U = \langle {\cal U} \rangle_0 &= {\bf 1} +
  \langle {\cal U}^{(1)} \rangle_0
  + {1 \over 2} \, \langle {\cal U}^{(2)} \rangle_0 + ... \cr
  &= {\bf 1} + U^{(1)} + {1 \over 2} \, U^{(2)} + ... \cr
  W = \langle {\cal W} \rangle_0 &= -4 + {\rm Tr} \, U  \cr
  &= {\rm Tr} \, U^{(1)} + {1 \over 2} {\rm Tr} \, U^{(2)} + ...
  \cr
  &= W^{(1)} + {1 \over 2} W^{(2)} + ...
\cr}$$

The Wilson loop $W$ is given by
$$\eqalignno{
  W &= {1 \over 2} \, {\rm P} \oint dx^\mu \oint dy^\nu \,
  \langle \Gamma^\alpha_{\mu \beta}(x) \Gamma^\beta_{\nu \alpha}(y) \rangle
  + o(\kappa^2), &(5.16)
\cr}$$
where the brackets denote the bare propagator of $\Gamma$.
Using (5.2), (5.11) we obtain
$$\eqalignno{
  \langle \Gamma^\alpha_{\mu \beta}(x)
  \Gamma^\beta_{\nu \alpha}(y) \rangle &=
  - {\kappa^2 \over (2\pi)^2} \left\{ a_1 \,
  {\partial \over {\partial x_\mu}}  {\partial \over {\partial x_\nu}}
  + a_2 \delta_{\mu \nu} \, \partial^2 \right\}
  {1 \over (x-y)^2} = \cr
  &= - {\kappa^2 \over (2\pi)^2} \left\{ a_1 \,
  {\partial \over {\partial x_\mu}}  {\partial \over {\partial x_\nu}}
  {1 \over (x-y)^2} + a_2
  \delta_{\mu \nu} \, \delta^4(x-y) \right\} , &(5.17)
\cr}$$
where $a_1$ and $a_2$ are two numerical coefficients.
 From (5.16), (5.17) one concludes that $W$
vanishes to this order, except for some divergent perimeter
contributions due to the $\delta$-function.

In the next chapter we shall illustrate in detail the physical meaning
of this vanishing and its consequences.
About the causes, it is essentially due to symmetry
reasons and to the absence of massive propagating modes.
If one writes down the most general form a massless graviton
propagator may have while respecting Poincar\'e invariance\footnote\ddag{A
non-vanishing contribution proportional to $\hbar$ may arise on
a non-flat background [M., 1993 c].}, and repeats
the calculation of the Wilson loop, one still finds that it
vanishes to order $\hbar$.

For instance, the situation in $(R$+$R^2)$-gravity
(see Section 7.1) is different.
Here $R$ and $R_{\mu \nu}$, unlike in Einstein gravity,
can propagate (cfr.\ Section 5.3). There can be some non-vanishing
invariant correlations of the curvature and eq.\ (5.17) will be replaced by
$$\eqalignno{
  \langle \Gamma^\alpha_{\mu \beta}(x)
  \Gamma^\beta_{\nu \alpha}(y) \rangle  &=
  \left\{ a_1 \,
  {\partial \over {\partial x_\mu}}  {\partial \over {\partial x_\nu}}
  + a_2 \delta_{\mu \nu} \, \partial^2 \right\}
  \left( - {\kappa^2 \over (2\pi)^2 (x-y)^2}
  + {\cal D}_M(x-y) \right),
\cr}$$
where ${\cal D}_M$ is the propagator of a massive mode.
Therefore, the term proportional to $a_2$ is not ultra-local any more.
However, we recall that only the lightest
(massless) states with spin two should dominate at large distances
[see for ex.\ Hamber, 1992a].

\medskip
As it is well known, Einstein's gravity written in the
first order formalism is a gauge theory of the Lorentz
group (i.e., the action is invariant under local Lorentz
transformations), but not of the whole Poincar\'e group $ISO(3,1)$. A
consistent gauge formulation can be obtained only introducing some auxiliary
fields $q^a$ [Grignani and Nardelli, 1992].

So it is not possible to consider in (3+1) dimensions, like in (2+1)-gravity
[Witten, 1988], the holonomies of the Lie algebra valued connection
$$\eqalignno{
  {\cal A}_\mu(x) &= e^a_\mu(x) P_a + \Gamma^{ab}_\mu(x) \omega_{ab} ,
\cr}$$
where $P_a$ and $\omega_{ab}$ are the generators of the translations and of the
Lorentz transformations.
The holonomies of ${\cal A}_\mu$ may have
more content than the holonomies of $\Gamma_\mu$ alone. For instance,
it can be easily verified that the term
$$\eqalignno{
  {\rm Tr} \oint dx^\mu \oint dy^\nu \langle
  e^a_\mu(x) P_a \, e^b_\nu(y) P_b \rangle_0 &=
  -2 \delta_{ab} \oint dx^\mu \oint dy^\nu \langle
  e^a_\mu(x) \, e^b_\nu(y) \rangle_0
\cr}$$
is not trivial to leading order, unlike the corresponding term containing
the connection. However, this term does not
respect the invariance of the action.

\vfill\eject

\centerline{\bf 6. \ FURTHER PROPERTIES OF THE WILSON LOOPS.} \bigskip

\bigskip \noindent
{\bf 6.1 \ Geometrical and physical interpretation.} \bigskip

As we have seen in Section 5.4, the Wilson loop of the Christoffel
connection vanishes to leading order in Einstein's gravity. Also,
we remind that this loop is equivalent to that of the gauge
connection $\Gamma^a_{\mu b}$ (see eq.\ (5.6) and the ensuing
discussion).

In order to understand the physical meaning of this vanishing,
a group-theoretical analysis of the matrix ${\cal U}$ is needed.
We shall see that in the euclidean theory the vanishing of its trace amounts to
a very strong geometrical statement.

Let us first consider, for illustrative purposes, the case of a Yang-Mills
theory of the group $SO(3)$. The gauge connection has the form
$$\eqalignno{
  A_\mu(x) &= A^i_\mu(x) \, L_i \, ; \qquad i=1,2,3 ,
\cr}$$
where the matrices $L_i$ constitute a representation of the Lie algebra of the
group. In particular, to fix the ideas, let us choose the adjoint
representation; in this case the matrices $L_i$ have elements $(L_i)^m_n$
($m,n=1,2,3$), which are related to the structure constants $\varepsilon_{imn}$
of the group. The connection $A_\mu(x)$ performs the parallel transport of a
3-dimensional vector $V^n$ in the ``internal'' space according to the formula
(compare eq.\ (5.1))
$$\eqalignno{
  dV^m &= A^i_\mu(x) \, (L_i)^m_n \, V^n \, dx^\mu .
\cr}$$
The vector is rotated during the transport, but its length remains unchanged.
Let us consider the matrix ${\cal O}(C)$ which describes the parallel transport
along a closed curve $C$. ${\cal O}(C)$ is defined by a P-exponential, through
a formula similar to eq.\ (5.3). Suppose that we take a vector $V$ in a
point $P$ of $C$, and parallel-transport it along $C$, returning to $P$; let us
denote by $V'$ the new vector we obtain in this way. The vectors $V$ and $V'$
have the same length, that is
$$\eqalignno{
  \delta_{mn} V^m V^n &= \delta_{mn}{V'\, }^m {V'\, }^n , &(6.1)
\cr}$$
but they differ by an angle $\theta$, which is related to the trace of ${\cal
O}(C)$. For small angles, we have, by a proper choice of the coordinate axes in
the internal space
$$\eqalignno{
  {\cal O}(C) &= \left(
  \pmatrix{
    1 - {1 \over 2} \theta^2 & \theta & 0 \cr
    - \theta & 1 - {1 \over 2} \theta^2 & 0 \cr
    0 & 0 & 1
  \cr}
  \right) , &(6.2)
\cr}$$
that means
$$\eqalignno{
  {\rm Tr} \, {\cal O}(C) &= 3 - \theta^2 .
\cr}$$

More generally, we remind that the Lie algebra of $SO(3)$ has just one Casimir
invariant, namely the operator
$$\eqalignno{
  L^2 &= L_1^2 + L_2^2 + L_3^2 .
\cr}$$
This operator commutes with each of the $L_i$'s, so we can in general rotate
our coordinate system as to have $L^2 = L_3^2$, and the rotation matrix takes
in this case the form (6.2), i.e.\ we have
$$\eqalignno{
  {\cal O}(C) &= {\bf 1}+\theta L_3 + {1 \over 2} \theta^2 L_3^2 + ... &(6.3)
\cr}$$
Taking the trace of (6.3), remembering that ${\rm Tr}\, L_i=0$
and using the normalization condition of the Lie generators
$$\eqalignno{
  {\rm Tr} \, L_i L_j &= - 2 \delta_{ij} ,
\cr}$$
we find that $\theta^2$ is the coefficient of the Casimir invariant in the
expansion of the exponential.

\medskip
Next we come to consider the group $SO(4)$. Intuitively, adding a new dimension
we can make an independent rotation. Multiplying two 4-dimensional matrices
similar to (6.2), the first representing a rotation by an angle
$\theta_I$ perpendicular to one plane and the second a rotation by another
angle $\theta_{II}$ perpendicular to another plane, we find that
$$\eqalignno{
  {\rm Tr} \, {\cal O}(C) &= 4 - ( \theta_I^2 + \theta_{II}^2 ) &(6.4)
\cr}$$
Also we know that $SO(4)=[SO(3)]_I \, \times \, [SO(3)]_{II}$ and that
we have two Casimirs now [see for instance Wybourne, 1974],
corresponding to $(L_I^2 + L_{II}^2)$,
whose ``eigenvalue'' appears in (6.4), and $(L_I^2 - L_{II}^2)$,
which is not of interest in this case.

\medskip
The group $SO(4)$ is the relevant one for euclidean quantum gravity. In fact,
the geometrical interpretation of the matrix ${\cal U}(C)$ is the following.
During the parallel transport of a vector $V$ in spacetime, its length, given
by
$$\eqalignno{
  |V|^2 &= V^a V^b \delta_{ab} = V^\mu V^\nu g_{\mu \nu}(x),
\cr}$$
does not change. If we transport $V$ along a closed curve $C$, returning
to the starting point, we obtain another vector $V'$, which has the same
length of $V$ , and differs from it only in the orientation. Hence we have
for any vector
$$\eqalignno{
  V^a V^b \delta_{ab} &= {V'\,}^a {V'\,}^b \delta_{ab} =
  {\cal U}^a_c(C) V^c \, {\cal U}^b_d(C) V^d \, \delta_{ab} ,
\cr}$$
or, in matrix notation,
$$\eqalignno{
  {\cal U}^T(C) \, {\cal U}(C) &= {\bf 1} .
\cr}$$
The matrix ${\cal U}$ belong then to $SO(4)$ and its trace has the form
(6.4).

\medskip
If the variance of the angles $\theta_I$ and $\theta_{II}$ is zero
to order $\hbar$ (because $W^{(2)}$ vanishes), the angles
themselves have to vanish identically in any configuration, that is
$$\eqalignno{
  {\cal U}(C) &= {\bf 1} \ \ {\rm for \ any} \ C.
\cr}$$
This is a very strong geometrical statement, as it implies that, still
to order $\hbar$, all the weak field configurations which effectively enter
the functional integral
$$\eqalignno{
  z &= \int d[h] \, \exp \left\{ - \hbar^{-1} \, S[h] \right\} &(6.5)
\cr}$$
have no curvature. In other words, the curved configurations -- which possibly
dominate in other regimes -- are in this approximation totally suppressed.

\medskip
This unexpected situation should be compared with what happens, for instance,
in a ordinary $SO(3)$ or $SO(4)$ gauge theory. In this case the leading term
$W^{(2)}(C)$ does not vanish and the variance of the rotation angles is not
zero to order $\hbar$.  For instance, if the curve $C$ has the form of a
rectangle of sides $L$ and $T$, with $L \ll T$, the quantity $\, -(\hbar
T)^{-1} \log \langle \theta^2 \rangle_0$ is
the potential energy of two non-abelian
charges kept at rest at a distance $L$ each from the other.

So the matrices of the parallel transport in the ``internal'' gauge manifold,
considered configuration by configuration, are not equal to the identity
matrix. Interpreting $\hbar$ as the temperature $\Theta$ of an equivalent
statistical system, we see that when $\Theta$ grows from zero to some small
value -- such that we may disregard $\Theta^2$ or higher orders -- the
Yang-Mills fields develop ``localized excitations'', i.e.\ regions of various
sizes where the Yang-Mills curvature is not vanishing.

All this does not happen for the gravitational field, which remains essentially
in a ``flat'' state. Such a picture also explains the absence in this
approximation of any invariant correlation of the curvature (compare
Section 5.3).

The curved configurations to order $\hbar$ in the functional integral
could have been interpreted in a natural way as ``virtual gravitons'',
since they represent gauge-invariant excitations which are off-shell
and localized in space and time. (Virtual ``gravitons of the metric''
do exist in perturbation theory, but they do not necessarily
represent physical objects.) Their absence is a very peculiar property
of gravity, which also forces us to imagine a peculiar mechanism
for the gravitational interaction (see Section 6.4).

\bigskip \noindent
{\bf 6.2 \ The dumbbell correlation function.} \bigskip

In this section all the matrix elements of $U$ will be computed to order
$\kappa^2$, for a dumbbell-like contour (see fig.\ 2).
To this order we have $U^{(1)}=0$, while $U^{(2)}$ is given by
$$\eqalignno{
  {U^{(2)}}^\alpha_\beta &= {\rm P} \oint_C dx^\mu
  \oint_C dy^\nu \, \langle \Gamma^\alpha_{\mu \gamma}(x) \,
  \Gamma^\gamma_{\nu \beta}(y) \rangle, &(6.6)
\cr}$$
where $C$ is the dumbbell. $U^{(2)}$ is
invariant to order $\kappa^2$, by virtue of
(5.10). Although not invariant to higher orders,
it constitutes an interesting example of two-point
correlation function.

Disregarding gradient terms in the integrand and ultra-local
terms of the form $\delta^N(x-y)$, eq.\ (6.6) becomes, in the
Feynman-De Witt gauge and in any dimension $N>2$
$$\eqalignno{
  {U^{(2)}}^\alpha_\beta &= {1 \over 4} \, \delta^{\gamma \gamma '} \,
  \delta^{\alpha \alpha '} \, {\rm P}
  \oint_C dx^\mu \oint_C dy^\nu
  \left< [\partial_\gamma h_{\alpha ' \mu}(x) -
  \partial_{\alpha'} h_{\mu \gamma}(x)]
  [\partial_\beta h_{\gamma ' \nu}(y) -
  \partial_{\gamma'} h_{\nu \beta}(y)]
  \right> \cr
  &= {1 \over 4} \, c_N \kappa^2 \  {{3N-N^2} \over {N-2}} \,
  {\rm P} \oint_C dx^\mu \oint_C dy_\mu \,
  {{\partial} \over {\partial x_\alpha}} {{\partial} \over {\partial y^\beta}}
\,
  {1 \over (x-y)^{N-2}}. &(6.7)
\cr}$$
It is immediate to verify that the trace of ${U^{(2)}}$ is
an ultra-local term. Eq.\ (6.7) is remarkable under two aspects.
First, it shows that the matrix ${U^{(2)}}$ is symmetric.
Second, we see that it vanishes identically in dimension $N=3$;
this happens because the Riemann curvature in an
empty space-time of dimension 3 is always zero (compare the
discussion following eq.\ (5.12).

When $C$ is a dumbbell the integral breaks down in 16 parts
(fig.\ 3), which may be denoted as follows
$$\eqalignno{
  {U^{(2)}}(r,D) &= u_{AA} + u_{BB} + u_{CC} + u_{DD} + \cr
  & + u_{AB} + u_{BA} + u_{AC} + u_{CA} + u_{AD} + u_{DA} + \cr
  & + u_{BC} + u_{CB} + u_{BD} + u_{DB} + u_{CD} + u_{DC} = \cr
  &= u_{AA}(r) + u_{CC}(r) + \cr
  & + u_{BB}(d) + u_{DD}(d) + 2 u_{BD}(d) + \cr
  & + 2 u_{AC}(r,D).
\cr}$$
In this equation we have put $d=D-2r$ and omitted the indices
$\alpha$ and $\beta$. The ``transport'' terms depending on
$d$ can be easily shown to vanish in the limit
of zero-width $\delta$ of the strip (this holds only for the linearized
theory, which behaves like an abelian one).

The most interesting case is that of $r\ll D$.
We remind that when a vector is transported around an infinitesimal
surface $dx^\mu \wedge dx^\nu$, its variation is given by the matrix
$$\eqalignno{
  M^\rho_\gamma &= R^\rho_{\gamma \mu \nu} \, dx^\mu \wedge dx^\nu,
\cr}$$
where $R^\rho_{\gamma \mu \nu}$ is the Riemann curvature.
Like in Section 5.3, let us
denote by $\cal{R}^\rho_\gamma$ the matrix $M^\rho_\gamma$ divided
by the area, in the case of the surface being a small geodesic
circle. $\cal{R}^\rho_\gamma$ is a kind of ``regularized curvature'';
the regularization is non-local and gauge invariant, at the
linearized level.
Thus the integral ${U^{(2)}}^\alpha_\beta(r,D)$ may symbolically be written
in this limit as
$$\eqalignno{
  {U^{(2)}}^\alpha_\beta(r,D) &= 2\pi^2 r^4 \left\{
  \langle {\cal{R}}^\alpha_\gamma(0) {\cal{R}}^\gamma_\beta(0) \rangle +
  \langle {\cal{R}}^\alpha_\gamma(0) {\cal{R}}^\gamma_\beta(D) \rangle
  \right\}.
\cr}$$
The first term on the r.h.s.\ is divergent but independent on $D$
and the second one is the ``dumbbell function'' (fig.\ 4)
$$\eqalignno{
  \langle {\cal{R}}^\alpha_\gamma(0) {\cal{R}}^\gamma_\beta(D) \rangle =
  \lim_{a \rightarrow 0} \
  {{-c_N \, \kappa^2 \, (3N-N^2)} \over {4 \pi^2 a^2 \, D^{N+2}}}
  \int_0^{2\pi} ds \int_0^{2\pi} dt \, \, {1 \over 2} \cos (s-t) \,
  k^N \, \phi^\alpha_\beta(s,t,a),
\cr}$$
where $ = r/D$ and the non-zero matrix elements are
$$\eqalignno{
  & \phi^1_1(s,t,a)  =  1 - N k^2 (1-a u)^2 ; \cr
  & \phi^2_2(s,t,a) = 1-N k^2 a^2 v^2 ; \cr
  & \phi^1_2(s,t,a) = \phi^2_1(s,t,a) =
  -N \, k^{2} a v (1- a u); \cr
  & \phi^\mu_\mu(s,t,a) = 1;
  \qquad \mu=3,4,... \cr
\cr}$$
with
$$\eqalignno{
  & k = 1 / \sqrt{1-2a u + 2a^2 w}; \cr
  & u  =  \cos s - \cos t; \qquad
  v  =  \sin s - \sin t; \qquad
  w  =  1-\cos(s-t).
\cr}$$

Expanding these elements in powers of $a$ and
integrating\footnote\dag{This is a quite long calculation, which can be easily
performed using for instance the symbolic program {\it Mathematica}.},
one finds that the first non-zero coefficients are those of $a^2$.
In this way, denoting by $\Phi$ the integral of $\phi$
with respect to $s$, $t$, one finally obtains
$$\eqalignno{
  & \langle {\cal{R}}^\alpha_\gamma(0) {\cal{R}}^\gamma_\beta(D) \rangle =
  {{-c_N \, \kappa^2 \, (3N-N^2)} \over {4 \pi^2 \, D^{N+2}}}
  \ \Phi^\alpha_\beta;
\cr}$$
$$\eqalignno{
  & \Phi^1_1 = {1 \over 2} N(1+N)(N-2); \cr
  & \Phi^2_2 = -{1 \over 2} N(N-2); \cr
  & \Phi^1_2 = \Phi^2_1 = 0; \cr
  & \Phi^\mu_\mu = - {1 \over 2} \, N^2, \qquad \mu=3,4,... &(6.8)
\cr}$$
In eq.\ (6.8) one can verify one more time
that the trace vanishes. Actually, this is the point where
this fact was noticed for the first time.

\bigskip \noindent
{\bf 6.3 \ Geodesic corrections.} \bigskip

This section will be concerned with a problem raised
by the presence of a ``dynamic metric'' in quantum gravity.
The idea of a dynamic geometry is inherent to General Relativity,
but it could not be applied to the quantized theory as long
as it was considered as a field theory on a fixed background.
A recent realization of this idea can be found in the Montecarlo
simulations of Regge calculus (see Chapter 7).

\medskip
Our attention still focuses on the loops of the connection.
A natural question arises: how can they be defined
in the absence of a flat background?

Let us first consider the case of a
classical gravitational field. We assume spacetime to be described
by a manifold M, endowed with a metric $g$. Let us consider
on M a closed smooth curve $C$; $C$ must be defined in an
intrinsic fashion, without reference to any coordinate system.
In particular, it is important to specify the {\it form}
and the {\it size} of $C$, as in the two following
examples:

\noindent
(1) the circle of geodesic radius $r$ (fig.\ 5). It is defined
  as follows: given a center point $O$ and a plane through $O$,
  we go along the geodesic lines which start from $O$ tangent
  to the plane and we stop when the invariant distance from $O$
  is $r$. The points we reach in this way belong to the
  circle.

\noindent
(2) the (planar) ``dumbbell'' (fig.\ 6). It consists of two
  circles of geodesic radius $r$, whose centers are placed
  along a geodesic line, the second lying an invariant distance $D$
  apart from the first. The circles are joined by a
  strip of small width $\delta$ ($\delta \rightarrow 0$).

\medskip
In a curved space with arbitrary metric the geodesic circles
and the geodesic dumbbells may ``look'' quite different
from those in the flat space. When we compute the
vacuum average of a quantity like $W$ through a numerical
simulation of Regge calculus, what happens, intuitively, is the
following: the computer produces an arbitrary field
configuration, with a probability weighted by the exponential
of the euclidean action; on this configuration $W$
is ``measured'', along a contour defined in a similar way
as we did for the geodesic circle or the geodesic dumbbell;
these steps are repeated for many configurations and finally
the average of the results is computed.

On the other hand, the attitude one takes in the perturbative
calculations, like in Chapter 5, is different.
One works on a flat background, looking at small fluctuations
of the metric around the background, but disregarding the
effect of these fluctuations on the definition of the contours.

It is possible to reconcile the two views
in the case of a weak field, by introducing ``geodesic
corrections'' in the perturbative calculations on a flat
background [M., 1993a]. We start recalling some useful properties of the
normal coordinates $\{ \xi^\mu \}$. Chosen an origin $O$,
they can be defined in a neighbourhood of $O$ imposing the
condition
$$\eqalignno{
  \xi^\mu \xi^\nu {\Gamma}_{\mu \nu}^\rho (\xi) &= 0. &(6.9)
\cr}$$
As a consequence of (6.9), the lines of the form
$\xi^\mu(\lambda)=\lambda v^\mu$, where $v$ is a fixed vector,
represent geodesic lines through $O$. Furthermore, the
invariant distance between a point $\xi$ and the origin is
simply given by $\sqrt{\xi^\mu \xi^\nu \delta_{\mu \nu}}$.
It follows that a geodesic circle of radius
$r$ and center $O$ has the familiar form
$$\eqalignno{
  \xi(s) &= (r \cos s, r \sin s, 0,...); \qquad 0 \leq s \leq 2\pi. &(6.10)
\cr}$$

Given any coordinate system $\{ x^\mu \}$ with origin $O$, the
normal coordinates $\xi$ can be computed
from the coordinates $x$ by solving the system
(3.11), (3.12). We are interested in the first-order
perturbative solution given by (3.13), namely
$$\eqalignno{
  x^\mu(\xi) &= \xi^\mu + X^\mu(\xi) + O(\Gamma^2), &(6.11)
\cr}$$
where
$$\eqalignno{
  X^\mu (\xi) &= - \xi^\rho \xi^\sigma \int_0^1 dt \, (1-t) \,
  \Gamma^\mu_{\rho \sigma} (t \xi) , &(6.12)
\cr}$$
being $\Gamma$ the connection in the coordinates $x$.

The corrections due to (6.11) involve in the term
${\cal U}^{(1)}$, which on a flat background gives no
contributions to order $\kappa^2$.
We choose $\{ x^\mu \}$ to be the harmonic coordinates fixed by
the Feynman-De Witt gauge. In order to fix the ideas we
also make the hypotesis that $C$ is a geodesic
circle, although this hypotesis is in fact irrelevant.
The function $x^\mu(s)$ which represents the circle in the
coordinates $x$ can be obtained from eq.s (6.10), (6.12).
Since we are interested in terms up to order $\Gamma^2$ (corresponding
to order $\kappa^2$ in the quantum perturbative series), the approximation
(6.12) will be sufficient. We thus have
$$\eqalignno{
  {{\cal U}^{(1)}}^\alpha_{\beta} &=
  \oint_C dx^\mu (\xi) \Gamma^\alpha_{\mu \beta}[x(\xi)] = \cr
  &= \oint_C d\xi^\mu \Gamma^\alpha_{\mu \beta}(\xi) +
  \oint_C dX^\mu(\xi) \Gamma^\alpha_{\mu \beta}(\xi) + \cr
  & + \oint_C d\xi^\mu \partial_\nu \Gamma^\alpha_{\mu \beta} (\xi)
  X^\nu (\xi) + O(\Gamma^3), &(6.13)
\cr}$$
where $\xi$ is given by (6.10). In eq.\ (6.13),
the first integral represents the naive contribution, which
vanishes when it is averaged on the vacuum; the second
integral is a kind of ``index'' correction; the third one is
a correction to the ``argument''.
Using (6.12), one finds for the average value of
${\cal U}^{(1)}$ on the vacuum state
$$\eqalignno{
  {U^{(1)}}^\alpha_{\beta} &=
  - \int_0^1 dt \, (1-t) \, f^\alpha_\beta (t) + o(\kappa^2), &(6.14)
\cr}$$
where
$$\eqalignno{
  f^\alpha_\beta (t) &= \oint_C d(\xi^\rho \xi^\sigma)
  \langle \Gamma^\alpha_{\mu \beta}(\xi)
  \Gamma^\mu_{\rho \sigma}(t \xi) \rangle + \cr
  & + t \oint_C d\xi^\nu \, \xi^\rho \xi^\sigma \langle
  \Gamma^\alpha_{\mu \beta}(\xi) \partial_\nu
  \Gamma^\mu_{\rho \sigma}(t \xi) \rangle + \cr
  & + \oint_C d\xi^\mu \, \xi^\rho \xi^\sigma \langle \partial_\nu
  \Gamma^\alpha_{\mu \beta}(\xi) \Gamma^\nu_{\rho \sigma}(t \xi) \rangle .
\cr}$$
Disregarding gradient terms in the integrand we obtain
$$\eqalignno{
  f^\alpha_\beta (t) &= \delta^{\alpha \alpha '}
  \delta^{\nu \mu}_{\gamma \tau}
  \oint_C d\xi^\tau \, \xi^\rho \xi^\sigma \left< \left[
  \partial_\beta \partial_\nu
  h_{\alpha ' \mu}(\xi) \right]_{[\alpha ' \beta]}
  \Gamma^\gamma_{\rho \sigma}(t \xi) \right> . &(6.15)
\cr}$$

The antisymmetry of $f^\alpha_\beta$ causes its trace
to vanish. It follows that the geodesic contribution of order
$\kappa^2$ to $W$ vanishes, like the naive contribution
${\rm Tr} \, U^{(2)}$. We notice that the antisymmetry of
$f$ in (6.15) does not depend on the form of the
propagator, and is thus a geometric property. On the contrary,
we remind that the trace of the naive contribution is
sensitive to the dynamics of the field (compare Section 5.4).
We finally point out that eq.\ (6.15)
holds for any closed curve $C$, not just for the geodesic
circle.

Returning to (6.15), let us now compute all the matrix
elements. We notice that eq.\ (6.15), substituted into
(6.14), produces a correction of the form
$$\eqalignno{
  U^{(1)} &= \int_0^1 dt \, (1-t) \oint_C d\xi \,
  \langle \Psi(\xi) \, \Psi(t\xi) \rangle, &(6.16)
\cr}$$
where $\Psi$ is a field. This expression shows that the
geodesic contribution is related to the necessity of fixing
a scale for the distances. Since (6.16) diverges
for $t \sim 1$, in the following we shall regularize it
by stopping the integrations at $t=1-\epsilon$.

Substituting the propagator (5.11) into (6.15)
we find, after some steps, in any dimension $N>2$
$$\eqalignno{
  & {U^{(1)}}^\alpha_{\beta} = - c_N \, \kappa^2 \,
  \delta^{\alpha \alpha '} \int_0^{1-\varepsilon} {dt \over (1-t)^N}
  \oint_C d\xi^\tau \, \xi^\rho \xi^\sigma \times \cr
  & \qquad \times \left\{ \left( {{3N-N^2} \over {N-2}} \right) \,
  \delta_{\alpha ' \rho} \, \partial_\beta \partial_\tau \partial_\sigma
  {1 \over \xi^{N-2}}
  - \left( {{2} \over {N-2}} \right) \, \delta_{\alpha ' \tau} \,
  \partial_\beta \partial_\rho \partial_\sigma
  {1 \over \xi^{N-2}}
  \right\}_{[\alpha ' \beta], \ (\rho \sigma)} .
\cr}$$
Finally, disregarding gadient terms in the integrand we obtain
$$\eqalignno{
  {U^{(1)}}^\alpha_{\beta} &= -c_N \, \kappa^2 \, \chi^{(1)}_N(\epsilon)
  \oint_C {1 \over 2} \, \left[ d\xi^\alpha \partial_\beta -
  d\xi_\beta \partial^\alpha \right]
  \, {1 \over \xi^{N-2}} = \cr
  &= -c_N \, \kappa^2 \, \chi^{(2)}_N(\epsilon) \oint_C
  d(\xi^\alpha \wedge \xi_\beta) \, {1 \over \xi^{N}}, &(6.17)
\cr}$$
where $\chi^{(i)}_N(\epsilon)$ diverges like $\epsilon^{1-N}$
when $\epsilon \rightarrow 0$.
For a geodesic circle of radius $r$ in the plane 1-2, the only
non zero elements of ${U^{(1)}}$ are simply
$$\eqalignno{
  {U^{(1)}}^1_2 &= -{U^{(1)}}^2_1 = -c_N \, \kappa^2 \, \chi^{(2)}_N(\epsilon)
\,
  {\pi \over r^{N-2}}.
\cr}$$
For a geodesic dumbbell we have, in the limit $r \ll D$,
$$\eqalignno{
  {U^{(1)}}^1_2 &= -{U^{(1)}}^2_1 \simeq -c_N \, \kappa^2 \,
  \chi^{(2)}_N(\epsilon) \,
  {{\pi \, r^2} \over {\left( {D \over 2} \right)^N}}.
\cr}$$
This correction should be added to (6.7).
Although it is formally quite clear, a further analysis
seems to be necessary in order to understand its physical
meaning.

We also remind that the matrix (6.17) is not invariant
under gauge transformations. Nevertheless, its simplicity
suggests us that it could have a quite general geometrical
meaning.

\bigskip \noindent
{\bf 6.4 \ The static potential energy.} \bigskip

It is known that the static potential energy $U(L)$ of two
sources of a gauge field is related to the Wilson loop of
temporal size $T$ and spatial size $L << T$ by the formula
$$\eqalignno{
  e^{- \hbar^{-1} T U(L)} &= W(L,T) .
\cr}$$
In the strong coupling limit this expression leads to the confining potential
$U(L)=kL$, while in the weak coupling limit one easily recovers the Coulomb
potential $U(L)=-e^2/L$.

The vanishing of $W$ to leading order in gravity raises the problem of
finding another invariant expectation value of the quantized field which gives
the static potential energy between two masses. This problem has a well defined
solution indeed [M., 1993b]. One starts from the following known formula of
Euclidean field theory
$$\eqalignno{
  {\cal E} &= \lim_{T \to \infty} - {{1} \over {\hbar T}} \log
  {{\int d[\phi] \, \exp \{ - \hbar^{-1} ( S_0[\phi] +
  S_{\rm Inter.}[\phi,J] ) \} } \over
  {\int d[\phi] \, \exp \{ - \hbar^{-1} S_0[\phi] \} } }, &(6.18)
\cr}$$
where $\phi$ is a quantum field and $J$ is a classical source coupled to
$\phi$,
which is switched off outside the interval $(-{1 \over 2} T,\,
{1 \over 2} T)$. ${\cal E}$
represents the energy of the ground state of the system.
We shall only show here how this formula works in the case of
a weak gravitational field on a flat background.
Replacing $\phi$ with
the gravitational field, $S_0$ with Einstein's action and $J$ with two
particles of masses $m_1$, $m_2$, following the trajectories
$$\eqalignno{
  x^\mu(t_1) &= \left( t_1,\, -{L \over 2},\, 0,\, 0 \right); \qquad
  y^\mu(t_2) = \left( t_2,\, {L \over 2},\, 0,\, 0 \right) , &(6.19)
\cr}$$
we find (for a weak field)
$$\eqalignno{
  {\cal E} &= \lim_{T \to \infty} - {{1} \over {\hbar T}} \times \cr
  & \times \log
  {{\int d[h] \, \exp \hbar^{-1} \left\{ - S_{\rm Einst.}[h] -
  m_1 \int dt_1 \, \sqrt{1-h_{00}[x(t_1)]} -
  m_2 \int dt_2 \, \sqrt{1-h_{00}[y(t_2)]} \right\}} \over
  {\int d[h] \, \exp \{ - \hbar^{-1} S_{\rm Einst.}[h] \} } }.
\cr}$$
It is easy to verify that to leading perturbative order this gives
the correct result
$$\eqalignno{
  {\cal E} = m_1 + m_2 - {{m_1 m_2 G} \over {L}} .
\cr}$$

It is interesting to make a
comparison with electrodynamics. In that case the analogue of the functional
integral which appears in the logarithm of (6.18) has the form
$$\eqalignno{
  & \left< \exp \left\{ e \int_{-{T \over 2}}^{{T  \over 2}} dt_1
  A_0[x(t_1)] - e \int_{-{T \over 2}}^{{T \over 2}} dt_2
  A_0[y(t_2)] \right\} \right> . &(6.20)
\cr}$$
(The two charges have been chosen to be opposite: $q_1=e$, $q_2=-e$.) Reversing
the direction of integration in the second integral and closing the contour at
infinity, one is able to show that the quantity (6.21) coincides with the
Wilson loop of a single charge $g$, thus giving a gauge invariant expression
for the potential energy.

In gravity this is not possible: we may imagine that an expression like
(6.21) could be obtained in the first-order formalism (with $A_0$ replaced
by the tetrad $e^0_0$), but the masses necessarily have the same sign,
so the loop cannot be closed.

It is possible to give a definite meaning to the expression
(6.18) for the energy also in the case when no background is present.
It can also be proven that the gravitational interaction energy
is always negative [M., 1993b]. A very recent work based on this formula is
that of Hamber and Williams [1994].

\vfill\eject

\centerline{\bf 7. LATTICE GRAVITY.} \bigskip

For quantum field theories with relevant non-perturbative effects,
the lattice discretization is often quite important, since it
allows numerical evaluations through the Montecarlo simulation
technique. This is also the case for quantum gravity, which is
usually thought to be very strong at small distances, where the theory
should somehow determine the small scale ``quantized'' structure of
spacetime [see Rovelli, 1991 and ref.; Smolin, 1992 and ref.].

In the discretized theory the correlation length is a crucial quantity.
For this reason we shall briefly recall here some advances made in
lattice gravity during the last years, especially from Hamber,
Caracciolo and Pelissetto, Ambj\o rn and co-workers and Greensite.

The point of view of the lattice [see for instance
Kogut and Wilson, 1974; Kogut, 1979; Seiler, 1982; Creutz, 1983]
is very different from that of continuum field theory; the two
are connected, however, by the ``continuum limit''.

Let us consider the equivalent statistical system of a
field theory discretized on a euclidean lattice.
We remind that if the transition between two phases of the system
is of the second order, then the appearence of long range correlations
causes the details of the lattice to become irrelevant.
The original (renormalizable) continuum theory can thus be
reobtained, at the transition point, when the lattice spacing
goes to zero. If the original continuum theory is not
renormalizable in the usual sense (like Einstein gravity), a more general
procedure, called ``asymptotic safety'' [Weinberg, 1979]
can be possibly applied.

The chapter is organized as follows. As schematically shown in
Table 1, the study of a field theory on the lattice starts choosing
one specific version of the theory and a discretization scheme.
Some possibilities for gravity are displayed in the table for
illustration purposes, with no attempt to completeness; in fact,
we are not interested here to make a review of all the
different approaches and to compare the results (which should be
the same, at least for physical quantities!); we are interested
here to the vacuum correlations at geodesic distance. So we have
chosen to analyze in more details the way which seems to be the
most promising under this respect, namely: $(R+R^2)$-gravity
(Section 7.1), discretized through Regge calculus (Section 7.2),
as it appears in the simulations of Hamber (Section 7.3). The
remaining section, 7.4, is dedicated to a brief description of
other approaches.

Throughout the chapter we shall often quote
from the review of Menotti [1990], were many omitted
details can be found.

\bigskip \noindent
{\bf 7.1 \ $(R+R^2)$-gravity.} \bigskip

It is known that the scalar curvature $R$ is the only
scalar which contains second derivatives of the metric.
Further scalars can be constructed, however, which contain
higher derivatives, namely $R^2$, $R^{\mu \nu} R_{\mu \nu}$,
$R^{\mu \nu \rho \sigma} R_{\mu \nu \rho \sigma}$ (among the
three, only two are independent). The Einstein's action
can thus be modified as follows:
$$\eqalignno{
  S &= - \int d^4x \, \sqrt{g} \, \left[ {1 \over \kappa^2} \, R -
  {a \over 4} \, R^{\mu \nu \rho \sigma} R_{\mu \nu \rho \sigma} -
  {1 \over 3} \left( b - {a \over 4} \right) \, R^2 \right] . &(7.1)
\cr}$$
The terms quadratic in the curvature would not be observable
in macroscopic phenomena, but they are important
to the quantum theory, since the first radiative corrections
to Einstein's lagrangian contain just $R^2$ terms. As Stelle
showed in detail [1977], $(R+R^2)$--gravity can be considered
as a renormalizable, although non unitary, theory\footnote\dag{It has
been shown that in a
suitable range of the couplings $a$ and $b$ the theory is
asymptotically free with respect to them. Also,
asymptotic freedom with respect to $G$ may arise under certain
circumstances [Julve and Tonin, 1978]}.
Various attempts of recovering unitarity by graphs resummation,
suitable gauge fixing and other techniques [Antoniadis et al.,
1986; see also Weinberg, 1979] did not lead to any definitive
outcome.

The authors of numerical simulations on $(R+R^2)$
models believe usually that the loss of unitarity is
a problem confined to perturbation theory, as the whole
theory could still be unitary [Hamber, 1986].

\bigskip \noindent
{\bf 7.1 \ Regge calculus.} \bigskip

The basic idea of this formalism [Regge, 1961] is to replace the
smooth spacetime manifold with a simplicial manifold, i.e.,
a manifold made up of flat regions (``simplexes'') connected by ``edges''.
This is easy to figure out in two dimensions (see fig.\ 7).
The discontinuity in the parallel transport arises when
an edge is crossed, and the curvature concentrates in those
points where a few edges meet, called ``hinges''. If we
transport a vector along a closed path, we only notice a
difference between the original vector and the transported
one, if the path contains at least one hinge. In a generic
dimension $N$ the dimension of the hinges is $N-2$; hence
for $N=2$ they are points, but for $N=3$ they are lines and
for $N=4$ they are triangles.

Given an hinge and one simplex which touches it,
the opening angle $\theta_N$ of the simplex with respect to
the hinge can be defined. In two dimensions, the geometrical
meaning of such an angle is apparent. In general, $\theta_N$ is
given by the expression
$$\eqalignno{
  \sin \theta_N &= {N \over N-1} \, {{V_N \, V_{N-2}} \over {V_{N-1}} \,
  {V'}_{N-1}},
\cr}$$
where $V_N$ is the volume of the mentioned $N$-dimensional
simplex and $V_{N-1}$, ${V'}_{N-1}$ are the volumes of the
two $(N-1)$-dimensional simplexes which share the hinge
(in the two-dimensional case, they are the two edges which
meet at the hinge).

Let us now consider all the simplexes which meet at a given
hinge and sum their deficit angles referring to that hinge.
We obtain in this way a ``deficit angle''
$$\eqalignno{
  \delta_h &= 2\pi - \sum_{s \subset h} \theta_s .
\cr}$$

Regge postulated the following, simple action:
$$\eqalignno{
  S &= 2 \sum_h A_h \, \delta_h , &(7.2)
\cr}$$
where the sum ranges over all hinges and $A_h$ is the surface
of the hinge. In the continuum limit, the action (iliz)
really approaches Einstein's action (for references about the
exact meaning of this limit and the uniqueness of
Regge's action, see [Menotti, 1990] and [Williams and
Tuckey, 1991]).

The extension of Regge's action by a cosmological term of
the form $\lambda \sum_s V_s$ is straightforward. On the
contrary, the lattice version of the $R^2$ terms is quite
involved [for a review see
Hamber, 1986]. To write them, one must associate to each
hinge a suitably defined volume $V_h$. The simplest $R^2$
term is
$$\eqalignno{
  \sum_h {{A^2_h \, \delta^2_h} \over {V_h}} & \to
  {1 \over 4} \int d^N x \, \sqrt{g(x)} \,
  R_{\mu \nu \lambda \rho} R^{\mu \nu \lambda \rho} .
\cr}$$
The other terms are more complex, as they entail the introduction
of a lattice analogue of the Riemann tensor. All the lattice
translations are not unique; however, they are equivalent
in the continuum limit -- when the maximum lenght of the
edges vanishes, with respect to the characteristic radius of
curvature of the smooth manifold.

\bigskip \noindent
{\bf 7.3  \ The simulations of Hamber.} \bigskip

In a remarkable series of papers starting in 1985, Hamber has
developed a technique for simulating quantum Regge calculus of
$(R+R^2)$-gravity [see Hamber, 1986, 1992, 1993 and references
therein].

The typical quantities which were first computed in these simulations
were rather reminiscent of Ising model, namely the average curvature
$\bar{R}$ and the average curvature squared $\bar{R^2}$. It turned
out that they diverge along a line which goes across the parameters
space $(a,\kappa)$, thus dividing it into two phases, called
``smooth phase'' and ``rough phase''. The critical indices were
computed too.

Recently, the evaluation of two-point correlation functions at
geodesic distance has been addressed [Hamber, 1993 b]; the results
are interesting although rather preliminary.

In a typical simulation,
the simplicial lattice is built up of rigid hypercubes,
which can be subdivided into simplices by introducing
face diagonals, body diagonals and hyperbody diagonals.
This choice is not unique and is dictated by a criterion
of simplicity, with the advantage that such a lattice can
be used to study rather large systems with little modification.
The length of the edges $\l$ is individually varied
(by moving at random through the lattice), and a new trial
length is accepted with probability
$min\,(1,\ e^{-\Delta S})$, where $\Delta S$
is the variation of the action under the change in edge length.
If the triangle inequalities or their higher-dimensional
analogues are violated, the new edge length is rejected.
It should be noticed that in order to compute the variation
in the action under the change of one edge length, a large
number of adjoining triangles and their deficit angles has
to be considered.

Lattices of size between $4^4$ (with 3840 edges) and
$16^4$ (with 983040 edges) were considered. Periodic boundary
conditions (four-torus) were used, since it is expected that
for this choice boundary effects should be minimized. The lengths
of the runs typically varied between 10-40$k$ Montecarlo
iterations on the $4^4$ lattice, 2-6$k$ on the $8^4$ lattice
and 0.5$k$ on the $16^4$ lattice. On the larger lattices
duplicated copies of the smaller lattices are used as starting
configurations, allowing for additional equilibration sweeps
after duplicating the lattice in all four directions.

Like we said above, typical quantities which are ``measured''
in the simulations are the average curvature
$$\eqalignno{
  \bar{R} &= \langle \l^2 \rangle \,
  {{\langle 2 \sum_h \delta_h A_h \rangle} \over
  {\langle \sum_h V_h \rangle}} \sim
  {{\langle \int \sqrt{g} \, R \rangle} \over
  {\langle \int \sqrt{g} \rangle}}
\cr}$$
and the average of the curvature squared
$$\eqalignno{
  \bar{R^2} &= \langle \l^2 \rangle^2 \,
  {{\langle 4 \sum_h {{\delta^2_h A_h} \over {V_h}} \rangle} \over
  {\langle \sum_h V_h \rangle}}
  \sim {{\langle \int \sqrt{g} \, R^2_{\mu \nu \rho \sigma} \rangle} \over
  {\langle \int \sqrt{g} \rangle}} .
\cr}$$

One can also estimate the local fluctuation
$$\eqalignno{
  \chi_{\bar{R}} &= {{1} \over {\langle \sum_h V_h \rangle}}
  \left[ \langle \left( 2 \sum_h \delta_h A_h \right)^2 \rangle -
  \langle 2 \sum_h \delta_h A_h \rangle^2 \right] .
\cr}$$
A divergence in $\chi$ should be indicative of a second-order
phase transition (as it happens for the magnetic susceptibility
or the specific heat).

As the bare Newton constant and the coupling $a$ are varied
(compare eq.\ (7.1), where one sets in this case $b=a/4$), a
continuus phase transition is found, separating a ``smooth''
phase from a ``rough'' phase. In the first phase the curvature
is small and negative, and the fractal dimension is consistent
with four. In the second phase the simplices are collapsed, the
curvature is large and positive, and the fractal dimension
is much less than four, indicating the presence of finger-like
structures. Approaching the critical point from the only
physically acceptable phase, the smooth one, it was found that
the curvature vanishes with an exponent $\delta = 0.62(5)$. At the
critical point the curvature fluctuation $\chi$ diverges, leading
to the possibility of defining a non-trivial lattice continuum
(this can be then checked measuring the correlations: see below).

We notice thus that the vacuum expectation value of the curvature
can be used as an order parameter for the transition; moreover,
it can be used as a possible definition of the effective, long-distance
cosmological constant. Usually one adds to the action (7.1)
a cosmological term with some bare $\lambda$; the effective cosmological
constant is then given by
$$\eqalignno{
  \left( {4\lambda \over \kappa} \right)_{\rm eff} &=
  {{\langle \int \sqrt{g} \, R \rangle} \over
  {\langle \int \sqrt{g} \rangle}} .
\cr}$$
As one approaches the fixed point at $\kappa_c$ (with fixed $a$),
one finds as expected $(\lambda / \kappa)_{\rm eff} \to 0$.

The computation of the correlations at geodesic distance has been
recently addressed, in particular for the averages
$$\eqalignno{
  G_R(D) & \sim \langle \sum_{h(x)} \delta_h A_h
  \sum_{h'(y)} \delta_{h'} A_{h'} \, \delta(|x-y|-D) \rangle ,
\cr}$$
which corresponds to the correlation of the scalar curvature, namely
$$\eqalignno{
  & \sim \langle \sqrt{g(x)} R(x) \, \sqrt{g(y)} R(y) \,
  \delta(|x-y|-D) \rangle
\cr}$$
and
$$\eqalignno{
  G_V(D) & \sim \langle \sum_{h(x)} V_h
  \sum_{h'(y)} V_{h'} \, \delta(|x-y|-D) \rangle ,
\cr}$$
which corresponds to the volume correlations
$$\eqalignno{
  & \sim \langle \sqrt{g(x)} \, \sqrt{g(y)} \delta(|x-y|-D) \rangle .
\cr}$$

The computation proceeds as follows. First, the geodesic distance
between any two points $x$ and $y$ is determined in a fixed
background geometry. Next the correlations are computed for all
pairs of points within geodesic distance $D$ and $D+\Delta D$, where
$\Delta D$ is an interval slightly larger than the average
lattice spacing $\l_0=\sqrt{\langle \l^2 \rangle}$, but much smaller
than the distance $D$ considered. Finally, the correlations determined
for a fixed geodesic distance $D$ are averaged over all the metric
configurations.

In principle one could also compute correlations of vector and tensor
quantities introducing the matrix of the parallel transport
(compare Chapter 5), but this is quite complicated on a Regge lattice
and has not be done yet.

The details of the method are described in the mentioned paper.
The main result is that the volume correlations are negative
at large distances, while the curvature correlations are always
positive. If the correlations are fitted to an exponential decay,
one finds that the behavior is always consistent with a mass that
decreases when one approaches the critical point. This behavior is
precisely what is expected if the model is supposed to reproduce
the classical Einstein theory for distances which are very large
compared to the ultraviolet cutoff scale. Note that this happens
for a model of gravity which at short distances is far removed
from the pure Einstein theory, containing both a bare cosmological
term and bare higher derivative lattice terms!

\bigskip \noindent
{\bf 7.4 \ The gauge approach and others.} \bigskip

We shall describe the main features of the gauge formalism
in dimension $N=4$, referring to the literature for issues
like the reflection positivity, the invariance under
reparametrization transformations and the phenomenon of
graviton doubling (see Menotti [1990] and references; Smolin
[1979]; Menotti and Pelissetto [1986, 1987]).

In the gauge approach one exploits the (incomplete) analogy
between gravity and the gauge theories. The attitude is taken
of considering gravity as a field theory in Minkowski or
euclidean space, whose lagrangian is invariant under a group
of local transformations (reparametrization transformations).
Several people have introduced along these lines discretized
versions of gravity. We shall briefly describe here the
formulation due to Smolin, which can be considered as the
prototype of these gauge formulations. Essentially Smolin's
formulation is the discretized version of the Mac Dowell --
Mansouri gauge formulation of De Sitter gravity [1977].
Let us consider the De Sitter group $O(4,1)$, which goes
to $O(5)$ in the euclidean, and introduce the usual hypercubic
lattice, familiar to gauge theories. Associate now to each
link of the hypercubic lattice a finite element of $O(5)$
$$\eqalignno{
  U(n,n+\mu) &= \exp \left( {a \over 2} J_{bc} \, \omega^{bc}_\mu
  + a P_c \, e^c_\mu \right) ,
\cr}$$
where $J_{ab}$ and $P_a$ satisfy the following commutation
relations
$$\eqalignno{
  & [J_{ab} , J_{cd}] = \delta_{bc} J_{ad} - \delta_{ac} J_{bd}
  - \delta_{bd} J_{ac} + \delta_{ad} J_{bc} \cr
  & [J_{ab} , P_{c}]  =  \delta_{bc} P_a - \delta_{ac} P_b \cr
  & [P_a , P_b]  =  - J_{ab} .
\cr}$$

The action takes the form
$$\eqalignno{
  S &= \alpha \sum \epsilon^{\mu \nu \lambda \rho} \,
  U^{AB}_{\mu \nu}(n) \, U^{CD}_{\lambda \rho}(n) \, \epsilon_{ABCD5} ,
\cr}$$
where $\alpha={3 \over {8 \lambda \kappa^2}}$ and
$$\eqalignno{
  U^{AB}_{\mu \nu}(n) &= U^{AC}(n,n+\mu) \, U^{CD}(n+\mu,n+\mu+\nu) \,
  \times \cr
  & U^{DE}(n+\mu+\nu,n+\nu) \, U^{EB}(n+\nu,n)
\cr}$$
is the plaquette in the vector representation. It is clearly
invariant under local $O(4)$-transformations, but not under
$O(5)$. This is a consequence of the incomplete similarity
between gravity and gauge theories. In the formal continuum
limit we obtain the Mac Dowell-Mansouri action
$$\eqalignno{
  S &= - \int d^4 x \, e \, \left[\left( {1 \over \kappa}\, R - \lambda \right)
  - {3 \over {8 \lambda \kappa^2}} \, \epsilon^{\mu \nu \lambda \rho} \,
   R^{ab}_{\mu \nu} R^{cd}_{\lambda \rho} \, \epsilon^{abcd} \right] ,
\cr}$$
which is the familiar Einstein action, plus a cosmological
and a Gauss-Bonnet term. The discretized action is
reparametrization invariant only in the formal continuum
limit.

The extension of the procedure above to the Poincar\'e group
has been given by Menotti and Pelissetto [1987]. To this end
one has to introduce in the formalism the local tetrads,
which transform locally under $O(4)$. They are defined by
the expression
$$\eqalignno{
  E^{a}_\mu(n) &= - {1 \over 4a} \, {\rm tr} \left[
  H \, P^{a} \, U(n,n+\mu) \, H \, U(n+\mu , n) \right] ,
\cr}$$
where $H={\rm diag}(1,1,1,1,-1)$.

The gauge approach is strictly bounded to the program outlined
by Weinberg of finding a fixed point around which the theory
could possibly be renormalized.

The simulations were performed by Caracciolo
and Pelissetto [1987, 1988; see also Menotti, 1990]
on the De Sitter model in the gauge formulation.
They made extensive simulations on an $8^4$ lattice.
Due to the compact nature of the group $SO(5)$ and the
presence of the lattice cut-off, the action is limited and
thus the theory must exibit a well defined ground state.
For small values of $\alpha$ the vacuum is similar to the
QCD vacuum in the strong coupling region.
The large $\alpha$ region is characterized by the vanishing
of the vierbein, so that the action becomes dominated by the
topological term. An interesting order parameter is given by
$$\eqalignno{
  P &= {{R^{ab}_{\mu \nu} \tilde{R}^{ab}_{\mu \nu}} \over
  {R^{ab}_{\mu \nu} R^{ab}_{\mu \nu}} },
\cr}$$
whose distribution is peaked around zero in the small $\alpha$
region and around the extremes $\pm 1$ in the large $\alpha$
region. At the value ${\alpha \over 4}=0.08 \pm 0.01$
a sudden phase transition is observed. Around this value
all measured quantities, like the mean value of the action
per site, the trace of the metric, the $O(4)$ curvature and
$P$, exibit very large hysteresis cycles, thus showing that the
transition is first order. For $\alpha > \alpha_c$
the system ends up in non reproducible states showing the
existence of many metastable states above $\alpha_c$ and
recalling a situation similar to a spin glass.

These results are very different, also qualitatively, from
those of Hamber. Unfortunately, such disagreements are not
unusual in the numerical simulations of very complex models.
In a third approach [Ambj\o rn et al., 1992] a transition
is found, that could be first- or second-order; furthermore,
the average curvature at the transition is non zero, which
makes the continuum limit quite difficult to understand.

Finally, it is worth mentioning that recently Greensite has given
a new interesting approach to euclidean quantum gravity
[Greensite, 1993 and references]. It is known that the euclidean
Einstein action is not bounded from below, and this means that
the euclidean functional integral is not well defined. There have
been various attempts to solve this problem [see for instance
Mazur and Mottola, 1990 and references]. The method of Greensite
is in practice a prescription which can be related to
stochastic quantization and which prevents the field configurations
in the functional integral from running into the singular configurations
which make the action unbounded. This method leads to an effective
action which reproduces Einstein's equations but is non-local
beyond first order. Although higher order calculations are difficult,
it can be proved in a nice way [Greensite, 1992] that the effective
cosmological constant generated by this model is vanishing. The
Montecarlo simulations are still at an initial stage.

\vfill\eject

\centerline{\bf REFERENCES.} \bigskip

\medskip \noindent
Alvarez, E., Rev. Mod. Phys. {\bf 61} (1989) 561.

\medskip \noindent
Ambj\o rn, J., J. Jurkiewicz and C.F. Kristjansen, {\it Quantum gravity,
dynamical triangulations and higher derivative regularization}, Copenhagen
preprint NBI-HE-95-53, july 1992.

\medskip \noindent
Antoniadis, I. and E.T. Tomboulis, Phys. Rev. {\bf D 33} (1986) 2756.

\medskip \noindent
Antoniadis, I., J. Iliopoulos and T.N. Tomaras, Nucl. Phys. {\bf 267} (1986)
497.

\medskip \noindent
Bassetto, A., G. Nardelli and R. Soldati, {\it Yang-Mills theories in algebraic
non covariant gauges}, World Scientific, Singapore, 1991.

\medskip \noindent
Boulware, D.G. and S. Deser, Nuovo Cim. {\bf 31 A} (1976) 498.

\medskip \noindent
Caracciolo, S. and A. Pelissetto, Phys. Lett. {\bf B 193} (1987)
237; Nucl. Phys. {\bf B 299} (1988) 693; Nucl. Phys. {\bf B 4}
(Proc. Suppl.) (1988) 78; Phys. Lett. {\bf B 207} (1988) 468.

\medskip \noindent
Clement, G., Int. J. Theor. Phys. {\bf 24} (1985) 267; Ann. Phys. (N.Y.)
{\bf 201} (1990) 241.

\medskip \noindent
Collins, J.C., {\it Renormalization}, Cambridge University Press, 1984.

\medskip \noindent
Creutz, M., {\it Quarks, gluons and lattices}, Cambridge Monographs on
Mathematical Physics, Cambridge, 1983.

\medskip \noindent
Deser, S., R. Jackiw and G. t'Hooft, Ann. Phys. (N.Y.) {\bf 152} (1984) 220.

\medskip \noindent
Deser, S. and R. Jackiw, Ann. Phys. (N.Y.) {\bf 192} (1989) 352.

\medskip \noindent
De Witt, B., Phys. Rev.\ {\bf 160} (1967a) 1113; Phys. Rev.\ {\bf 162}
(1967b) 1195; Phys. Rev. {\bf 162} (1967c) 1239.

\medskip \noindent
Durand, L. and E. Mendel,  Phys. Rev. {\bf D 26} (1982) 1368.

\medskip \noindent
Erd\'ely, A. et al., {\it Higher trascendental functions}, Vol. I, Ch.s. 2 e 3,
McGraw-Hill, New York, 1953.

\medskip \noindent
Fock, V.A., Sov. Phys. {\bf 12} (1937) 404.

\medskip \noindent
Gaigg, P. and W. Kummer, eds., {\it Physical and non standard gauges},
Lecture Notes in Physics {\bf 361}, Springer Verlag, Berlin, 1990.

\medskip \noindent
Greensite, J., Phys. Lett. {\bf B 281} (1992) 219;
Nucl. Phys. {\bf B 390} (1993) 439.

\medskip \noindent
Grignani, G. and C. Lee, Ann. Phys. (N.Y.) {\bf 196} (1989) 386.

\medskip \noindent
Grignani, G. and G. Nardelli, Phys. Rev. {\bf D 45} (1992) 2719.

\medskip \noindent
Hamber, H.W., in {\it Critical phenomena, random systems, gauge theories -
Les Houches Summer School, Session XLIII (1984)}, North-Holland,
Amsterdam, 1986; Phys. Rev. {\bf D 45} (1992a) 507;
Nucl. Phys. {\bf B} (Proc. Suppl.) {\bf 25 A} (1992b) 150;
Nucl. Phys. {\bf B 400} (1993a) 347; {\it Invariant correlations
in simplicial gravity}, report UCI-TH-93-37, October 1993 (1993b).

\medskip \noindent
Hamber, H.W. and R.M. Williams, {\it Newton potential in quantum Regge
gravity}, preprint CERN-TH.7314/94, DAMTP-94-49, June 1994.

\medskip \noindent
Hawking, S.W., in {\it General relativity: an Einstein centenary survey},
S.W. Hawking and W. Israel eds., Cambridge University Press, 1979.

\medskip \noindent
Hehl, F. et al., Rev. Mod. Phys. {\bf 48} (1976) 393.

\medskip \noindent
Jackiw, R., {\it Topics in planar physics}, MIT preprint
CTP-1824, December 1989.

\medskip \noindent
Julve, J. and M. Tonin, Nuovo Cim. {\bf 46 B} (1978) 137.

\medskip \noindent
Kobayashi, S. and K. Nomizu, {\it Foundations of differential geometry},
Wiley, New York, 1969.

\medskip \noindent
Kogut, J., Rev. Mod. Phys. {\bf 51} (1979) 659.

\medskip \noindent
Kogut, J. and K.G. Wilson, Phys. Rep. {\bf C 12} (1974) 75.

\medskip \noindent
Mac Dowell, S.W. and F. Mansouri, Phys. Rev. Lett. {\bf 38} (1977) 739;
Phys. Rev. Lett. {\bf 38} (1977) 1376.

\medskip \noindent
Mandelstam, S., Ann. Phys. (N.Y.) {\bf 19} (1962) 25;
Phys. Rev. {\bf 175} (1968) 1580, 1604.

\medskip \noindent
Mazur, E. and E. Mottola, Nucl. Phys. {\bf B 341} (1990) 187 and references.

\medskip \noindent
Menotti, P., Nucl. Phys. (Proc. Suppl.) {\bf B 17} (1990) 29.

\medskip \noindent
Menotti, P. and A. Pelissetto, Ann. Phys. (N.Y.) {\bf 170} (1986) 287; Phys.
Rev. {\bf D 35} (1987) 1194.

\medskip \noindent
Menotti, P. and D. Seminara, Ann. Phys. {\bf 208} (1991a) 449; Nucl. Phys.
{\bf B 376} (1992) 411.

\medskip \noindent
Menotti, P. and D. Seminara, {\it The radial gauge propagator}, Pisa preprint
IFUP-TH 49/91 (1991b); to appear on Il Nuovo Cim. {\bf A}.

\medskip \noindent
Menotti, P. and D. Seminara, Phys. Lett. {\bf B 301} (1993) 25.

\medskip \noindent
Menotti, P., G. Modanese and D. Seminara, Ann. Phys. {\bf 224} (1993) 110.

\medskip \noindent
Modanese, G. and M. Toller, J. Math. Phys. {\bf 31} (1990) 452.

\medskip \noindent
Modanese, G., J. Math. Phys. {\bf 33} (1992a) 1523.

\medskip \noindent
Modanese, G., Phys. Lett. {\bf B 288} (1992b) 69.

\medskip \noindent
Modanese, G., J. Math. Phys. {\bf 33} (1992c) 4217.

\medskip \noindent
Modanese, G., Phys. Rev. {\bf D 47} (1993a) 502.

\medskip \noindent
Modanese, G., {\it A formula for the static potential energy in
quantum gravity}, report CTP \# 2217, June 1993 (1993b).

\medskip \noindent
Modanese, G., Phys. Rev. {\bf D 49} (1994) 6534.

\medskip \noindent
Modanese, G., Phys. Lett. {\bf B 325} (1994) 354.

\medskip \noindent
Papapetrou, A., Proc. Roy. Soc. {\bf A 209} (1951) 248.

\medskip \noindent
Regge, T., Nuovo Cimento {\bf 19} (1961) 558.

\medskip \noindent
Rovelli, C., Class. Q. Grav. {\bf 8} (1991) 1613.

\medskip \noindent
Schwinger, J., Phys. Rev. {\bf 82} (1952) 684; {\it
Particles, sources and fields}, vol. I, \S 3.8, Addison Wesley,
New York, 1989.

\medskip \noindent
Seiler, E., {\it Gauge theories as a problem of constructive quantum
field theory and statistical mechanics}, Lecture Notes in Physics
{\bf 159}, Springer Verlag, Berlin, 1982.

\medskip \noindent
Smolin, L., Nucl. Phys. {\bf B 148} (1979) 333; {\it Recent
developments in non perturbative quantum gravity}, Syracuse preprint,
1992.

\medskip \noindent
Souriau, J.M., Ann. Inst. Henri Poincar\'e, {\bf A XX} (1974) 315.

\medskip \noindent
Stelle, K.S., Phys. Rev. {\bf D 16} (1977) 953.

\medskip \noindent
Teitelboim, C., Nucl. Phys. {\bf B 396} (1993) 303.

\medskip \noindent
Toller, M., J. Math. Phys. {\bf 24} (1983) 613.

\medskip \noindent
Toller, M. and R. Vaia, J. Math. Phys. {\bf 25} (1984) 1039.

\medskip \noindent
Toller, M., private communication, 1988.

\medskip \noindent
Tsamis, N.C. and R.P. Woodard, Ann. Phys. (N.Y.) {\bf 215} (1992) 96.

\medskip \noindent
Veltman, M. J. G., in {\it Methods in Field Theory - Les Houches Summer
School, 1975}, edited by R. Balian and J. Zinn-Justin, North-Holland,
Amsterdam, 1976.

\medskip \noindent
Vilenkin, A., Phys. Rep. {\bf 121} (1985) 263.

\medskip \noindent
Weinberg, S., in {\it General relativity: an Einstein centenary survey},
S.W. Hawking and W. Israel eds., Cambridge University Press,
1979.

\medskip \noindent
Weinberg, S., {\it Gravitation and cosmology: principles and applications
of the general theory of relativity}, J. Wiley, New York, 1972.

\medskip \noindent
Wybourne, B.G., {\it Classical groups for physicists},
J. Wiley, New York, 1973.

\medskip \noindent
Wightman, A.S., Phys. Rev. {\bf 101} (1956) 860.

\medskip \noindent
Williams, R.M. and P.A. Tuckey, Cl. Q. Grav. {\bf 9} (1992) 1409.

\medskip \noindent
Witten, E., Nucl. Phys. {\bf B 311} (1988) 46.

\medskip \noindent
Zakharov, V.D., {\it Gravitational waves in Einstein's theory},
J. Wiley, New York, 1974.

\end